\documentclass[pre,twocolumn,nobalancelastpage,amsmath,amssymb,showpacs,
nofootinbib, superscriptaddress]{revtex4-1}

\usepackage{graphics}      
\usepackage{graphicx}      
\usepackage{longtable}     
\usepackage{url}           
\usepackage{bm}            
\usepackage{xcolor}
\usepackage{amsmath}
\usepackage{dcolumn}

\begin{document}
\title{Shear-Driven Flow of Athermal, Frictionless, Spherocylinder Suspensions in Two Dimensions: Stress, Jamming, and Contacts}
\author{Theodore A. Marschall}
\affiliation{Department of Physics and Astronomy, University of Rochester, Rochester, NY 14627}
\author{S. Teitel}
\affiliation{Department of Physics and Astronomy, University of Rochester, Rochester, NY 14627}
\date{\today}

\begin{abstract}
We use numerical simulations to study the flow of a bidisperse mixture of athermal, frictionless, soft-core two dimensional spherocylinders driven by a uniform steady state  shear strain applied at a fixed finite rate.  Energy dissipation occurs via a viscous drag with respect to a uniformly sheared host fluid, giving a simple model for flow in a non-Brownian suspension and resulting in  a Newtonian rheology.  We study the resulting pressure $p$ and deviatoric shear stress $\sigma$ of the interacting spherocylinders as a function of packing fraction $\phi$, strain rate $\dot\gamma$, and a parameter $\alpha$ that measures the asphericity of the particles; $\alpha$ is varied to consider the range from nearly circular disks to elongated rods.  We consider the direction of anisotropy of the stress tensor, the macroscopic friction $\mu=\sigma/p$, and the divergence of the transport coefficient $\eta_p=p/\dot\gamma$ as $\phi$ is increased to the jamming transition $\phi_J$.  From a phenomenological analysis of Herschel-Bulkley rheology above jamming, we estimate $\phi_J$ as a function of asphericity $\alpha$ and show that the variation of $\phi_J$ with $\alpha$ is the main cause for differences in rheology as $\alpha$ is varied; when plotted as $\phi/\phi_J$ rheological curves for different $\alpha$ qualitatively agree.  However a detailed scaling analysis of the divergence of $\eta_p$ for our most elongated particles suggests that the jamming transition of spherocylinders may be in a different universality class than that of circular disks.  We also compute the number of contacts per particle $Z$ in the system and show that the value at jamming $Z_J$ is a non-monotonic function of $\alpha$ that is always smaller than the isostatic value.  We measure the probability distribution of contacts per unit surface length $\mathcal{P}(\vartheta)$ at polar angle $\vartheta$ with respect to the spherocylinder spine, and find that as $\alpha\to 0$ this distribution seems to diverge at $\vartheta=\pi/2$, giving a finite limiting probability for contacts on the vanishingly small flat sides of the spherocylinder.  Finally we consider the variation of the average contact force as a function of location on the particle surface.
\end{abstract}
\maketitle

\section{Introduction}
\label{sec:intro}

In a system of athermal granular particles with only repulsive contact interactions, as the packing fraction of particles $\phi$ increases, the system undergoes a jamming transition \cite{OHern,LiuNagel} at a critical $\phi_J$.  For $\phi<\phi_J$ the system behaves similarly to a liquid, while for $\phi>\phi_J$ the system behaves like a rigid but disordered solid.  Since one is dealing with athermal particles ($T=0$), details of the jamming transition may in principle depend on the physical protocol by which the system jams.  It is useful to distinguish between two different types of jamming, {\em compression-driven} and {\em shear-driven} jamming.  

In compression-driven jamming \cite{OHern,Wyart0} the particle packing $\phi$ is increased by isotropically compressing the system.  As $\phi$ increases, particles come into contact with each other.  At a critical $\phi_J$ a mechanically stable rigid backbone of particles percolates across the system, the system pressure $p$ becomes finite, and the system jams.  For frictionless particles, $p$ increases continuously \cite{OHern} from zero as $\phi$ increases above $\phi_J$.  Since the compression is isotropic, the total shear stress $\sigma$ in the system, even in the solid state,  in principle vanishes.  It is known that the precise value of $\phi_J$ can vary somewhat with the details of the compression protocol, in particular the ensemble of configurations from which compression begins, and the rate of compression \cite{Chaudhuri,Vagberg.PRE.2011,Ozawa}.

Shear-driven jamming \cite{OlssonTeitelPRL,OlssonTeitelPRE,VagbergOlssonTeitel} occurs when the system is sheared, at constant volume or constant pressure, with a uniform shear strain rate $\dot\gamma$.
In a steady state at sufficiently small $\phi$ the system will have shear flow like a liquid.  The shear stress $\sigma$ in this liquid is finite for finite $\dot\gamma$, but vanishes as $\dot\gamma\to 0$, resulting in a finite transport coefficient, $\lim_{\dot\gamma\to 0} [\sigma/\dot\gamma^m]$ (with $m=1$ for a system with Newtonian rheology \cite{OlssonTeitelPRE}, and $m=2$ for a system with Bagnoldian rheology \cite{VagbergOlssonTeitel}).
As $\phi$ increases, a critical packing $\phi_J$ is reached such that for $\phi>\phi_J$  the system develops a finite yield stress $\sigma_0$, defined by  $\lim_{\dot\gamma\to 0}[\sigma]=\sigma_0>0$.  This $\phi_J$ is the {\em shear-driven} jamming transition.  For frictionless particles,  shear-driven jamming behaves like a continuous phase transition \cite{OlssonTeitelPRL}:  the transport coefficient 
diverges continuously as $\phi\to\phi_J$ from below, and $\sigma_0$ increases continuously from zero as $\phi$ increases above $\phi_J$.  For $\phi>\phi_J$, if $\sigma<\sigma_0$ the system is in a static jammed solid phase, while for $\sigma>\sigma_0$ the system is in a yielded flowing plastic phase.  The precise value of $\phi_J$ is independent of the initial configuration from which the system begins to be sheared \cite{Vagberg.PRE.2011}.  Our work in this paper will concern this shear-driven jamming transition.

Most numerical studies of the jamming transition, and granular materials more generally, have used spherical shaped particles for simplicity.  
It is therefore interesting to ask how behavior may be modified if the particles have shapes with a lower rotational symmetry \cite{Borzsonyi.Soft.2013}.  
Several recent numerical and experimental works have explored the effect of non-spherical shape on  compression-driven jamming. 
Such works have included studies of monodisperse distributions of aspherical ellipsoids \cite{Donev.PRL.2004,Donev.Science.2004,Man.PRL.2005,Donev.PRE.2007}, oblate ellipsoids \cite{Donev.Science.2004,Man.PRL.2005,Donev.PRE.2007}, and prolate ellipsoids \cite{Donev.Science.2004,Man.PRL.2005,Donev.PRE.2007,Sacanna.JPhysC.2007,Zeravcic.EPL.2009,Schreck.PRE.2012} in three dimensions (3D), and bidisperse distributions of ellipses \cite{Donev.PRE.2007,Mailman.PRL.2009,Schreck.PRE.2012,VanderWerf} in two dimensions (2D).
Spherocylinders,  consisting of cylindrical tubes with hemispherical endcaps, have been used to model rod-shaped particles in 3D \cite{Williams2003PRE,Wouterse.JPCM.2007,Azema2010,Azema2012,Zhao} and in 2D \cite{VanderWerf,MarschallCompress}. Other work has considered cut spheres \cite{Wouterse.JPCM.2007} in 3D, as well as particles with non-convex shapes \cite{VanderWerf,MarschallStaples,Saint-Cyr}.  For compression-driven jamming of elongated particles, such as ellipses, ellipsoids and spherocylinders, these works find several common features: (i) the critical jamming packing fraction $\phi_J$ is a non-monotonic function of the particle aspect ratio, increasing as the particle deviates from a sphere, and then decreasing as the particle gets increasingly elongated \cite{Donev.PRL.2004,Donev.Science.2004,Man.PRL.2005,Donev.PRE.2007,Sacanna.JPhysC.2007,VanderWerf,Williams2003PRE,Wouterse.JPCM.2007,Azema2010,Zhao,MarschallCompress}; (ii) particle packings at $\phi_J$ are hypostatic, with the average number of contacts per particle $Z_J < 2d_f$ where $d_f$ is the  number of degrees of freedom of a particle, as determined by its rotational symmetries \cite{Donev.Science.2004,Donev.PRE.2007,Zeravcic.EPL.2009,Schreck.PRE.2012,Mailman.PRL.2009,VanderWerf,Williams2003PRE,Wouterse.JPCM.2007,Azema2010,Zhao,MarschallCompress}; (iii) unlike particles in thermal equilibrium \cite{Onsager,Bolhuis}, isotropically compressed athermal particles show no long range orientational order upon jamming \cite{Donev.Science.2004,Man.PRL.2005,Sacanna.JPhysC.2007,Zhao,MarschallCompress}

The question of aspherical particles in steady state shear flow has only been considered more recently.  Unlike uniformly compressed systems, uniformly sheared systems do show orientational ordering due to torques induced on the particles by the shear flow.  Several numerical works focused on this shear-induced orientational ordering of ellipsoids \cite{Campbell} and rod-shaped particles \cite{Guo1,Guo2} of different aspect ratios in 3D approaching, but staying below, jamming.  They found that orientational order increased with increasing packing $\phi$, and particles were oriented at a finite angle $\theta_2>0$ with respect to the direction of the shear flow.   Experiments and simulations of rod-shaped particles in 3D \cite{Borzsonyi1,Borzsonyi2,Wegner,Wegner2} found similar results, while also studying the rotation of particles in steady state shear, and the transient approaches to the steady state.  Other experimental works have studied the transient behavior of orientational ordering and pressure $p$ of  ellipses in 2D under quasistatic shearing \cite{Farhadi,Wang}.  Numerical simulations measuring  the dependence of the jamming packing  $\phi_J$, the average number of contacts per particle $Z_J$, and particle orientation as a function of particle aspect ratio, and the rheological macroscopic friction $\mu=\sigma/p$ as a function of inertial number $I=\dot\gamma d/\sqrt{p/\rho}$ in the hard-core limit below jamming, have been carried out for frictional 3D spherocylinders  sheared by biaxial compression \cite{Azema2010, Azema2012}, frictionless 3D spherocylinders  in steady state simple shear \cite{Nagy}, and  both frictionless and frictional 2D ellipses in steady state simple shear \cite{Trulsson}.  The rheology of 3D frictional and frictionless spherocylinders in steady simple shear has also recently been simulated \cite{Nath}.

In this work  we consider the uniform steady state  shearing of a system of 2D spherocylinders with varying aspect ratio.  The above previous works \cite{Campbell, Guo1, Guo2, Borzsonyi1,Borzsonyi2,Wegner,Wegner2,Azema2010, Azema2012,Nagy,Trulsson,Nath}  modeled dry granular materials, in which energy is dissipated in particle collisions, the rheology is Bagnoldian, and there may be microscopic inter-particle Coulombic friction.  The presence of microscopic inter-particle friction, in particular, is known to have a significant effect on many features of dry granular particle rheology \cite{Cruz,Bi,Otsuki,Pica,Bouzid,Saw}.

In contrast, here we model particles in suspension where rheology is Newtonian.
We use a simple model consisting of frictionless, soft-core, elastic particle interactions, with a viscous drag with respect to the suspending medium, and overdamped motion in which inertial effects are ignored.  This is a simplification compared to real physical suspensions, which may include hydrodynamic forces \cite{hydro}, lubrication forces \cite{lub1,lub2,lub3}, and inertial effects \cite{inertia}.  More recently, frictional contact interactions have been proposed to become important when the lubrication layer breaks down upon close contact of particles near jamming, and this has been proposed as a possible mechanism for  shear thickening \cite{DST0,DST1,DST2,DST3,DST4,DST5,DST6}.   

The model in our present work ignores these complications.  
However just as frictionless models have played an important theoretical role in the study of granular systems of spherical particles
\cite{OHern,LiuNagel,Wyart0,Chaudhuri,Vagberg.PRE.2011,Ozawa,OlssonTeitelPRL,OlssonTeitelPRE,VagbergOlssonTeitel,Peyneau,Andreotti,Wyart1,Vagberg.PRL.2014,Wyart2,Berthier},
it is of interest to see what results when the same model is applied to non-spherical particles.
The greater simplicity of our model allows a more thorough investigation over a wide range of the parameter space, in particular  going to lower values of the strain rate $\dot\gamma$ and smaller values of the particle asphericity $\alpha$.  At the same time, our use of a common simple model  allows direct comparison with our earlier work on compression-driven jamming in this same system \cite{MarschallCompress}.  Our work is carried out in the spirit that it is useful to first understand the behavior of simple models before adding more realistic complexities.

In the present paper we  focus on rheological and structural aspects of our system as a function of particle asphericity $\alpha$, packing fraction $\phi$, and shear strain rate $\dot\gamma$.   In a companion paper \cite{MT2} we will focus on the orientational and translational ordering of particles and particle rotations; some of our results on this latter topic have already been presented \cite{MKOT}.
Among other results  we carry out a critical scaling analysis that suggests the shear-driven jamming transition for Newtonian spherocylinders  may be in a different universality class than that of spherical particles.  We compute the packing fraction for shear-driven jamming  $\phi_J$, as well as the average number of contacts per particle at jamming $Z_J$, as a function of particle asphericity and make a direct comparison to results for compression-driven jamming.  We find that the system is always hypostatic with a number of contacts smaller than the isostatic value, $Z_J < Z_\mathrm{iso}$.  
We consider the Herschel-Bulkley rheology for $\phi>\phi_J$ and show that the empirically determined  exponent $n$, which characterizes the $\dot\gamma$ dependence, varies with both $\phi$ and particle asphericity $\alpha$, and that $n$ in general takes different values for the pressure $p$ and the deviatoric shear stress $\sigma$.  We compute the  viscosities $p(\phi)/\dot\gamma$ and $\sigma(\phi)/\dot\gamma$ and show that the main effect of differing  particle asphericities can be explained in terms of the  shift in $\phi_J$ as $\alpha$ varies. 
We also consider the distribution of particle contact locations around the surface of the particle, and find that for small $\alpha$ this distribution strongly peaks along the particle's flat sides; the total probability for the contact to lie somewhere on the flat sides stays constant even as $\alpha\to 0$ and the particles become circular, thus indicating that the $\alpha\to0$ limit is singular.  The remainder of this paper is organized as follows.  In Sec.~II we define our model and the quantities to be computed.  In Sec.~III we present our numerical results.  In Sec.~IV we summarize our conclusions.

\section{Model and Simulation Method}
\label{sec:modelMethod}

\subsection{Model}
\label{sec:Model}

\begin{figure}
\centering
\includegraphics[width=2.5in]{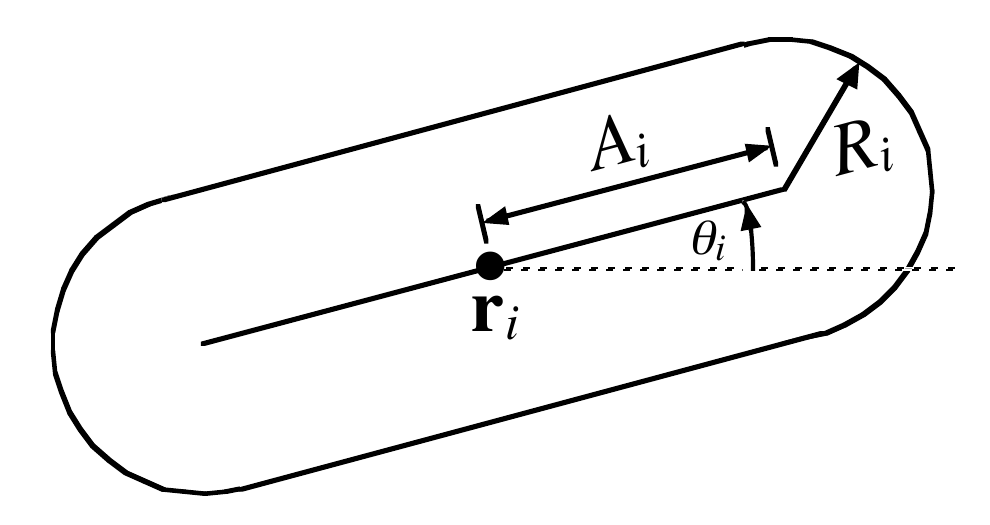}
\caption{ An isolated spherocylinder indicating the spine half-length $A_i$, end cap radius $R_i$, center of mass position $\mathbf{r}_i$, and angle of orientation $\theta_i$. 
}
\label{sphero} 
\end{figure}

A two dimensional spherocylinder consists of a rectangle with two circular end caps, as illustrated in Fig.~\ref{sphero}.  We  denote the half length of the rectangular part of spherocylinder $i$ by $A_i$, and the radius of the end cap, which is also the half width of the rectangle, by $R_i$. We will refer to   the axis of length $2A_i$, which goes down the center of the rectangle, as the ``spine" of the spherocylinder. For every point on the perimeter of the spherocylinder, the shortest distance from the spine is $R_i$.   We define the asphericity of the spherocylinder as,
\begin{equation}
\alpha_i=A_i/R_i
\label{ealpha}
\end{equation}
so that $\alpha=0$ describes a circular particle, and the length-to-width aspect ratio is $1+\alpha$.  We define the center of mass position of the particle as $\mathbf{r}_i=(x_i,y_i)$, and the orientation of the particle with respect to the flow direction along $\mathbf{\hat x}$ as $\theta_i$, as shown in Fig.~\ref{sphero}.

\begin{figure}
\centering
\includegraphics[width=2.3in]{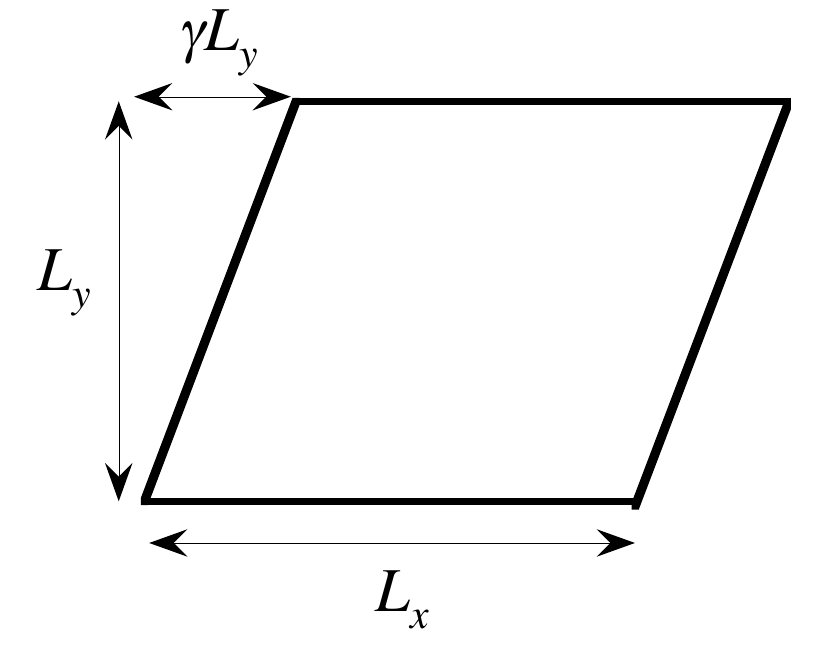}
\caption{Unit cell of our numerical system of length $L_x$ and height $L_y$.  Periodic boundary conditions are taken along $\mathbf{\hat x}$ while Lees-Edwards boundary conditions with shear strain $\gamma$ are taken along $\mathbf{\hat y}$.
}
\label{box} 
\end{figure}

Our system consists of $N$ spherocylinders in a box of fixed total area with length $L_x$ and height $L_y$.   In the flow direction $\mathbf{\hat x}$ we use periodic boundary conditions, while in the transverse direction $\mathbf{\hat y}$ we use Lees-Edwards boundary conditions \cite{LeesEdwards} to introduce a simple shear strain $\gamma$. Our system can therefore be viewed as a periodic tiling of space with the rhombic unit cell shown in Fig.~\ref{box}.   If $\mathcal{A}_i$ is the area of particle $i$, then the packing fraction of the system is,
\begin{equation}
\phi = \frac{1}{L_x L_y} \sum_{i=1}^{N} \mathcal{A}_i,
\label{ephi}
\end{equation}
where for spherocylinders, 
\begin{equation}
\mathcal{A}_i=\pi R_i^2+4A_iR_i = R_i^2(\pi + 4\alpha_i).
\label{eArea}
\end{equation}
In this work we take $L_x=L_y\equiv L$, and consider only systems in which all of the particles have the same asphericity $\alpha$.  
We take a bidisperse  distribution of particle sizes to prevent crystallization,  using equal numbers of big and small particles where the ratio of the big radius to the small radius is $R_b / R_s = 1.4$.

\begin{figure}
\centering
\includegraphics[width=3.5in]{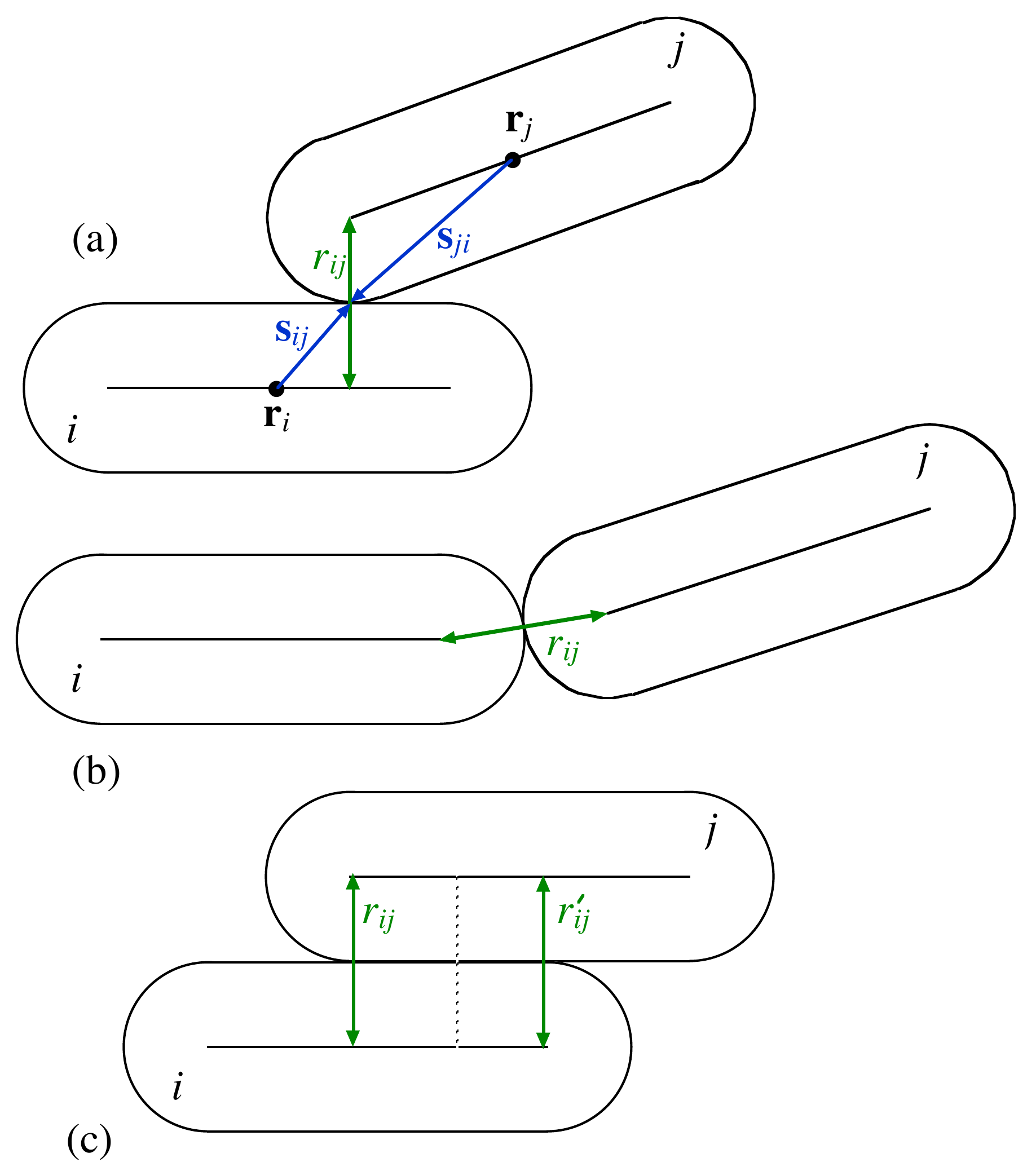}
\caption{ Geometry of  spherocylinder contacts:  (a) Two spherocylinders in tip-to-side contact, indicating the minimal spine separation $r_{ij}$ and the moment arms $\mathbf{s}_{ij}$ and $\mathbf{s}_{ji}$.
(b) Two spherocylinders in tip-to-tip contact. (c) Two spherocylinders in side-to-side contact; if the lengths $r_{ij}$ and $r^\prime_{ij}$ are equal within our numerical accuracy, then we take the location of the contact to be midway between, as illustrated by the dashed line.
}
\label{rodConfigs}
\end{figure}

Our particles will move under the influence of elastic soft-core contact forces and a viscous drag force.  The elastic forces arise when particles come into physical contact with each other.
Two spherocylinders $i$ and $j$ come into contact when the shortest distance between their spines, $r_{ij}$, is less than the sum of their radii $d_{ij} = R_i + R_j$.
An efficient algorithm for determining this distance $r_{ij}$ is given in Ref.~ \cite{Pournin.GranulMat.2005}.
When $r_{ij} < d_{ij}$, the contact between the spherocylinders may be classified as one of three types, as illustrated in Figs. \ref{rodConfigs}(a), \ref{rodConfigs}(b), and \ref{rodConfigs}(c), respectively: (i) tip-to-side, (ii) tip-to-tip, or (iii) side-to-side contact. 
We regard a contact as being side-to-side whenever the  distances of two spine tips on opposite ends of the spherocylinders to the spine of the other spherocylinder, indicated as $r_{ij}$ and $r^\prime_{ij}$ in Fig.~\ref{rodConfigs}(c),  are both less than $d_{ij}$, so that there is overlap down an extended length of the spherocylinders' flat side. If one of these lengths is measurably smaller than the other, say $r_{ij}<r^\prime_{ij}$, we take the point of contact to be at that position; if to our numerical accuracy these lengths are the same, which occurs when the two spherocylinders are parallel to an accuracy $ | \theta_i - \theta_j | \lesssim 10^{-8}$, then we put the 
point of contact to be midway between, as illustrated by the dashed line in Fig.~\ref{rodConfigs}(c).

Once the contacts have been identified, we define the elastic energy in our system using a one-sided repulsive harmonic interaction.
The total elastic energy is given by,
\begin{equation}
U^{\mathrm{el}} =\frac{1}{2}\sum_{i=1}^N\sideset{}{^\prime} \sum_{j} k_e (1 - r_{ij} / d_{ij})^2,
\end{equation}
where the primed sum is over only particles $j$ in contact with $i$,  i.e., with $r_{ij} < d_{ij}$.

The elastic force on particle $i$ due to contact with $j$  is given by,
\begin{equation}
\mathbf{F}_{ij}^\mathrm{el}= -\frac{\partial U^\mathrm{el}}{\partial \mathbf{r}_i} = (k_e / d_{ij}) (1 - r_{ij} / d_{ij}) \hat{\mathbf{r}}_{ij},
\label{eFij}
\end{equation}
where $\mathbf{\hat r}_{ij}$ is the normal pointing inwards to particle $i$ along the bond $r_{ij}$, and the force acts at the point of contact, specifically at a distance $(R_i/d_{ij})r_{ij}$ from the spine of particle $i$.  
The total elastic force on the center of mass of particle $i$ is then,
\begin{equation}
\mathbf{F}_i^{\mathrm{el}} = \sideset{}{^\prime} \sum_j \mathbf{F}_{ij}^{\mathrm{el}}.
\label{eFel}
\end{equation}
The elastic forces also give a torque on particle $i$,
\begin{equation}
\tau^{\mathrm{el}}_{i} = \mathbf{\hat z}\cdot \sideset{}{^\prime} \sum_j \mathbf{s}_{ij} \times \mathbf{F}_{ij}^{\mathrm{el}},
\label{etauel}
\end{equation}
where $\mathbf{s}_{ij}$ is the moment arm from the center of mass $\mathbf{r}_i$ of spherocylinder $i$ to the point of contact with spherocylinder $j$, as illustrated in Fig.~\ref{rodConfigs}(a).

Our model is one of non-Brownian particles in suspension.  We will take the local average velocity of the host medium at position $\mathbf{r}$ to be $\mathbf{v}_\mathrm{host}(\mathbf{r})$.  Using a simple model \cite{MarschallCompress,MarschallStaples}, that ignores hydrodynamic interactions but is expected to be good for large  particle densities, we assume a  local dissipative drag force per unit area acting at position $\mathbf{r}$ on particle $i$ to be,
\begin{equation}
\mathbf{f}_i^\mathrm{dis}(\mathbf{r})=-k_d[\mathbf{v}_i(\mathbf{r}) - \mathbf{v}_\mathrm{host}(\mathbf{r})],
\label{efrdis}
\end{equation}
where $\mathbf{v}_i(\mathbf{r})$ is the local velocity of the particle at position $\mathbf{r}$,
\begin{equation}
\mathbf{v}_i(\mathbf{r})=\dot{\mathbf{r}}_i+\dot\theta_i \mathbf{\hat z}\times (\mathbf{r}-\mathbf{r}_i).
\label{evr}
\end{equation}
Here $\dot{\mathbf{r}}_i=d\mathbf{r}_i/dt\equiv\mathbf{v}_i$ is the center of mass velocity of particle $i$ and $\dot\theta_i$ is the angular velocity about the center of mass. 
Integrating over the area of the particle, we then get the total dissipative force on particle $i$,
\begin{equation}
\mathbf{F}_i^{\mathrm{dis}} = \int_i d^2r\, \mathbf{f}_i^{\mathrm{dis}}(\mathbf{r}) ,
\label{efdis}
\end{equation}
and the total dissipative torque,   
\begin{equation}
\tau_i^{\mathrm{dis}} = \mathbf{\hat z} \cdot \int_i d^2r\left[ ( \mathbf{r}-\mathbf{r}_i) \times \mathbf{f}_i^{\mathrm{dis}}(\mathbf{r}) \right] .
\label{etaudis}
\end{equation}

In this work we study behavior in a simple shear flow under uniform constant shear strain rate $\dot\gamma$.  We therefore take the shear strain $\gamma$ that enters our Lees-Edwards boundary conditions and increase it with time according to $\gamma(t)=\dot\gamma t$, and assume a simple shear form for the velocity of the host medium,
\begin{equation}
\mathbf{v}_\mathrm{host}(\mathbf{r}) = \dot\gamma y \mathbf{\hat x}.
\label{evhost}
\end{equation}
For this case the dissipative force of Eq.~(\ref{efdis}) simplifies to,
\begin{equation}
\mathbf{F}_i^{\mathrm{dis}} = -k_d \mathcal{A}_i [\dot{\mathbf{r}}_i - \dot{\gamma} y_i  \mathbf{\hat{x}}], \\
\label{efdisShear}
\end{equation}
just as in the mean-field Durian bubble model \cite{Durian}.   Such a dissipative force has been used in many previous works  \cite{OlssonTeitelPRL,OlssonTeitelPRE,Vagberg.PRE.2011,Andreotti,Wyart1,Vagberg.PRL.2014,Wyart2,Berthier} to study shear driven jamming of spherical particles.   

In the Appendix we show that the dissipative torque on particle $i$, given by Eq.~(\ref{etaudis}), can be written in terms of the components of its moment of inertia tensor.  If $\theta_i$ is the orientation of the eigenvector, corresponding to the smaller eigenvalue of the moment of inertia tensor, with respect to the flow direction (this is just the orientation of the spine for spherocylinders), then
\begin{equation}
\tau_i^\mathrm{dis}=-k_d\mathcal{A}_i I_i\left[\dot\theta_i +\dot\gamma f(\theta_i)\right],
\label{etaudisShear}
\end{equation}
with
\begin{equation}
f(\theta)=\frac{1}{2}\left[1-\left({\Delta I_i}/{I_i}\right)\cos 2\theta\right],
\label{eftheta}
\end{equation}
where $I_i$ is the sum of the two eigenvalues of the moment of inertia tensor of particle $i$, and $\Delta I_i$ is the absolute value of their difference.  The values of $I_i$ and $\Delta I_i$ for  spherocylinders of asphericity $\alpha$ are given in the Appendix; here we note that $\Delta I_i=0$ for circular particles with $\alpha=0$, as required by symmetry.

The above elastic and dissipative forces are the only forces included in our model; there are no inter-particle dissipative or frictional forces.  We will carry out our simulations in the overdamped (low particle mass) limit, where the total force and torque on each particle are damped to zero, 
\begin{align}
\mathbf{F}_i^{\mathrm{el}} + \mathbf{F}_i^{\mathrm{dis}} &= 0, 
\label{eFitotal}\\
\tau_i^{\mathrm{el}} + \tau_i^{\mathrm{dis}} &= 0.
\label{etautotal}
\end{align} 
Using Eqs.~(\ref{efdisShear}) and (\ref{etaudisShear}) we then get for the translational and orientational equations of motion for particle $i$,
\begin{align}
\dot{\mathbf{r}}_i &=    \dot{\gamma}y_i{\mathbf{\hat x}}+\dfrac{\mathbf{F}_i^{\mathrm{el}}}{k_d \mathcal{A}_i},
\label{eq:ri_eom} \\
\dot{\theta}_i &= - \dot{\gamma} f(\theta_i)+ \dfrac{\tau_i^{\mathrm{el}}}{k_d I_i \mathcal{A}_i}.
\label{eq:theta_eom}
\end{align}

Note that, from the above equations of motion, an isolated particle with $\mathbf{F}_i^\mathrm{el}=0$ and $\tau_i^\mathrm{el}=0$, will just translate at the local shear flow velocity $\dot\gamma y_i\mathbf{\hat x}$, while rotating clockwise  with non-uniform angular velocity $\dot\theta_i=-\dot\gamma f(\theta_i)$.  For a circular particle with $\Delta I_i= \alpha=0$, the rotation is uniform with $\dot\theta_i=-\dot\gamma/2$.  For non-circular particles with $\Delta I_i\ne 0$, the particle will tumble non-uniformly, and rotational motion is analogous to the Jeffery orbits of a non-spherical particle in suspension in a shear flow \cite{Jeffery.RSPA.1922}; rotation is slowest when the particle is aligned parallel to the flow direction with $\theta_i=0$,  fastest when the particle is aligned transverse to the flow direction with $\theta_i=90^\circ$, and the steady state probability to find the particle at orientation $\theta$ is $\mathcal{P}(\theta)\propto 1/f(\theta)$.

For our simulations we will take $2 R_s = 1$ as the unit of distance, $k_e = 1$ as the unit of energy, and $t_0 = (2 R_s)^2 k_d / k_e = 1$ as the unit of time.
We numerically integrate the equations of motion (\ref{eq:ri_eom}) and (\ref{eq:theta_eom}) using a two-stage Heun method with a step size of $\Delta t = 0.02$.
Unless stated otherwise, we begin each shearing run in a finite energy configuration at the desired packing fraction $\phi$ with random initial positions and orientations.
To generate such initial configurations we place the spherocylinders in the system one-by-one, while rejecting and retrying any time a new placement would lead to an unphysical overlap where the spines of two spherocylinders  intersect.  We use $N=1024$ particles.
Most of our simulations extend to total strains of $\gamma\gtrsim 150$;  for our slowest $\dot\gamma=10^{-7}$  we strain only to $\gamma\approx 50$.  Such large strains are desirable to make sure that the rotational degrees of freedom are well equilibrated.  Discarding an   initial $\Delta\gamma\approx 20$ of the strain  from the averaging so as to eliminate transients effects, we find that our steady state averages are generally insensitive to the particular starting configuration \cite{Vagberg.PRE.2011}.
Note that we restrict the strain coordinate $\gamma$ used in our Lees-Edwards boundary condition  to the range $\gamma\in \left(-\frac{L_x}{2L_y}, \frac{L_x}{2L_y}\right]$; whenever it exceeds this maximum it is reset by taking $\gamma \to \gamma - \frac{L_x }{ L_y}$, allowing us to shear to arbitrarily large total strains.  Our simulations use a range of strain rates from $\dot\gamma=10^{-4}$ down to $10^{-6}$ for all $\alpha$; for $\alpha=0.03$ and $4$ we go down to $\dot\gamma=4\times 10^{-7}$, and for $\alpha=0.01$ and $0.001$ we go to $\dot\gamma=10^{-7}$.

\subsection{Stress}
\label{sec:Stress}

In this work we will be concerned with the stress that results from shearing the system.  We will ignore the constant isotropic pressure from the host medium and  consider only the stress arising from the particles.  There will be a contribution to the particle stress tensor from both the elastic and the dissipative forces.  

The elastic part of the stress tensor $\mathbf{p}$ is defined as usual \cite{Ball},
\begin{equation}
\mathbf{p}^\mathrm{el}=-\frac{1}{L_xL_y}\sum_{i=1}^N \boldsymbol{\Sigma}_i^\mathrm{el},\quad
\boldsymbol{\Sigma}_i^\mathrm{el}={\sum_j}^\prime\mathbf{s}_{ij}\otimes \mathbf{F}_{ij}^\mathrm{el},
\label{eSigeli}
\end{equation}
where the primed sum is over all particles $j$ in contact with $i$.  The dissipative part can be written as,
\begin{equation}
\mathbf{p}^\mathrm{dis}=-\frac{1}{L_xL_y}\sum_{i=1}^N \boldsymbol{\Sigma}_i^\mathrm{dis},\quad
\boldsymbol{\Sigma}_i^\mathrm{dis}=\int_id^2r\,(\mathbf{r}-\mathbf{r}_i)\otimes \mathbf{f}_i^\mathrm{dis}(\mathbf{r}),
\end{equation}
where the integral is over the area of particle $i$.  In the Appendix we show that, 
\begin{equation}
\boldsymbol{\Sigma}_i^\mathrm{dis}
=\kappa
\left[
\begin{array}{cc}
 (\dot\theta_i+\dot\gamma)\frac{\Delta I_i}{I_i}\sin 2\theta_i &  -\dot\theta_i (1+\frac{\Delta I_i}{I_i}\cos 2\theta_i)   \\[10pt]
 (\dot\theta_i+\dot\gamma)(1-\frac{\Delta I_i}{I_i}\cos 2\theta_i) & -\dot\theta_i \frac{\Delta I_i}{I_i}\sin 2\theta_i     
\end{array}
\right],
\label{eSigdisi}
\end{equation}
with $\kappa=k_d\mathcal{A}_i I_i/2$.

We note that since the torques $\tau_i^\mathrm{el,dis}$ are related to the force moment tensors $\boldsymbol{\Sigma}_i^\mathrm{el,dis}$ by
\begin{equation}
\tau_i^\mathrm{el,dis}=\boldsymbol{\Sigma}_{i,xy}^\mathrm{el,dis}-\boldsymbol{\Sigma}_{i,yx}^\mathrm{el,dis},
\end{equation}
and since $\tau_i^\mathrm{el}$ and $\tau_i^\mathrm{dis}$ in general do not separately vanish, then $\mathbf{p}^\mathrm{el}$ and $\mathbf{p}^\mathrm{dis}$ are not separately  symmetric tensors; however because of our overdamped equation of motion (\ref{etautotal}), the total torque $\tau_i^\mathrm{el}+\tau_i^\mathrm{dis}$ does vanish and so the total stress $\mathbf{p}=\mathbf{p}^\mathrm{el}+\mathbf{p}^\mathrm{dis}$ is symmetric.

While we include the dissipative part of the stress in our calculations, we note that it  
is generally small, around $\| \mathbf{p}^{\mathrm{dis}} \| \lesssim 10^{-7}$ for all densities, shear rates, and aspect ratios that we study.  This is generally smaller than the elastic part except for very dilute systems; near to the jamming transition it is negligible compared to the elastic part.

Measuring the stress tensor $\mathbf{p}$ for individual configurations using Eqs.~(\ref{eSigeli}--\ref{eSigdisi}), we average it over our ensemble of sheared states  to compute $\langle\mathbf{p}\rangle$.  From this we find the pressure,
\begin{equation}
p = [\langle p_{xx}\rangle + \langle p_{yy}\rangle]/2.
\end{equation}
Since $p$ is  the trace of $\mathbf{p}$, it is an invariant of the stress tensor under rotation of the coordinate system.
We are also interested in the shear stress $\sigma_{xy} =- \langle p_{x y}\rangle$.
However, since the shear stress is not an invariant of the coordinate system, it is  useful to look at the deviatoric shear stress, $\sigma$, which is defined as half the difference between the eigenvalues of the stress tensor.
The deviatoric stress is given by
\begin{equation}
\sigma = \sqrt{\frac{1}{4}[\langle p_{x x}\rangle - \langle p_{y y}\rangle]^2 +  \langle p_{x y}\rangle^2}.
\end{equation}

In the flowing liquid-like phase below jamming, where the stress vanishes as $\dot\gamma\to 0$, it will be useful to characterize the state of the system by considering the transport coefficients, the shear viscosity $\eta$ and its analog for pressure $\eta_p$,
\begin{equation}
\eta \equiv \sigma/\dot\gamma,\quad \eta_p\equiv p/\dot\gamma .
\end{equation}
Since our rheology is Newtonian, with $\mathbf{p}\propto\dot\gamma$ at sufficiently small $\dot\gamma$, we expect that as $\dot\gamma\to 0$ below jamming ($\phi<\phi_J$), $\eta\to \eta(\phi)$ and $\eta_p\to \eta_p(\phi)$ become functions only of the packing $\phi$  and that they diverge as $\phi\to\phi_J$ from below.  Above jamming ($\phi>\phi_J$), as $\dot\gamma\to 0$,
we expect $p\to p_0(\phi)$ and $\sigma\to\sigma_0(\phi)$ the finite yield stresses, and that these vanish as $\phi\to\phi_J$ from above.

It is also useful to consider the macroscopic friction coefficient,
\begin{equation}
\mu\equiv\sigma/p.
\end{equation}
Even though our particles have no microscopic frictional interactions, the macroscopic friction is nevertheless finite.  Above the jamming $\phi_J$, $\lim_{\dot\gamma\to 0}\mu=\sigma_0/p_0$ measures the ability of the jammed solid to support a finite shear stress without flowing.  Below jamming, $\mu$ measures the ratio of shear stress to pressure in the flowing liquid state, where both $\sigma$ and $p$  are proportional to $\dot\gamma$.  As $\dot\gamma\to 0$, the macroscopic friction takes a finite value $\mu_J$ exactly at the jamming $\phi_J$, which we will see depends on the asphericity $\alpha$ of the particles.

Finally, we note that the eigen-directions of the stress tensor will not in general align with those of the imposed strain tensor, i.e. along the diagonals $\mathbf{\hat x}\pm\mathbf{\hat y}$ of the system.  If if $\mathbf{\hat e_\pm}$ are the  orthonormal eigenvector directions corresponding to the two eigenvalues of the stress tensor $p_\pm=p\pm\sigma$, then we define the orientation of the minimal stress axis $\mathbf{\hat e}_-$ with respect to the flow direction $\mathbf{\hat x}$ by the angle $\theta_-$, where $\cos\theta_-=\mathbf{\hat e}_-\cdot\mathbf{\hat x}$, and
\begin{equation}
\theta_{-}\!  =\tan^{-1} \!\! \left( \frac{ \frac{1}{2}[\langle p_{yy}\rangle \! - \langle p_{xx}\rangle] - \sigma}{ \langle p_{x y}\rangle}\right) \!\! .
\label{ethetaminus}
\end{equation}

The quantity $N_1\equiv\langle p_{yy}\rangle - \langle p_{xx}\rangle$ is referred to as the normal stress difference, and the rheology can be expressed by giving $p$, $\sigma_{xy}$ and $N_1$.  Instead, we will describe the rheology by computing $p$, $\sigma$ and $\theta_-$.  We can relate these by,
\begin{equation}
\frac{N_1}{\sigma_{xy}}= \frac{1}{\tan\theta_-}-\tan\theta_- .
\label{eN1}
\end{equation}
If our system were a uniform continuum, then we would have $N_1=0$ and $\theta_-=45^\circ$ since  our simple shear in the $\hat{\mathbf{x}}$ direction corresponds to  
a compression along $-45^{\circ}$ and expansion along $45^{\circ}$, each at rate $\dot\gamma/2$ so that the area remains constant.

\section{Results}
\label{sec:Results}

\subsection{Stress}
\label{sec:StressR}

\subsubsection{Pressure, Shear Stress, and Transport Coefficients}
\label{secP}

We first consider the pressure $p$ and deviatoric shear stress $\sigma$, and the corresponding transport coefficients $\eta_p$ and $\eta$.  We will consider here behavior for two typical cases: spherocylinders with $\alpha=0.01$, representing particles that are only slightly deviating from circles, and spherocylinders with $\alpha=4$, representing moderately extended rods.  

\begin{figure}
\centering
\includegraphics[width=3.3in]{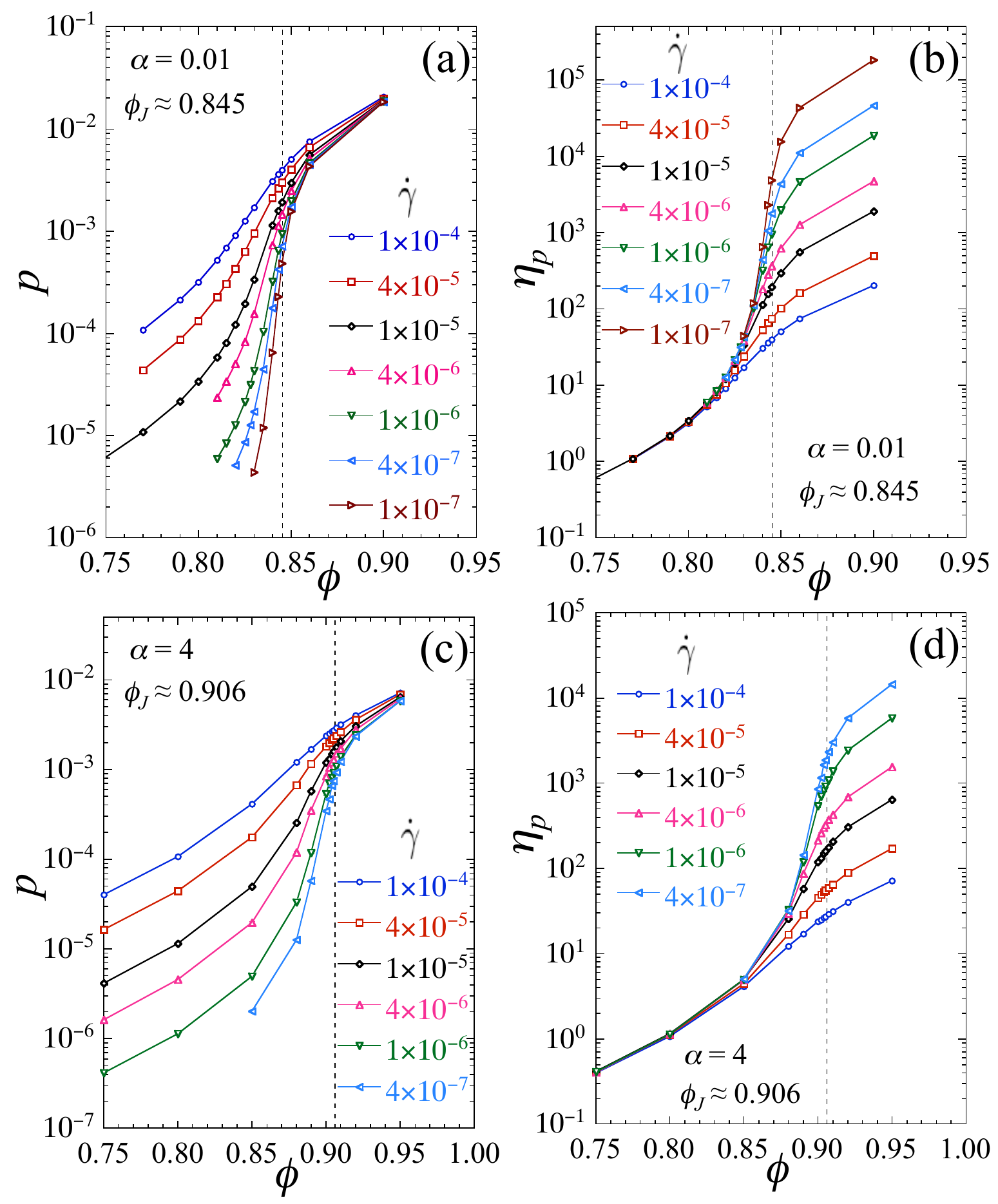}
\caption{(a) Pressure $p$ and (b) pressure transport coefficient $\eta_p\equiv p/\dot\gamma$   vs packing $\phi$ at different shear strain rates $\dot\gamma$ for particles with asphericity $\alpha=0.01$; (c) $p$  and (d) $\eta_p$ vs $\phi$ at different $\dot\gamma$ for particles with $\alpha=4$.  Vertical dashed lines indicate the jamming $\phi_J$.
}
\label{etap-vs-phi} 
\end{figure}

In Fig.~\ref{etap-vs-phi}(a) we plot the pressure $p$ vs $\phi$ for particles with $\alpha=0.01$; in Fig.~\ref{etap-vs-phi}(b) we plot the corresponding transport coefficient $\eta_p\equiv p/\dot\gamma$ vs $\phi$.  Results are shown for different shear strain rates $\dot\gamma$. In Figs.~\ref{etap-vs-phi}(c) and 4(d) we show similar results for $\alpha=4$. In each case the dashed vertical line locates the critical jamming density $\phi_J$, as determined by the analysis in Sec.~\ref{secHB} below.  
Here, and in subsequent plots, error bars represent one standard deviation of estimated statistical error; when error bars are not visible, they are smaller than the size of the symbol representing the data point.

\begin{figure}
\centering
\includegraphics[width=3.3in]{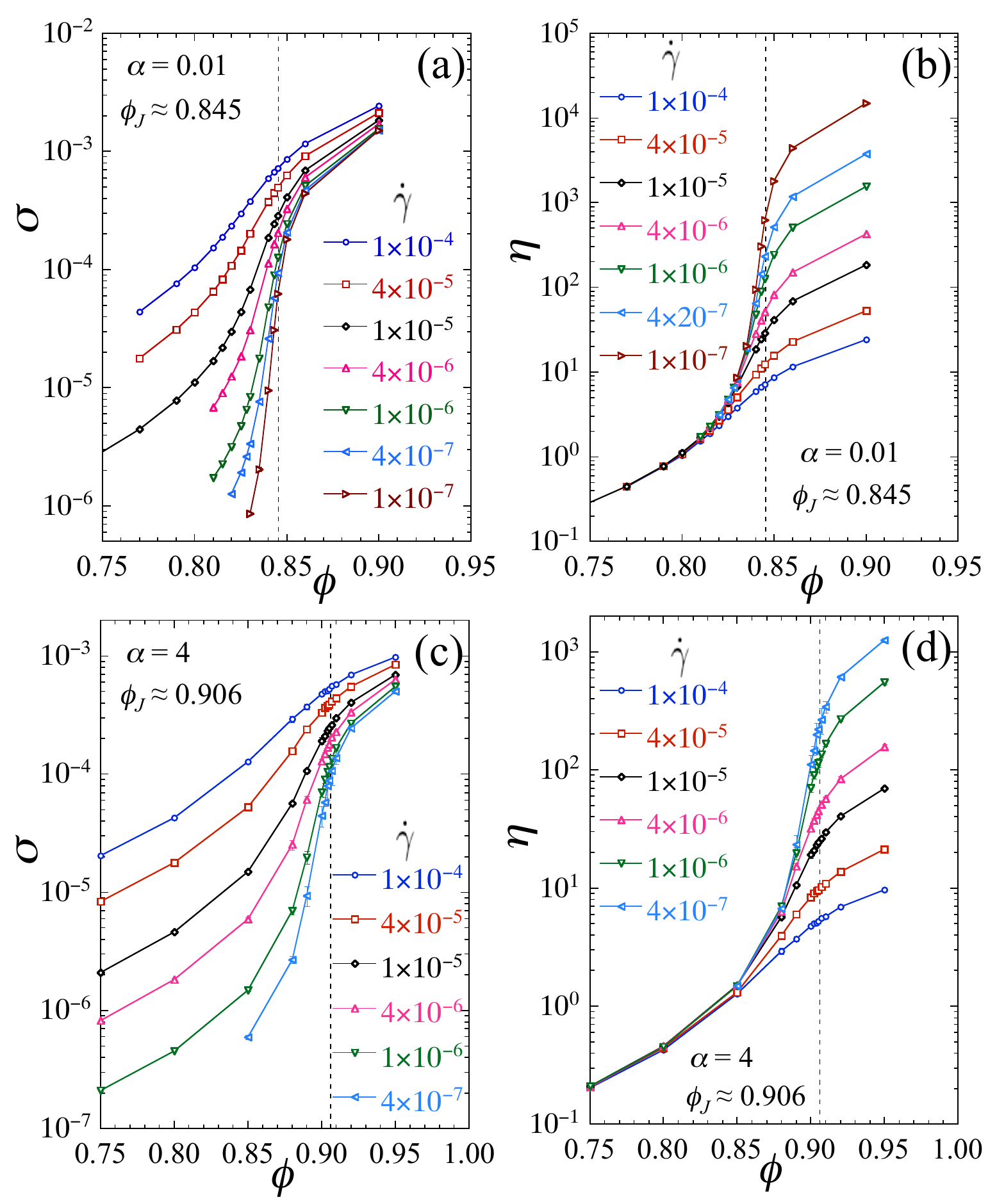}
\caption{(a) Deviatoric shear stress $\sigma$ and (b) shear viscosity $\eta\equiv\sigma/\dot\gamma$ vs packing $\phi$ at different shear strain rates $\dot\gamma$ for particles with asphericity $\alpha=0.01$; (c) $\sigma$  and (d) $\eta$ vs $\phi$ at different $\dot\gamma$ for particles with $\alpha=4$.  Vertical dashed lines indicate the jamming $\phi_J$.
}
\label{eta-vs-phi} 
\end{figure}

The behavior is as expected.  We see in Figs.~\ref{etap-vs-phi}(a) and \ref{etap-vs-phi}(c) that as $\dot\gamma\to 0$, the pressure $p$ appears to be vanishing for $\phi<\phi_J$, while   $p$ is approaching a finite constant, the yield stress $p_0(\phi)$, for $\phi >\phi_J$.  The transport coefficient $\eta_p$ in Figs.~\ref{etap-vs-phi}(b) and \ref{etap-vs-phi}(d) shows analogous behavior.  Since the rheology is Newtonian below jamming, with $p\propto\dot\gamma$ at sufficiently small $\dot\gamma$, for $\phi<\phi_J$ we see that  $\eta_p$ approaches a limiting curve as $\dot\gamma\to 0$, and that this limiting curve appears to be diverging as $\phi\to\phi_J$ from below.  This limiting curve $\eta_p^\mathrm{hc}(\phi)$, given by the upper envelope of the set of curves at finite $\dot\gamma$, represents the limit of hard-core particles where particle overlaps are prohibited.  
For each fixed $\dot\gamma$, the curve of $\eta_p$ vs $\phi$ departs from this limiting curve $\eta_p^\mathrm{hc}(\phi)$ at some particular $\phi_\times(\dot\gamma)$.   The region $\phi>\phi_\times(\dot\gamma)$ is where effects due to the soft-core nature of our particles can no longer be ignored, particle overlaps become measurable, and the divergence found in $\eta_p^\mathrm{hc}(\phi)$ as $\phi\to\phi_J$ gets cut off.   

As $\dot\gamma$ decreases, we see that $\phi_\times(\dot\gamma)$ moves closer to $\phi_J$.  Alternatively, we can invert $\phi_\times(\dot\gamma)$ to define the function $\dot\gamma_\times(\phi)$, which has the following physical meaning.  At fixed $\phi$, for $\dot\gamma<\dot\gamma_\times$ one is in the  region of linear Newtonian rheology with $p\propto\dot\gamma$; but for  $\dot\gamma > \dot\gamma_\times$ one enters a region of non-linear rheology.  We see that as $\phi$ approaches $\phi_J$ from below, $\dot\gamma_\times(\phi)$ decreases towards zero.
For $\phi>\phi_J$, we see from Figs.~\ref{etap-vs-phi}(b) and \ref{etap-vs-phi}(d) that $\eta_p$ steadily increases as $\dot\gamma$ decreases, reflecting the finite yield stress that exists above jamming, i.e., as $\dot\gamma\to 0$, $\eta_p\to p_0(\phi)/\dot\gamma$ diverges.  

In Fig.~\ref{eta-vs-phi} we show similar results, but now for the deviatoric shear stress $\sigma$.  We see the same qualitative behavior as found for the pressure $p$.

\subsubsection{Critical Scaling of Pressure}

The above behaviors of $p$ and $\sigma$, as well as the corresponding $\eta_p$ and $\eta$, can in principle be quantified by a critical scaling equation that describes the jamming point as a continuous phase transition \cite{OlssonTeitelPRE}.  For pressure the critical scaling equation is,

\begin{equation}
p(\phi,\dot\gamma) = \dot\gamma^q g\left(\frac{\phi-\phi_J}{\dot\gamma^{1/z\nu}}\right),
\label{escale}
\end{equation}
where $g(x)$ is a scaling function,  $\nu$ is the correlation length critical exponent, $z$ is the dynamic critical exponent, and $q$ is the exponent of the nonlinear rheology exactly at $\phi=\phi_J$: $p\sim \dot\gamma^q$.  The  strain rate scale $\dot\gamma_\times(\phi)$, which sets the crossover from linear to non-linear rheology below jamming, is given by,
\begin{equation}
\dot\gamma_\times\sim (\phi_J-\phi)^{z\nu}.
\label{egdotx}
\end{equation}

The condition that $p\to p_0(\phi)>0$ as $\dot\gamma\to 0$ above $\phi_J$ implies $\lim_{x\to +\infty}g(x)\sim x^y$, so that,
\begin{equation}
p_0(\phi)\sim (\phi-\phi_J)^y,\quad y=qz\nu,
\label{ep0}
\end{equation}
with $y$ the exponent that determines how the yield stress $p_0$ vanishes as $\phi\to\phi_J$ from above.
The condition that $\eta_p=p/\dot\gamma$ approaches a finite constant as $\dot\gamma\to 0$ below $\phi_J$ implies that $\lim_{x\to-\infty}g(x)\sim x^{-\beta}$, so that,
\begin{equation}
\eta_p\sim (\phi_J-\phi)^{-\beta},\quad \beta=z\nu-y,
\label{ebeta}
\end{equation}
with $\beta$ the exponent that determines the divergence of the transport coefficient $\eta_p$ as $\phi\to\phi_J$ from below.
Fitting the data for $p(\phi,\dot\gamma)$ to the scaling form of Eq.~(\ref{escale}) is in principle the best way to determine the values of the critical packing $\phi_J$ and the exponents $\beta$ and $y$ that describe behavior asymptotically close to $\phi_J$.  A similar scaling equation holds for the deviatoric shear stress $\sigma$.

Such a scaling analysis has been been carried out previously for circular disks ($\alpha=0$) \cite{OlssonTeitelPRE}.  There it was found that   {\em corrections-to-scaling} must be included, making the analysis significantly more complicated, and it was necessary to go to very small strain rates $\dot\gamma=10^{-8}$ in large systems with $N=65536$ particles to obtain consistent results.   Here we have not simulated such a large system, and with our smaller system of $N=1024$ we cannot probe such small strain rates without having to worry about finite size effects.  Thus we cannot attempt such a scaling analysis for small $\alpha$.  For larger $\alpha$, however, it is worthwhile to see how well such a scaling analysis might work, as the importance of corrections-to-scaling may vary with $\alpha$.  We therefore attempt a scaling analysis for our most elongated particles with $\alpha=4$, where we have data down to $\dot\gamma=4\times 10^{-7}$.  We choose to analyze pressure $p$ rather than shear stress $\sigma$, since prior results on circular disks \cite{OlssonTeitelPRE} indicate that corrections-to-scaling are significantly smaller for $p$ than for $\sigma$.  For our scaling analysis we use data from simulations with $N=1024$ particles for all but our smallest strain rate.  We have explicitly checked that for $\dot\gamma\ge 10^{-6}$, $N=1024$ is sufficiently large to avoid finite size effects; however for $\dot\gamma=4\times 10^{-7}$ a small finite size effect is observed for $N=1024$, and hence for this rate we use data from a larger system with $N=2048$.

To fit to the scaling form of Eq.~(\ref{escale}) we expand the logarithm of the a priori unknown scaling function $g(x)$ as a fourth order polynomial, i.e., $g(x)=\exp(c_0+c_1x+c_2x^2+c_3x^3+c_4x^4)$, and take as free fitting parameters $\phi_J$, $\beta$, $y$, and the $c_i$ (with $z\nu=\beta+y$ and $q=y/[\beta+y]$).  
Such a polynomial expansion for $\ln g(x)$ is expected to be a reasonable approximation  only for  {\em small} values of the scaling variable $x$, although the true scaling function $g(x)$ applies for the full range of $-\infty<x<\infty$.
Since scaling holds only asymptotically close to the critical point, we restrict the data to be used in our fit to packing fractions close to $\phi_J$, $0.88\le \phi\le 0.911$, and to strain rates $\dot\gamma\le\dot\gamma_\mathrm{max}$. We  then vary $\dot\gamma_\mathrm{max}$ to shrink the window of data closer to the critical point.  If our fits are to be regarded as good and stable we hope to find that the $\chi^2$ error per degree of freedom of the fit, $\chi^2/\mathrm{dof}\approx 1$, and that the fitted parameters stay constant, within the estimated statistical error, as $\dot\gamma_\mathrm{max}$ decreases.

\begin{figure}
\centering
\includegraphics[width=3.3in]{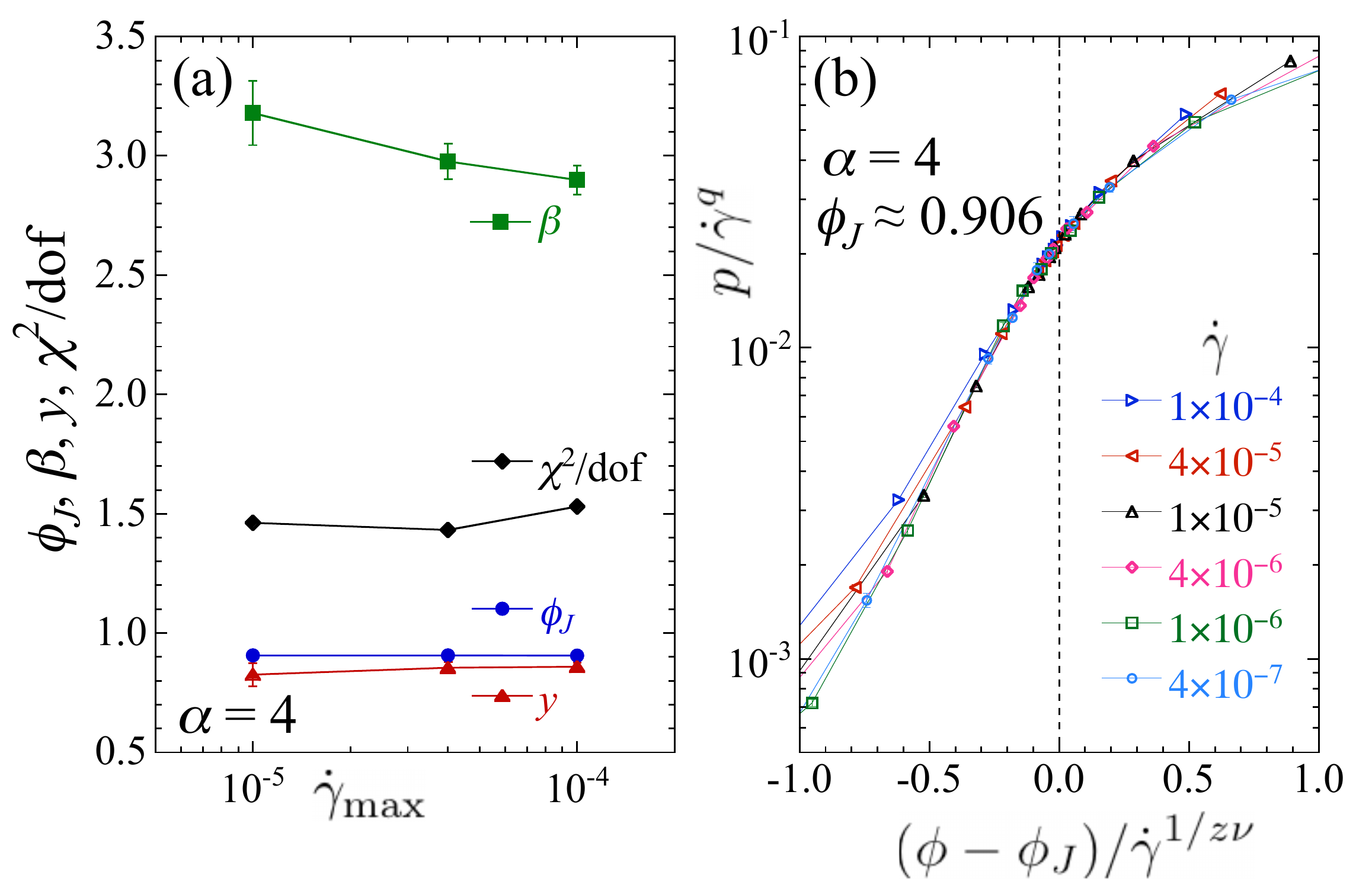}
\caption{For spherocylinders of asphericity $\alpha=4$:  (a) Fitting parameters of the scaling equation (\ref{escale}), $\phi_J$, $\beta$, and $y$, and $\chi^2/\mathrm{dof}$ of the fit, vs the maximum strain rate $\dot\gamma_\mathrm{max}$ used in the fit. (b) Scaling collapse of the data using the fitting parameters obtained from $\dot\gamma_\mathrm{max}=4\times 10^{-5}$.
}
\label{scale-a4} 
\end{figure}

In Fig.~\ref{scale-a4}(a) we plot our results for $\phi_J$, $\beta$, $y$, and  $\chi^2/\mathrm{dof}$ from this scaling fit vs $\dot\gamma_\mathrm{max}$, for $\dot\gamma_\mathrm{max}$ from $10^{-4}$ down to $10^{-5}$; we cannot use a smaller $\dot\gamma_\mathrm{max}$ as then the number of data points becomes too few.  
We find that $\phi_J$ and $y$ do appear to stay constant, within the estimated errors, but $\beta$ seems to systematically increase as $\dot\gamma_\mathrm{max}$ decreases.  The $\chi^2/\mathrm{dof}\approx 1.5$ stays roughly constant as $\dot\gamma_\mathrm{max}$ varies.
Taking the fitted parameters obtained from $\dot\gamma_\mathrm{max}=4\times 10^{-5}$, we have $\phi_J=0.9058\pm 0.0004$, $\beta=2.98\pm 0.07$, and $y=0.85\pm 0.02$, which give 
$q=y/(y+\beta)=0.222\pm 0.01$ and $1/z\nu =1/(y+\beta)= 0.26\pm 0.01$. 

In Fig.~\ref{scale-a4}(b) we show the  data collapse that results from these parameters, plotting $p/\dot\gamma^q$ vs $x=(\phi-\phi_J)/\dot\gamma^{1/z\nu}$.  With the above quoted fit parameters, the data used in obtaining the fit spans a range of the scaling variable $-1<x<0.2$.  In the scaling plot of Fig.~\ref{scale-a4}(b) we include data that lie outside this range, particularly  data for which $0.2<x<1$, as well as data for $\dot\gamma >\dot\gamma_\mathrm{max}=4\times 10^{-5}$.  One consistency check on our fit is to see if these latter data also collapses  well when plotted in terms of the scaled variables.
We see what appears to be a reasonable collapse.
As $|x|$ increases away from the critical point $x=0$, we start to see deviations from the common scaling curve for the larger $\dot\gamma$.  This is as expected since such data are too far from the critical point to lay in the scaling region.

We can compare the fitted exponents found here  to those found previously  \cite{OlssonTeitelPRE,OT3}  for circular disks ($\alpha=0$), $\beta=2.77\pm 0.02$, $y=1.08\pm 0.03$, $q=0.28\pm 0.02$, and $1/z\nu = 0.26\pm 0.02$.  
Comparing the critical exponents $\beta$ and $y$ for  $\alpha=0$ with those for $\alpha=4$, we  find that while the  exponents are close, they are nevertheless several standard deviations estimated statistical error different from each other. This suggests that the jamming of frictionless spherocylinders at finite $\alpha$ may be in a different universality class than the jamming of circular disks.  This might be expected since the universality class is generally determined from the symmetries of the system, and the $\alpha=0$ and $\alpha >0$ cases have different symmetries; sheared spherocylinders have a finite nematic orientational order $S_2>0$ \cite{MKOT}, while circular disks, by rotational symmetry, necessarily have  $S_2=0$.
However our conclusion on this issue should be regarded as tentative.  The increasing $\beta$ that we observe as $\dot\gamma_\mathrm{max}$ decreases suggests that corrections-to-scaling may not be negligible for our data, and so simulations of 
a larger system size $N$ at smaller strain rates $\dot\gamma$ may be needed to be more conclusive.  Nevertheless,  our result  that $\beta$ is {\em increasing} as $\dot\gamma_\mathrm{max}$ decreases, i.e., as we get closer to the critical point, would seem to suggest that the true asymptotic value of $\beta$ may be even further away from its $\alpha=0$ value than what we have found from our fits here.

Recently, a similar critical scaling analysis, for frictionless 3D spherocylinders of $\alpha=1$ in a model a of sheared dry granular material obeying a Bagnoldian rheology below jamming, has been presented in \cite{Nath}.  Although there remains controversy about the exact values of the critical exponents of such a Bagnoldian model for {\em spherical} particles \cite{VagbergOlssonTeitel,Peyneau,Wyart2,OH1,Hatano,OH2,Rahbari}, the values presented for spherocylinders in \cite{Nath} would seem to be clearly different from any of the proposed values for spheres.  However we note that the strain rates $\dot\gamma$ used in \cite{Nath} are at least two orders of magnitude larger than used in other works, and no details are given as to how the scaling analysis is carried out.  The authors of \cite{Nath} themselves say the following: ``However, care must be taken as the values of the exponents sensitively depend on the value for $\phi_c$ [our $\phi_J$]. Furthermore, our systems are rather small, and finite-size effects are likely to strongly influence these values."  Thus the analysis in \cite{Nath} cannot be taken as conclusive evidence that spherocylinders are in a different universality class from spheres for Bagnoldian systems.

\subsubsection{Herschel-Bulkley Rheology and Determination of $\phi_J$}
\label{secHB}

The critical scaling approach, discussed in the preceding section, is the most accurate way to determine the jamming packing $\phi_J$.  However for a general value of $\alpha$, as mentioned above, we do not have sufficient data for small enough $\dot\gamma$ and large enough $N$ to make such an analysis.  
To obtain the values of $\phi_J$ for our other values of $\alpha$ we  therefore resort to a different, more approximate, approach.  

\begin{figure}
\centering
\includegraphics[width=3.3in]{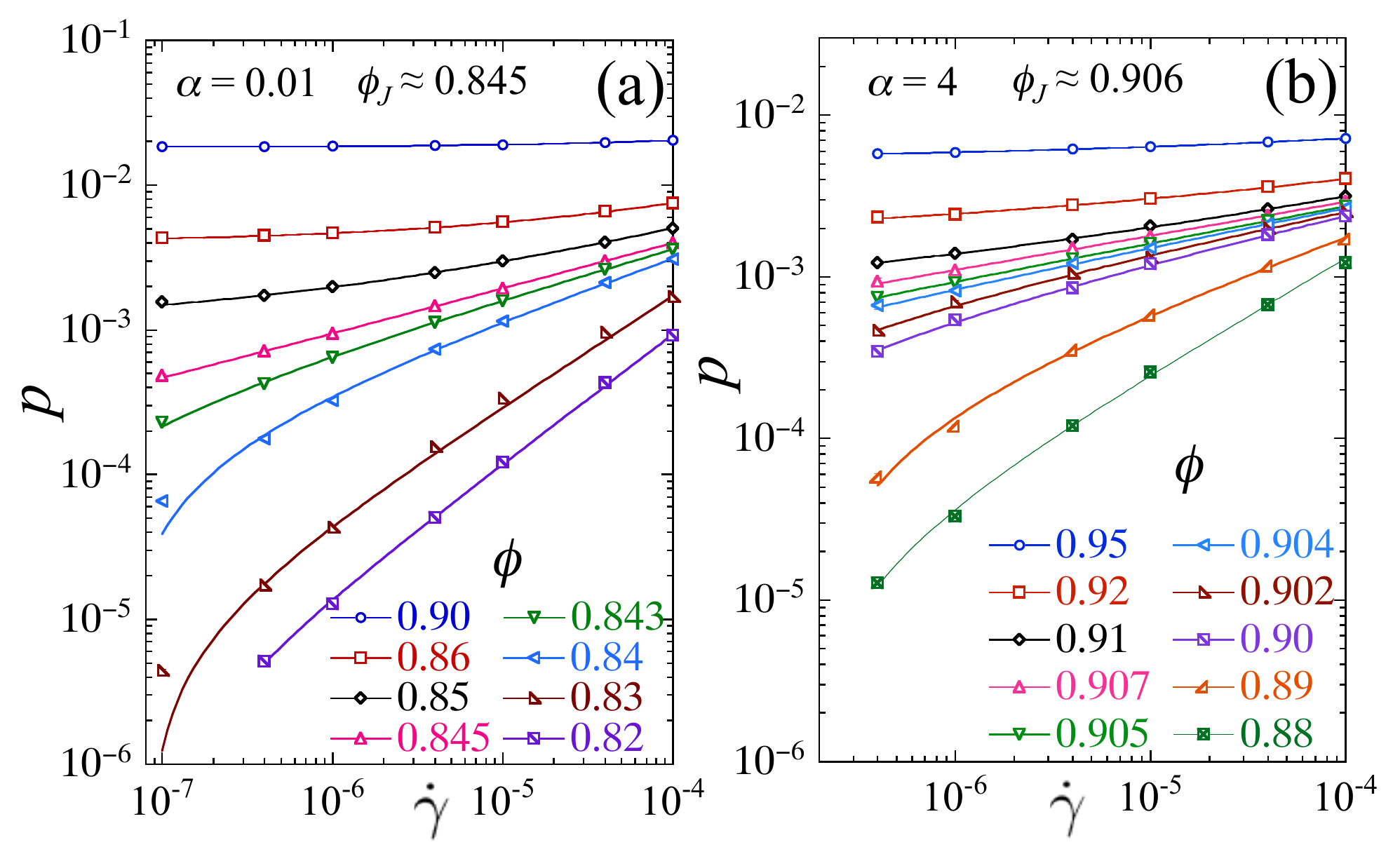}
\caption{Pressure $p$ vs shear strain rate $\dot\gamma$ at different packings $\phi$ for particles with asphericity (a) $\alpha=0.01$ and (b) $\alpha=4$. Solid lines are fits to the Herschel-Bulkley form of Eq.~(\ref{eHB}).
}
\label{p-vs-gdot} 
\end{figure}

For $\phi>\phi_J$ the rheological law is phenomenologically found to obey a Herschel-Bulkley (HB) form \cite{Larson,Hohler},
\begin{equation}
p = p_0 + c\dot\gamma^n.
\label{eHB}
\end{equation}
In Figs.~\ref{p-vs-gdot}(a) and (b) we plot $p$ vs $\dot\gamma$ at different $\phi$ for $\alpha=0.01$ and $4$, respectively.  Fitting to Eq.~(\ref{eHB}) gives the solid lines in Fig.~\ref{p-vs-gdot}.  

We see that above a certain value of $\phi$ the curves appear to saturate to a finite value $p_0$ as $\dot\gamma$ decreases, suggesting that these curves are at $\phi>\phi_J$ and obeying the HB form.  For lower $\phi$ the curves bend downwards as $\dot\gamma$ decreases, suggesting that $p\to 0$ and that these curves are at $\phi<\phi_J$.  For models of {\em frictional} particles, a shear thickening region is observed \cite{Saw,DST3} just below jamming, but such a complication is absent in frictionless systems.

For $\phi<\phi_J$ we do not expect Eq.~(\ref{eHB}) to be a good fit, and in Fig.~\ref{p-vs-gdot}(a) we see that the fit is indeed poor at the  smallest $\dot\gamma$ for the smaller $\phi$.  
In principle we know that below $\phi_J$ the rheology is Newtonian at small enough $\dot\gamma$, so one might expect to find a good fit to Eq.~(\ref{eHB}) in which $p_0=0$ and $n=1$.  However, as discussed in the two previous  sections, such a Newtonian rheology holds only for small $\dot\gamma <\dot\gamma_\times(\phi)\sim (\phi_J-\phi)^{z\nu}$, and  $\dot\gamma_\times(\phi)$ decreases to zero as one gets close to $\phi_J$.  Close to, but below, $\phi_J$ we do not have sufficient data in this Newtonian region.  Our fits to Eq.~(\ref{eHB}) in Fig.~\ref{p-vs-gdot} use the full range of our data, with $\dot\gamma$ extending up to $10^{-4}$, and so for $\phi<\phi_J$ include data that is outside the Newtonian region and into the non-linear region.  Such fits tend to give unphysical values of $p_0<0$.

\begin{figure}
\centering
\includegraphics[width=3.3in]{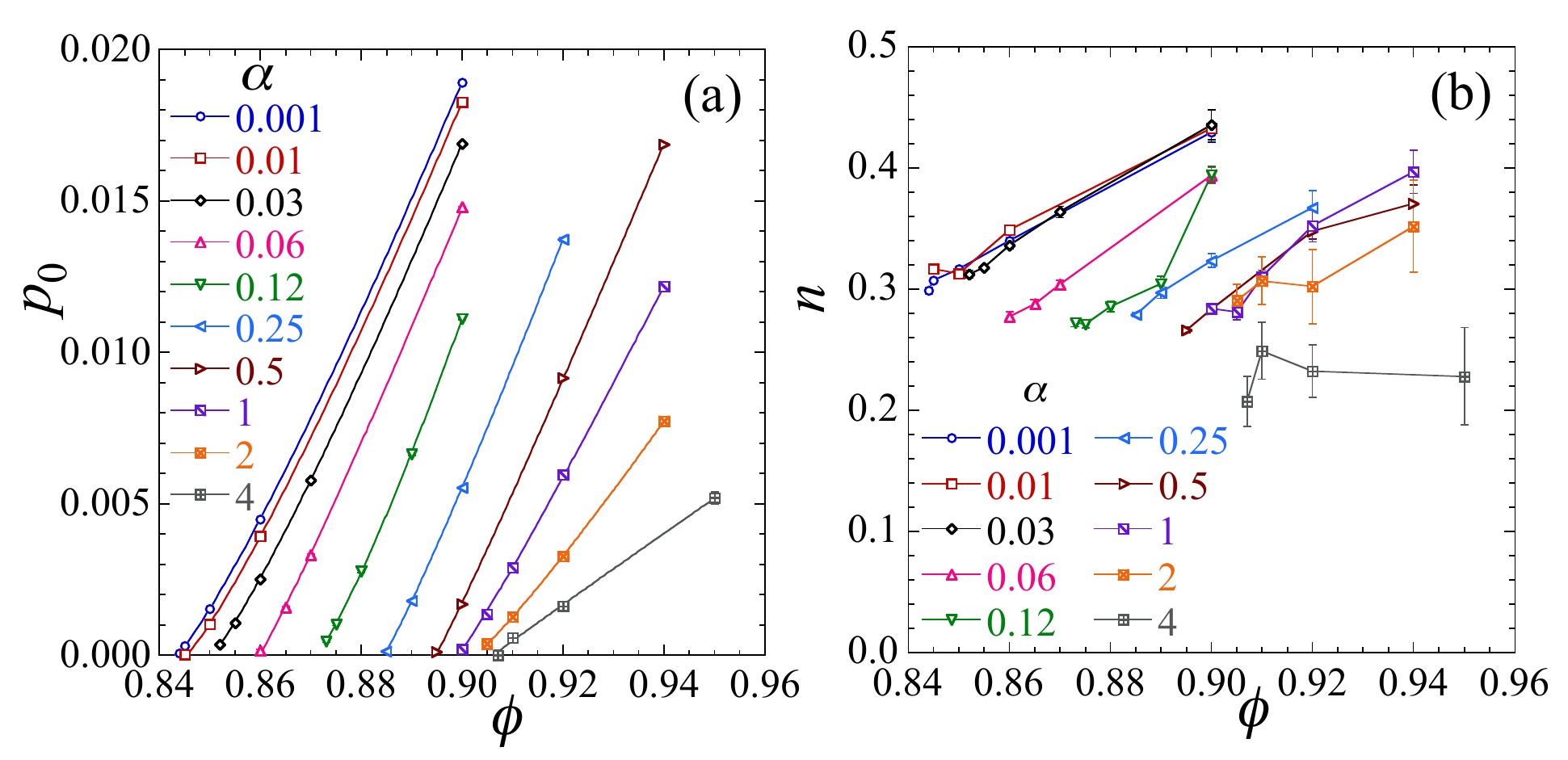}
\caption{(a) Yield pressure $p_0$ and (b) effective Herschel-Bulkley exponent $n$ vs packing $\phi$ for particles with different asphericities $\alpha$.
}
\label{p0-n-vs-phi} 
\end{figure}

However for larger $\phi>\phi_J$ the fits are reasonably good, and so in Fig.~\ref{p0-n-vs-phi}
we plot our results for $p_0$ and $n$ vs $\phi$ for different $\alpha$; we show only results which find $p_0\ge 0$.   We note that the Herschel-Bulkley form (\ref{eHB}) has, in principle,  a well defined value of $n$ in the limit $\dot\gamma\to 0$ \cite{OT3}; however we do not have results at enough values of $\dot\gamma$, nor small enough  $\dot\gamma$, to probe this asymptotic small $\dot\gamma$ limit.  Our results for the exponent $n$ in Fig.~\ref{p0-n-vs-phi} should therefore be regarded as only  {\em effective} exponents for the range of $\dot\gamma$ simulated;  we note, however, that for our particles with small $\alpha$, the range $0.3\lesssim n\lesssim 0.45$ that we find agrees with values typically found in the literature \cite{Schall}.

\begin{figure}
\centering
\includegraphics[width=3.3in]{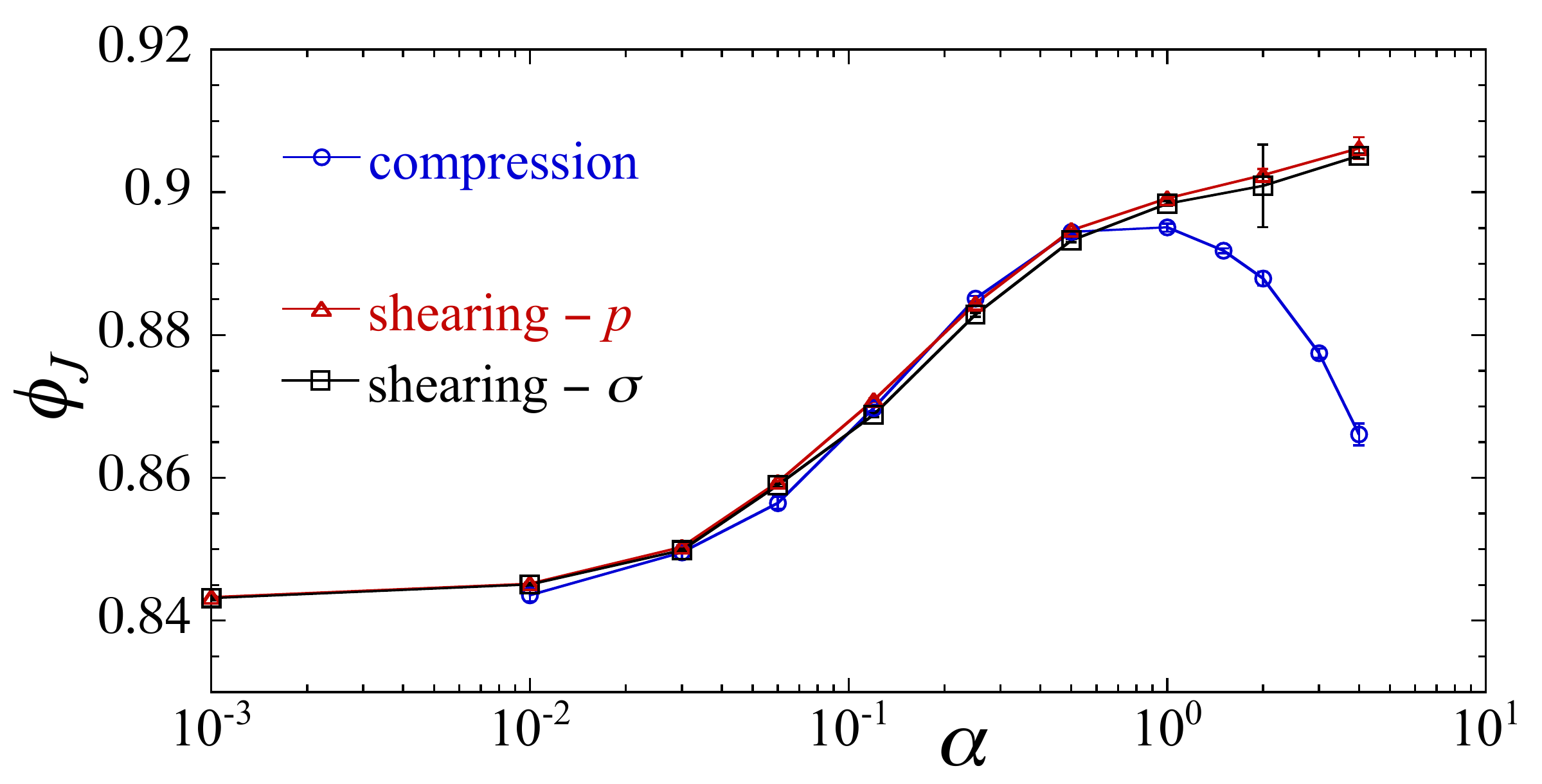}
\caption{Packing fraction at the jamming transition $\phi_J$ vs particle asphericity $\alpha$ for shear-driven jamming and for jamming by isotropic compression (from Ref.~\cite{MarschallCompress}).  For shear-driven jamming we show results from  our analysis of both the pressure $p$ and the deviatoric stress $\sigma$.
}
\label{phiJ-vs-alpha} 
\end{figure}

Using the values of $p_0(\phi)$ in Fig.~\ref{p0-n-vs-phi}(a) as estimates of the yield stress, we then extrapolate in $\phi$ to find the packing fraction at which $p_0$ vanishes, and take this as our estimate of the jamming point $\phi_J$.  Fitting to the form $p_0=c(\phi-\phi_J)^{\bar y}$, we plot the resulting $\phi_J$ in Fig.~\ref{phiJ-vs-alpha}.  For $\alpha=4$ this approach gives $\phi_J=0.906\pm 0.001$, in agreement with our result from the critical scaling analysis described in the preceding section.  In Fig.~\ref{phiJ-vs-alpha} we also plot, for comparison, the values of $\phi_J$ vs $\alpha$ that we have previously found \cite{MarschallCompress} for the  compression-driven jamming of this same system, when we isotropically compressed at a slow rate from random  configurations at an initial small $\phi_\mathrm{init}$.  We see that at small $\alpha\lesssim 0.5$ the $\phi_J$ from compression are quite close to, though systematically slightly smaller than, 
the $\phi_J$ from shearing.  However as $\alpha$ further increases, $\phi_J$ from compression reaches a peak and then decreases, while $\phi_J$ from shearing continues to slowly increase.  The greater $\phi_J$ for shearing as compared to compression, for the larger $\alpha$, is related to the nematic ordering that spherocylinders undergo when sheared \cite{Campbell, Guo1, Guo2, Borzsonyi1, Borzsonyi2, Wegner, Wegner2, Nagy, Trulsson,Nath,MKOT,MT2} as contrasted with the lack of such ordering when isotropically compressed \cite{Donev.Science.2004,Man.PRL.2005,Sacanna.JPhysC.2007,Zhao,MarschallCompress}.  The nematic ordering under shearing allows particles to pack more efficiently, and so increases the jamming $\phi_J$.  

The jamming packing $\phi_J(\alpha)$ for particles of different asphericity $\alpha$ has also been investigated for models of sheared dry granular materials obeying a Bagnold rheology below jamming.  The monotonically increasing behavior of $\phi_J$ vs $\alpha$ that we see here is qualitatively similar to what is seen in Ref.~\cite{Trulsson} for frictionless 2D ellipses; for frictionless 3D spherocylinders, Ref.~\cite{Nath} similarly sees a monotonic increasing behavior, however Ref.~\cite{Nagy} sees an odd non-monotonic dip followed by an increase as $\alpha$ increases above $\alpha\approx 0.7$.  However when inter-particle friction is added to such models \cite{Trulsson,Nath}, $\phi_J$ in general decreases just as is the case for spherical particles, but $\phi_J$ also becomes non-monotonic in $\alpha$, having a shape qualitatively similar to what we see for compression-driven jamming.

The exponent $\bar y$ in our above fits to $p_0(\phi)$ should not be regarded as the same as the true critical exponent $y$ of the scaling law  Eq.~(\ref{ep0}). The latter holds only asymptotically as $\phi\to\phi_J$, while $\bar y$ is obtained from the data in Fig.~\ref{p0-n-vs-phi}(a) by phenomenologically fitting over a relatively wide range of $(\phi-\phi_J)>0$.  We do not have data at enough values of $\phi$ closer to $\phi_J$ to probe the true asymptotic region; just as the exponent $n$ in Fig.~\ref{p0-n-vs-phi}(b) should be regarded as only an {\em effective} Herschel-Bulkley exponent for the range of $\dot\gamma$ used in the fit, so $\bar y$ must be regarded as only an {\em effective} exponent for the the range of $\phi$ used in our fit to $p_0(\phi)$.  As $\alpha$ varies, we find values of $\bar y$ that vary between 0.98 and 1.21.  For $\alpha=4$ we find $\bar y=0.98\pm 0.11$, which compares to the $y=0.85\pm 0.02$ found from our scaling analysis.  As a check on our method we have also tried a fit of $p_0(\phi)$ to a quadratic polynomial, $p_0=c_1(\phi-\phi_J)+c_2(\phi-\phi_J)^2$, and find the resulting $\phi_J$ to always be within $0.1\%$ of the $\phi_J$ found with the algebraic fit; the $\chi^2/\mathrm{dof}$ from this quadratic fitting is, however, usually an order of magnitude worse than from the algebraic fitting.

\begin{figure}
\centering
\includegraphics[width=3.3in]{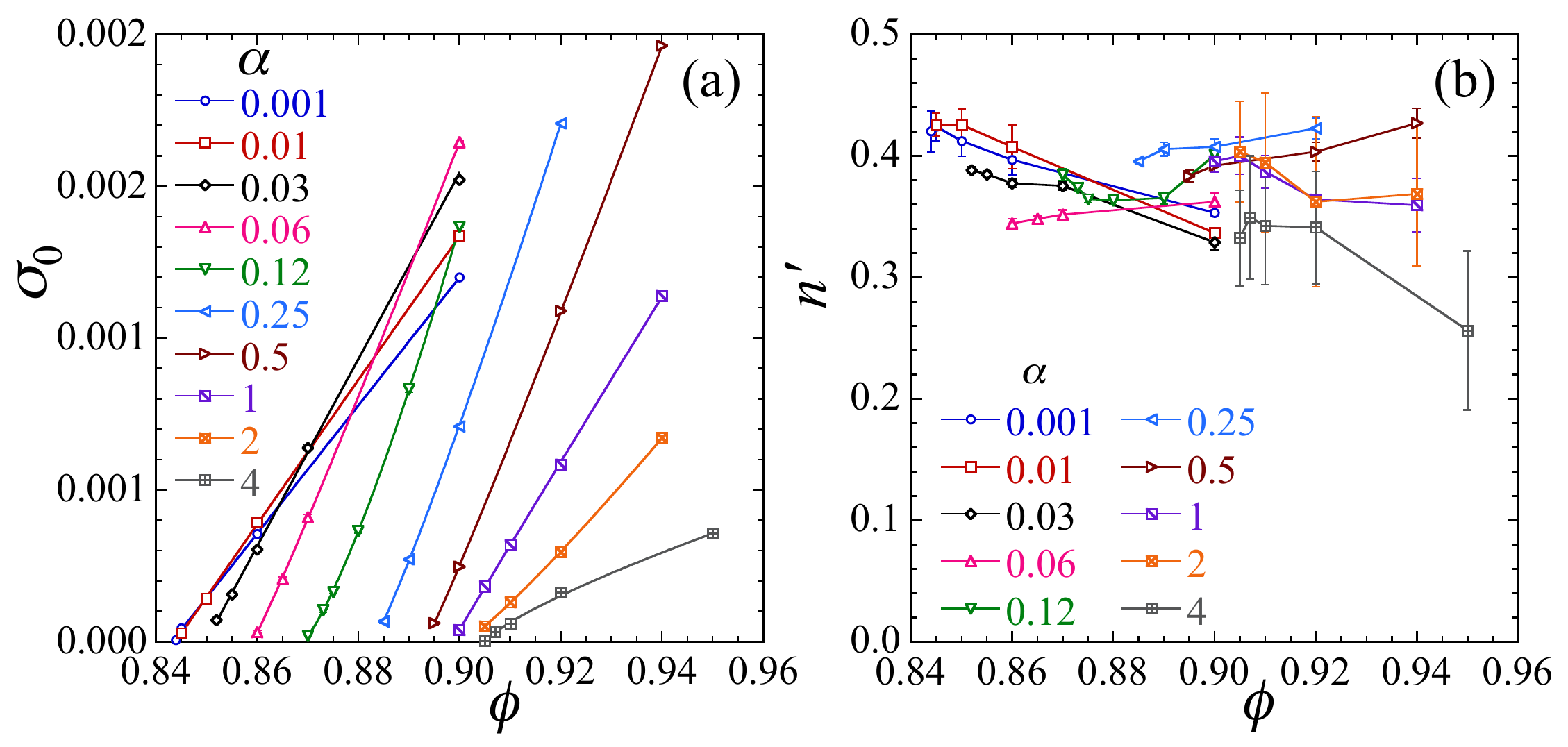}
\caption{(a) Deviatoric yield stress $\sigma_0$ and (b) effective Herschel-Bulkley exponent $n^\prime$ vs packing $\phi$ for particles with different asphericities $\alpha$.
}
\label{s0-n-vs-phi} 
\end{figure}

We can carry out a similar analysis as above, but using the deviatoric shear stress $\sigma$ rather than pressure $p$.  Fitting $\sigma(\dot\gamma)$ to a Herschel-Bulkley form at different $\phi$ for different $\alpha$,
\begin{equation}
\sigma=\sigma_0+c^\prime \dot\gamma^{n^\prime},
\end{equation}
we show  results for the deviatoric yield stress $\sigma_0$ and effective Herschel-Bulkley exponent $n^\prime$ in Fig.~\ref{s0-n-vs-phi}.  Extrapolating $\sigma_0$ to zero for each different $\alpha$ we arrive at an estimate for $\phi_J$ which we plot in Fig.~\ref{phiJ-vs-alpha}.  We see that this  estimate for $\phi_J$ agrees quite well with our earlier estimate from the analysis of pressure $p$; the $\phi_J$ obtained from $\sigma$ is just slightly smaller than that obtained from $p$, but the difference is always less than $0.25\%$.  

In Fig.~\ref{s0-n-vs-phi}(b) we plot our results for the exponent $n^\prime$ obtained from $\sigma$.  Comparing to similar results for the exponent $n$ from pressure  $p$ in Fig.~\ref{p0-n-vs-phi}(b), we see that $n$ and $n^\prime$ fall within the same general range of values, however it is clear that $n\ne n^\prime$, and for small $\alpha$ the trend as $\phi$ varies is opposite; the exponent $n$ from $p$ increases as $\phi$ increases, while the $n^\prime$ from $\sigma$ decreases.
This observation lends support to our assertion that $n$ and $n^\prime$ as computed here are only {\em effective} exponents for the range of $\dot\gamma$ we simulate, rather than being the true asymptotic $\dot\gamma\to 0$ values \cite{OT3}.

\subsubsection{Macroscopic Friction}
\label{secMu}

\begin{figure}
\centering
\includegraphics[width=3.3in]{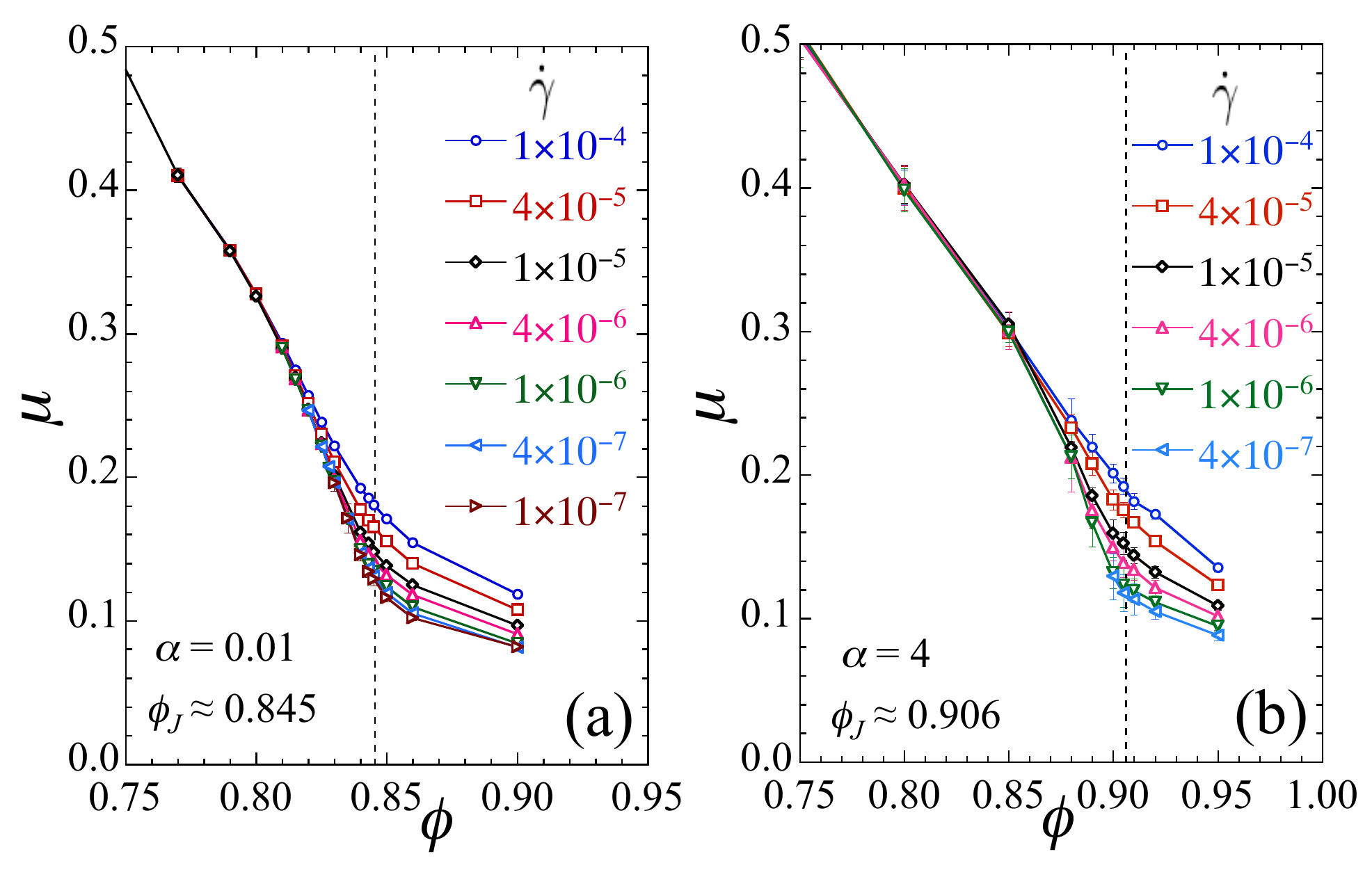}
\caption{Macroscopic friction $\mu=\sigma/p$ vs packing $\phi$ at different shear strain rates $\dot\gamma$ for particles with asphericity (a) $\alpha=0.01$ and (b) $\alpha=4$.  Vertical dashed lines indicate the jamming $\phi_J$.
}
\label{mud-vs-phi} 
\end{figure}

Next we look at the macroscopic friction, $\mu=\sigma/p$.  In Figs.~\ref{mud-vs-phi}(a) and \ref{mud-vs-phi}(b) we plot our results for $\mu$ vs $\phi$ at different strain rates $\dot\gamma$ for $\alpha=0.01$ and 4, respectively.  We see  at low $\phi$ that $\mu$ is nearly independent of $\dot\gamma$, however upon approaching $\phi_J$, and going above, the $\dot\gamma$ dependence becomes significant.  The low $\phi$ behavior is a consequence of the Newtonian rheology in the hard-core limit, where both $p$ and $\sigma$ are proportional to $\dot\gamma$ and so their ratio is a finite value independent of $\dot\gamma$.  
However, as discussed previously, this linear Newtonian region persists only for $\dot\gamma<\dot\gamma_\times(\phi)$, and $\dot\gamma_\times(\phi)\sim (\phi_J-\phi)^{z\nu}$.  Thus, as $\phi$ increases to $\phi_J$, $\dot\gamma_\times(\phi)$ decreases and goes to zero, and our results at finite $\dot\gamma$ are no longer small enough to be in the Newtonian region; we are in the non-linear region of soft-core behavior and so the $\dot\gamma$ dependencies of $p$ and $\sigma$ no longer cancel when computing $\mu$, and so $\mu$ develops the $\dot\gamma$ dependence seen in the figure.

Above $\phi_J$ we have $\lim_{\dot\gamma\to 0} \mu=\sigma_0/p_0$, and as $\sigma_0$ and $p_0$ are both components of the stress tensor we expect them both to scale $\sim (\phi-\phi_J)^y$ with the same exponent $y$ (as has been explicitly verified for circular disks \cite{OlssonTeitelPRE}).  Thus we expect that $\lim_{\dot\gamma\to 0}\mu$ is a finite constant.  However the Herschel-Bulkley form of the rheology above $\phi_J$, given by Eq.~(\ref{eHB}),  suggests that the $\dot\gamma$ dependencies of $p$ and $\sigma$ will not cancel, 
and so, just as found for $\phi$ close to but below $\phi_J$, we find a noticeable dependence of $\mu$ on $\dot\gamma$.

The above discussion has been framed in terms of simulations at constant volume, where the control parameters are packing fraction $\phi$ and shear strain rate $\dot\gamma$.   For dry particles with a Bagnoldian rheology, however, studies are often done at constant pressure rather than constant volume, and it has been common to introduce as a control parameter a quantity known as the {\em inertial number} $I$ \cite{Forterre},
\begin{equation}
I=d\dot\gamma \sqrt{\rho/p},
\end{equation}
where $d$ is a typical particle diameter and $\rho$ is the particle's mass density.  For hard-core particles (or soft-core particles at sufficiently small $\dot\gamma$), Bagnoldian rheology gives $p=B(\phi)\dot\gamma^2$ for $\phi<\phi_J$, and the inertial number $I\propto 1/\sqrt{B}$ does not depend on $\dot\gamma$ or $p$ separately, but only on the packing $\phi$ \cite{VagbergOlssonTeitel}.  The rheology is then described by the two ``constitutive equations,"  $\phi(I)$ and $\mu(I)$.

For Newtonian suspensions, an analogous quantity called the {\em viscous number} $J$ is defined as \cite{Boyer,Wyart2},
\begin{equation}
J=\eta_\mathrm{host}\dot\gamma/p,
\end{equation}
where $\eta_\mathrm{host}$ is the viscosity of the host medium.  With our units $\eta_\mathrm{host}=1$, and so we have $J=1/\eta_p$.  The hard-core limit below $\phi_J$ can then be described by $\phi(J)$ and $\mu(J)$.

\begin{figure}
\centering
\includegraphics[width=3.3in]{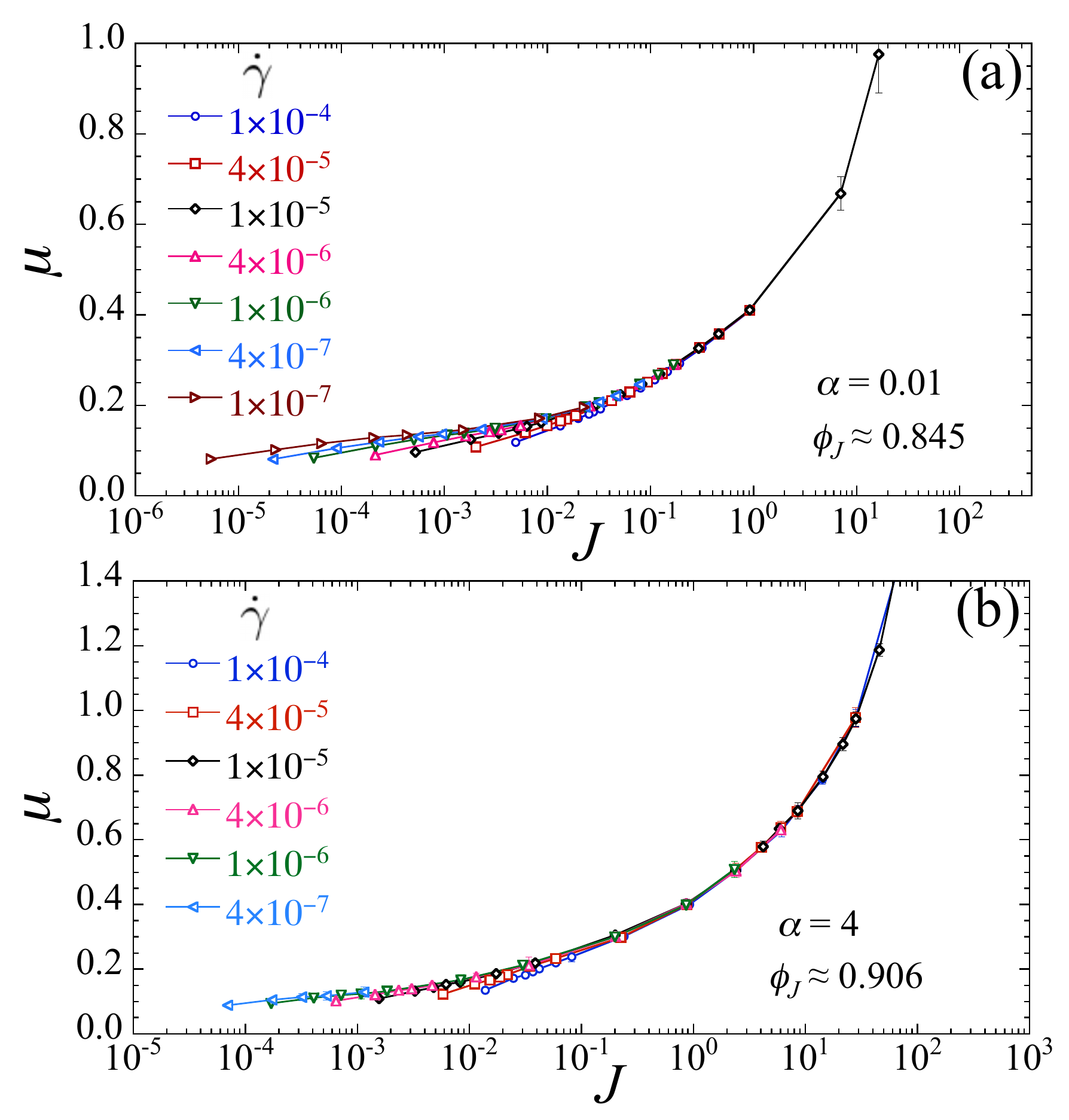}
\caption{Macroscopic friction $\mu=\sigma/p$ vs viscous number  $J$ at different shear strain rates $\dot\gamma$ for particles with asphericity (a) $\alpha=0.01$ and (b) $\alpha=4$.  
}
\label{mud-vs-J} 
\end{figure}

In Figs.~\ref{mud-vs-J}(a) and \ref{mud-vs-J}(b)  we plot $\mu$ vs $J$ at different fixed strain rates $\dot\gamma$, for particles with $\alpha=0.01$ and 4 respectively.  We see that the curves for different $\dot\gamma$ all collapse to a common $\dot\gamma\to 0$ limiting curve at large $J$, but that they depart from this curve as $J$ decreases; the smaller  the value of $\dot\gamma$ is, the smaller  the value of $J_\times(\dot\gamma)$ where this splitting off from the limiting curve occurs.  The limiting curve, given by the upper envelope of the set of curves at finite $\dot\gamma$, represents the hard-core limit below jamming.  The segments of the finite $\dot\gamma$ curves that lie below $J_\times(\dot\gamma)$ represent the soft-core region that one enters when approaching $\phi_J$ and going above.  Since by Eq.~(\ref{egdotx}) one enters the soft-core region when $(\phi_J-\phi)\sim \dot\gamma^{1/z\nu}$, and since by Eq.~(\ref{ebeta}) $J=1/\eta_p\sim (\phi_J-\phi)^\beta$, one has $J_\times(\dot\gamma)\sim \dot\gamma^{\beta/z\nu}$.  It is interesting to note that, while the crossover from hard-core to soft-core behavior as one approaches and goes above $\phi_J$ is immediately apparent in Figs.~\ref{etap-vs-phi}(b) and \ref{etap-vs-phi}(d)  for $\eta_p$ vs $\phi$, and in Figs.~\ref{eta-vs-phi}(b) and \ref{eta-vs-phi}(d)  for $\eta$ vs $\phi$, and in Fig.~\ref{mud-vs-phi} for $\mu$ vs $\phi$, the signature of this crossover is much  less apparent when plotting $\mu$ vs $J$ in Fig.~\ref{mud-vs-J}.

\subsubsection{Orientation of the Minimal Stress Axis}

\begin{figure}
\centering
\includegraphics[width=3.3in]{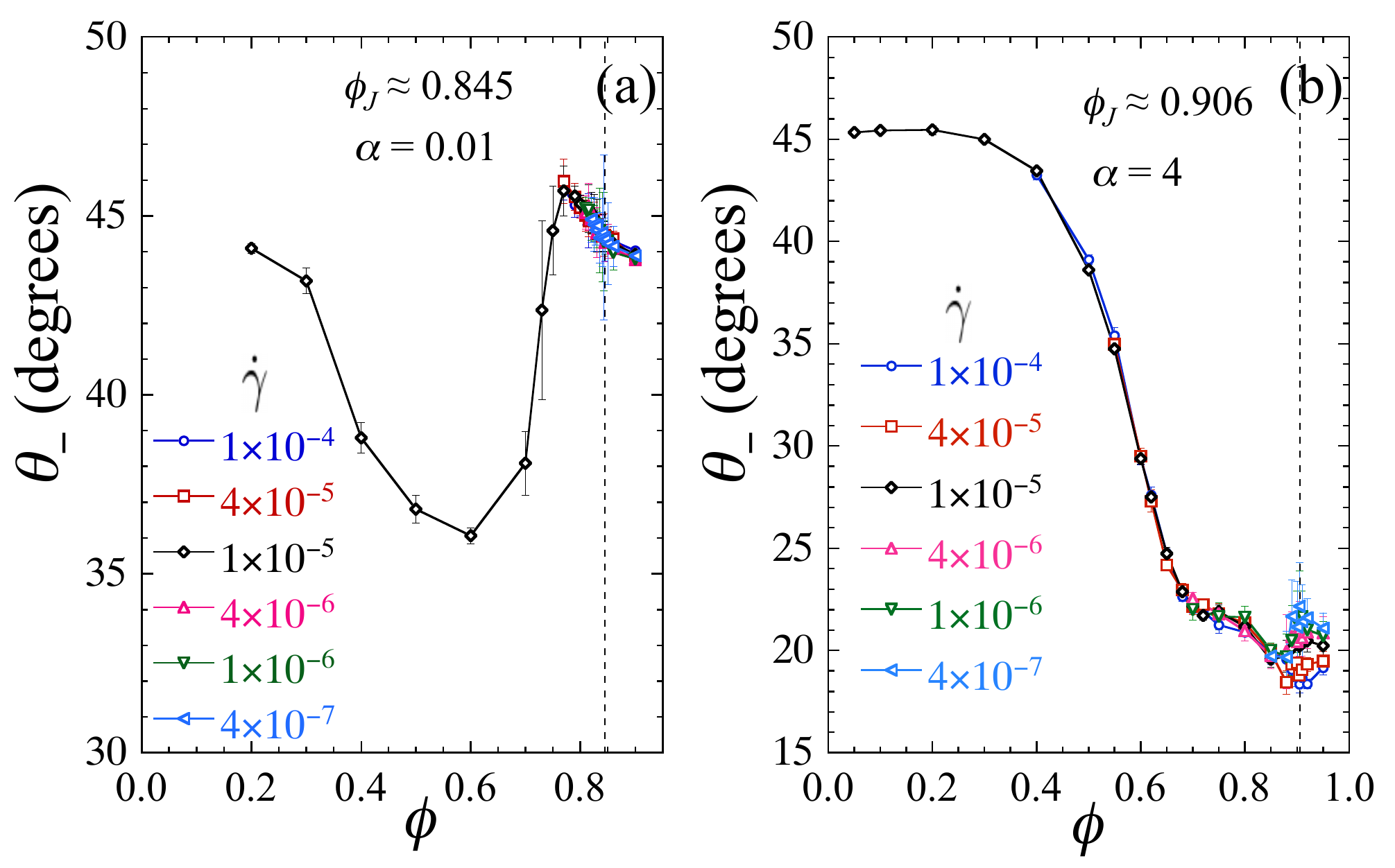}
\caption{Orientation $\theta_-$ of the minimal stress axis of the stress tensor $\mathbf{p}$ vs packing fraction $\phi$ for spherocylinders with asphericity (a) $\alpha=0.01$ and (b) $\alpha=4$, at different strain rates $\dot\gamma$.  The dashed vertical lines indicate the jamming transition at $\phi_J$.
}
\label{thetaminus-v-phi} 
\end{figure}

We now consider the orientation of the stress tensor $\mathbf{p}$, in particular the direction $\theta_-$ of the eigen-direction of minimal stress corresponding to the eigenvalue $p-\sigma$, given by Eq.~(\ref{ethetaminus}).  In Figs.~\ref{thetaminus-v-phi}(a) and \ref{thetaminus-v-phi}(b) we plot $\theta_-$ vs $\phi$ for different strain rates $\dot\gamma$, for spherocylinders of asphericity $\alpha=0.01$ and 4 respectively.  In both cases we see little significant dependence on $\dot\gamma$.

The limiting value of $\theta_-$ at small $\phi\to 0$ should be well approximated by the case of an isolated particle. In that case
the only contribution to the stress tensor is from the dissipative part $\mathbf{p}^\mathrm{dis}$, the rotational equation of motion is just $\dot\theta_i=-\dot\gamma f(\theta_i)$, and so the probability to be at angle $\theta$ is just $\mathcal{P}(\theta)\propto 1/f(\theta)$.  Using these in Eq.~(\ref{eSigdisi}) one sees that the diagonal elements of $\mathbf{p}^\mathrm{dis}$ vanish and  the off-diagonal elements are finite and equal, so  in this limit  $\theta_-=45^\circ$.

In Fig.~\ref{thetaminus-v-phi} we see agreement with this expectation at low $\phi$ for both $\alpha=0.01$ and 4.  However as $\phi$ increases we see different behaviors for these two cases.  For the nearly circular particles with $\alpha=0.01$, we see that $\theta_-$ decreases as $\phi$ increases, reaches a minimum, then increases again as $\phi_J$ is approached, and then decreases again as one goes above jamming.  
This non-monotonicity of $\theta_-$ is clearly a consequence of the elastic collisions.  We observe that where $\theta_-$ has its minimum, the majority of the particles in our systems are in contact with at least one other particle, but fewer than 2\% are in contact with more than two neighbors, so the dominant stresses here are the result of independent collisions which result from the shearing process.  Above $\phi_J$, where force chains span the system and collisions are no longer independent, we see that $\theta_-$ stays close to, but slightly smaller than, $45^\circ$.
For $\alpha=4$ we see that $\theta_-$ increases slightly as $\phi$ increases from low values,  then takes a large drop as $\phi_J$ is approached, and then increases slightly as $\phi$ goes above jamming.  The value $\theta_-\approx 20^\circ$ at high densities indicates a sizable normal stress difference; from Eq.~(\ref{eN1}) we get $N_1/\sigma_{xy}\approx 2.4$.

\subsubsection{Variation of Rheology with Asphericity $\alpha$}

\begin{figure}
\centering
\includegraphics[width=3.3in]{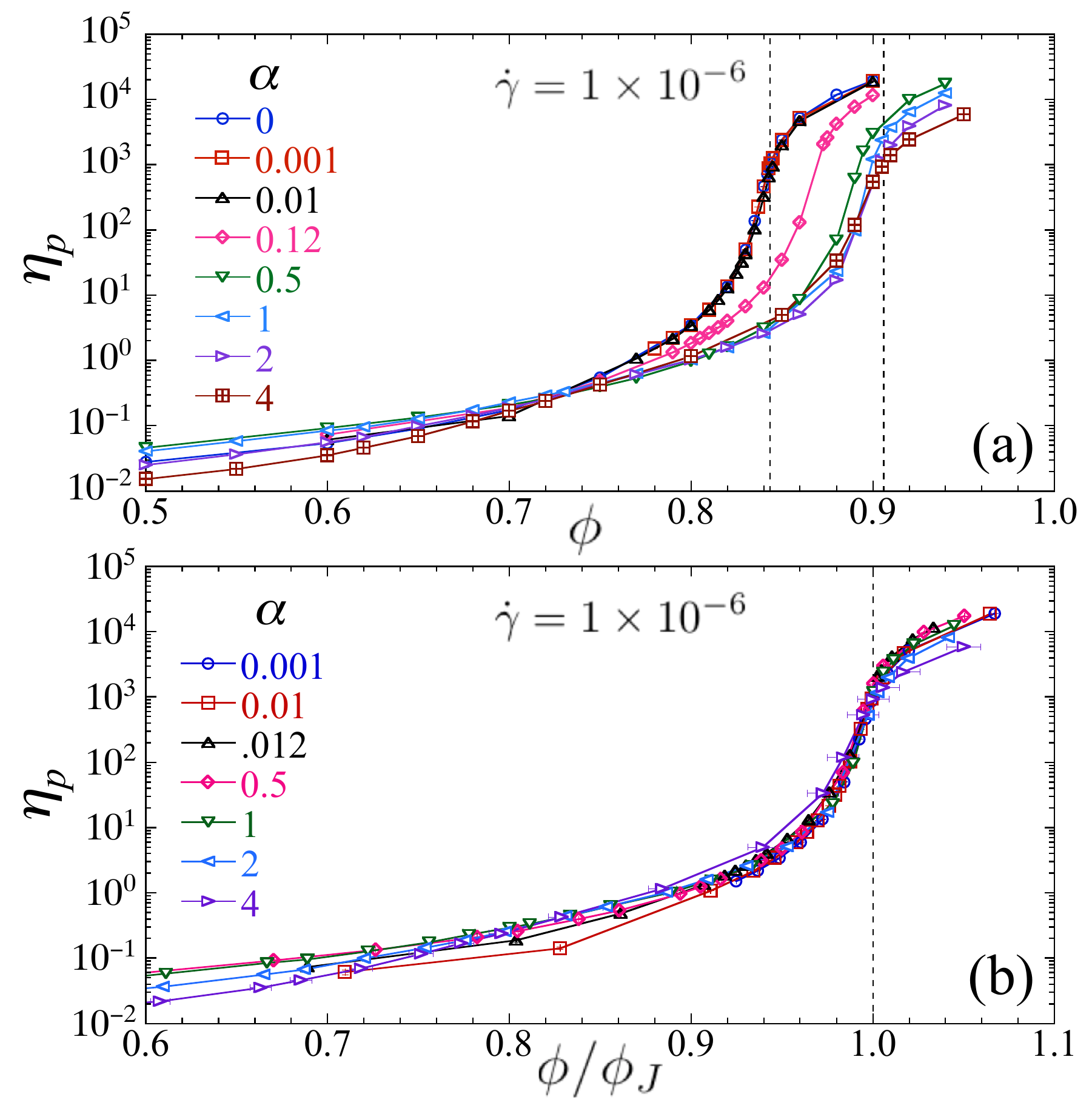}
\caption{Pressure transport coefficient $\eta_p=p/\dot\gamma$ vs (a) packing $\phi$ and vs (b) normalized packing $\phi/\phi_J$ for particles with different asphericity $\alpha$ at shear strain rate $\dot\gamma=10^{-6}$.  In (a) the lower vertical dashed line gives the jamming point $\phi_J\approx0.843$ for circular disks while the upper vertical dashed line gives the jamming point $\phi_J\approx 0.906$ for spherocylinders with $\alpha=4$.  In (b) the vertical dashed line gives $\phi/\phi_J=1$.
}
\label{etap-vs-phi-g1e-6} 
\end{figure}

Most of the previous sections dealt with the two representative cases of $\alpha=0.01$ and 4.  Here we wish to explore the rheology as $\alpha$ varies more generally.  To do this we will focus on results obtained at a fixed value of the shear strain rate\footnote{At low values of $\phi$ we use larger values of $\dot\gamma$, as we are in the hard-core limit where the stress becomes independent of $\dot\gamma$ for the $\dot\gamma$ we are considering.}  $\dot\gamma=10^{-6}$.  In Fig.~\ref{etap-vs-phi-g1e-6}(a) we plot the pressure transport coefficient $\eta_p=p/\dot\gamma$ vs $\phi$ for different $\alpha=0$ to 4.   We see that the largest variation between the curves of different $\alpha$ takes place for $0.01\lesssim\alpha\lesssim 1$, corresponding to the region where $\phi_J$ varies most rapidly (see Fig.~\ref{phiJ-vs-alpha}).  Not surprisingly, for $\phi\gtrsim 0.8$ we see that $\eta_p$ decreases as $\alpha$ increases; alignment of the elongated particles at high densities serves to reduce the stress.  However, in Fig.~\ref{etap-vs-phi-g1e-6}(b) we plot $\eta_p$ vs a normalize packing fraction $\phi/\phi_J$, where $\phi_J$ is the $\alpha$-dependent critical jamming packing fraction shown in Fig.~\ref{phiJ-vs-alpha}.  We see that the curves of $\eta_p$ for different $\alpha$ are now in large measure  the same, especially in the region approaching $\phi/\phi_J\approx 1$. At $\phi/\phi_J > 1$ we see that $\eta_p$ slightly decreases as $\alpha$ increases, while at low $\phi/\phi_J <1$ we find that $\eta_p$ slightly increases as $\alpha$ increases; howeer plotting vs $\phi/\phi_J$ we see that  the  difference in behavior of $\eta_p$  for the different $\alpha$, as seen in Fig.~\ref{etap-vs-phi-g1e-6}(a), is primarily a consequence of the variation of $\phi_J$ with $\alpha$.

\begin{figure}
\centering
\includegraphics[width=3.3in]{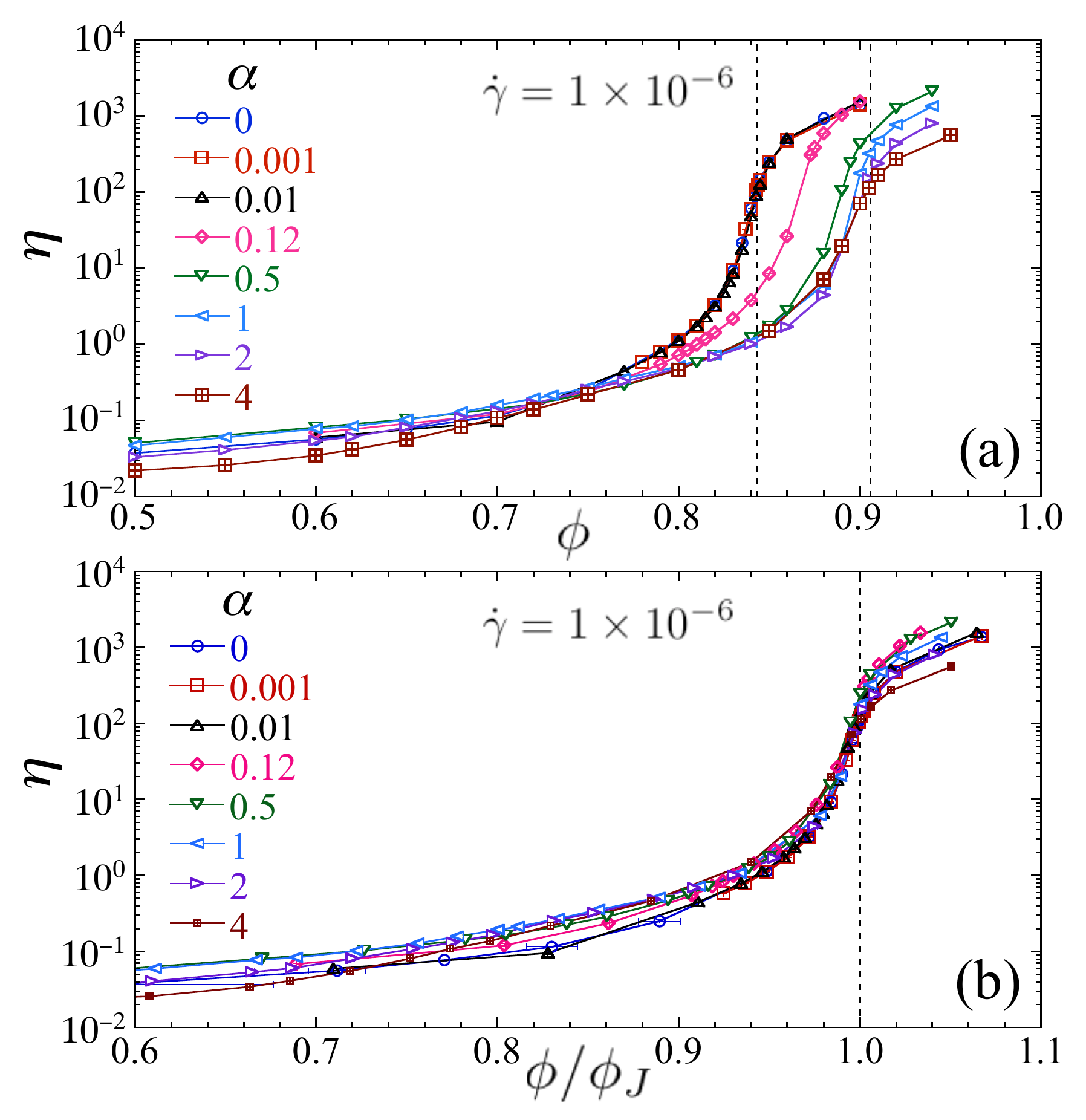}
\caption{Shear viscosity $\eta=\sigma/\dot\gamma$ vs (a) packing $\phi$ and vs (b) normalized packing $\phi/\phi_J$ for particles with different asphericity $\alpha$ at shear strain rate $\dot\gamma=10^{-6}$.  In (a) the lower vertical dashed line gives the jamming point $\phi_J\approx0.843$ for circular disks while the upper vertical dashed line gives the jamming point $\phi_J\approx 0.906$ for spherocylinders with $\alpha=4$.  In (b) the vertical dashed line gives $\phi/\phi_J=1$.
}
\label{etasd-vs-phi-g1e-6} 
\end{figure}

\begin{figure}
\centering
\includegraphics[width=3.3in]{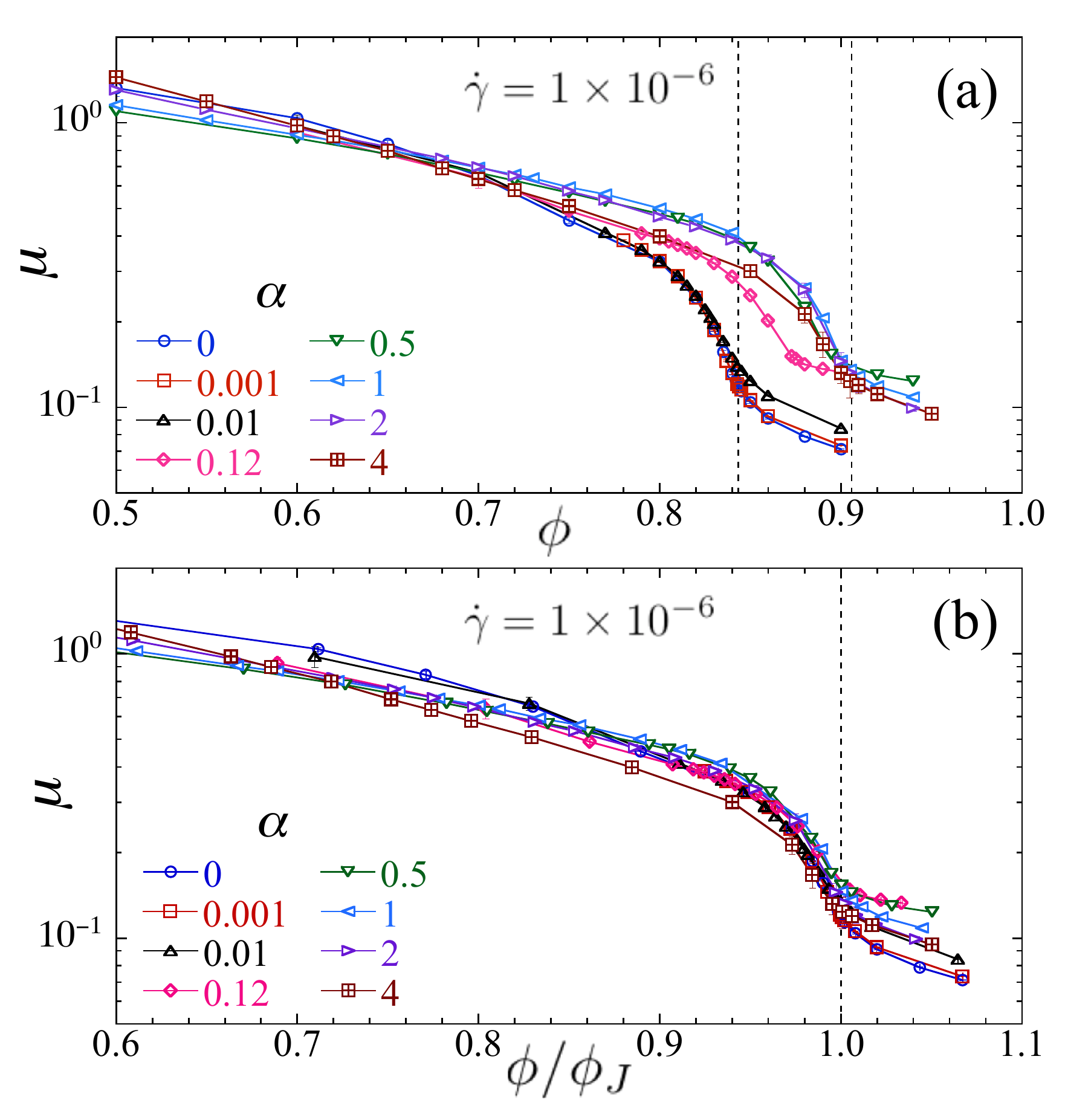}
\caption{Macroscopic friction $\mu=\sigma/p$ vs (a) packing $\phi$ and vs (b) normalized packing $\phi/\phi_J$ for particles with different asphericity $\alpha$ at shear strain rate $\dot\gamma=10^{-6}$.  In (a) the lower vertical dashed line gives the jamming point $\phi_J\approx0.843$ for circular disks while the upper vertical dashed line gives the jamming point $\phi_J\approx 0.906$ for spherocylinders with $\alpha=4$.  In (b) the vertical dashed line gives $\phi/\phi_J=1$.
}
\label{mu-vs-phi-g1e-6} 
\end{figure}

In Figs.~\ref{etasd-vs-phi-g1e-6}(a) and \ref{etasd-vs-phi-g1e-6}(b) we similarly plot the shear viscosity $\eta=\sigma/\dot\gamma$ vs $\phi$ and vs $\phi/\phi_J$, respectively, for different $\alpha$.  We find the same qualitative behavior as found for $\eta_p$.   In Figs.~\ref{mu-vs-phi-g1e-6}(a) and (b) we plot the macroscopic friction $\mu=\sigma/p$ vs $\phi$ and vs $\phi/\phi_J$, respectively.  Again we see that the curves of $\mu$ for different $\alpha$ tend to qualitatively agree when plotted vs $\phi/\phi_J$, though  the difference between the different $\alpha$ is more pronounced than for $\eta_p$ or $\eta$.

\begin{figure}
\centering
\includegraphics[width=3.3in]{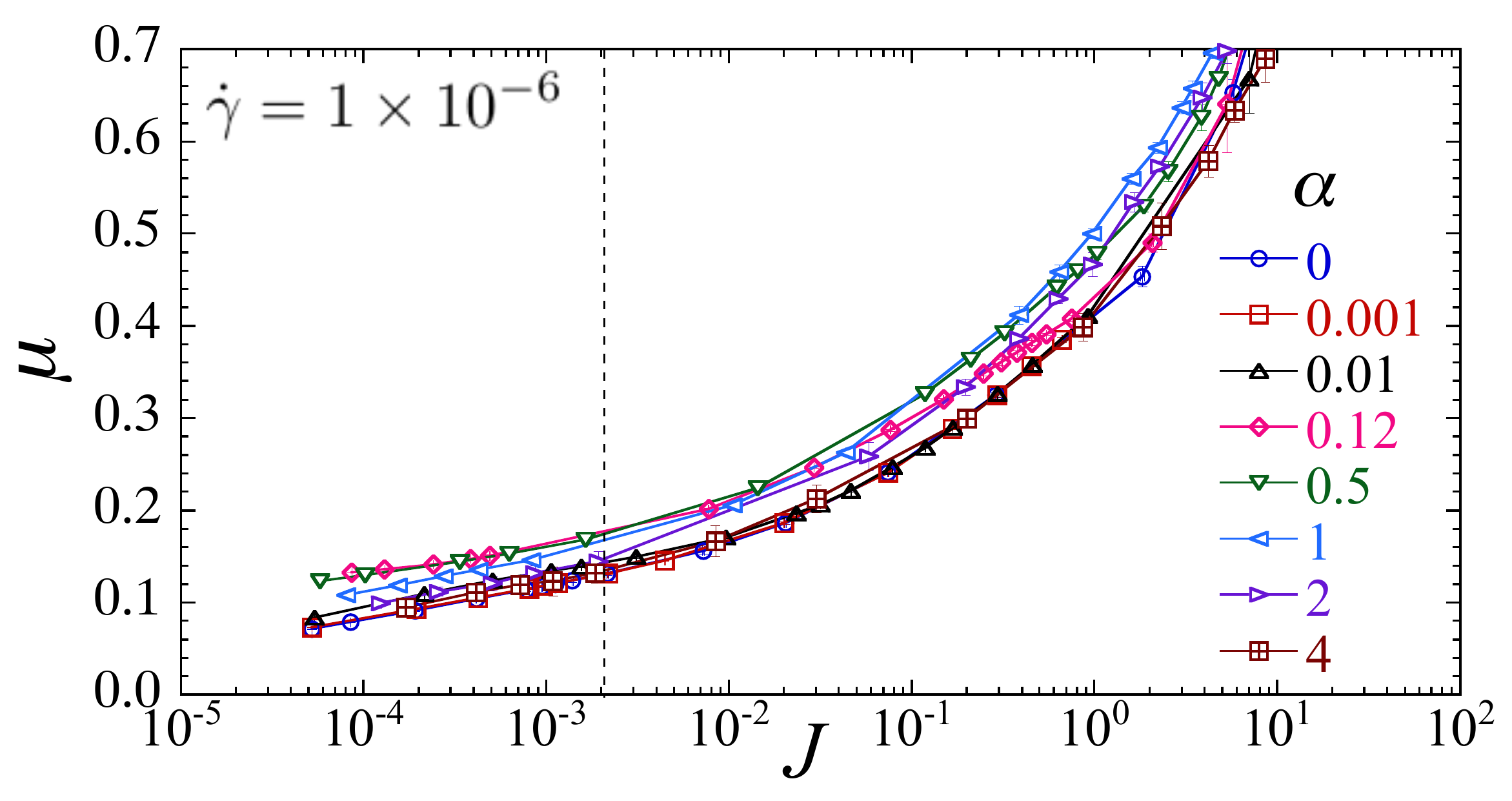}
\caption{Macroscopic friction $\mu=\sigma/p$ vs viscous number $J=\dot\gamma/p$ for particles with different asphericity $\alpha$ at fixed shear strain rate $\dot\gamma=10^{-6}$.  The vertical dashed line at $J_\times=0.002$ separates the region of hard-core behavior $J\gtrsim J_\times$ from soft-core behavior $J\lesssim J_\times$.  We include in this plot results for circular particles with $\alpha=0$.
}
\label{mu-vs-J-1e-6} 
\end{figure}

In Fig.~\ref{mu-vs-J-1e-6} we replot $\mu$ vs the viscous number $J$.  At the fixed strain rate $\dot\gamma=10^{-6}$ used here, we see from Fig.~\ref{etap-vs-phi-g1e-6} that  one is in the hard-core limit when $\eta_p\lesssim 500$.  When one goes above $\eta_p\approx 500$, the hard-core divergence of $\eta_p$ gets cut off by soft-core effects as one approaches and goes above $\phi_J$.  Thus for our system at $\dot\gamma=10^{-6}$, we are in the hard-core limit below $\phi_J$ when $J\gtrsim 0.002$, but we are in the soft-core region, close to and above $\phi_J$, when $J\lesssim 0.002$.  This crossover $J_\times = 0.002$ is indicated by the dashed vertical line in Fig.~\ref{mu-vs-J-1e-6}.  As mentioned earlier in Sec.~\ref{secMu}, it is difficult  to discern from the curves in Fig.~\ref{mu-vs-J-1e-6} alone just where one is crossing from the hard-core to soft-core regions.

Comparing the curves of $\mu$ vs $J$ in Fig.~\ref{mu-vs-J-1e-6} we see that they all follow the same qualitative shape, however there is clearly a spread in values as the asphericity $\alpha$ varies.  Looking carefully, we see that the variation with $\alpha$ is non-monotonic; the smallest and largest $\alpha$ give the smallest $\mu$, while intermediate $\alpha\approx 0.5$ have the largest $\mu$ at small $J>J_\times$.  A similar non-monoticity of $\mu$ with $\alpha$ has previously been reported in simulations of frictionless 3D spherocylinders \cite{Nagy} and frictionless and frictional 2D ellipses \cite{Trulsson} obeying Bagnoldian rheology.

\begin{figure}
\centering
\includegraphics[width=3.3in]{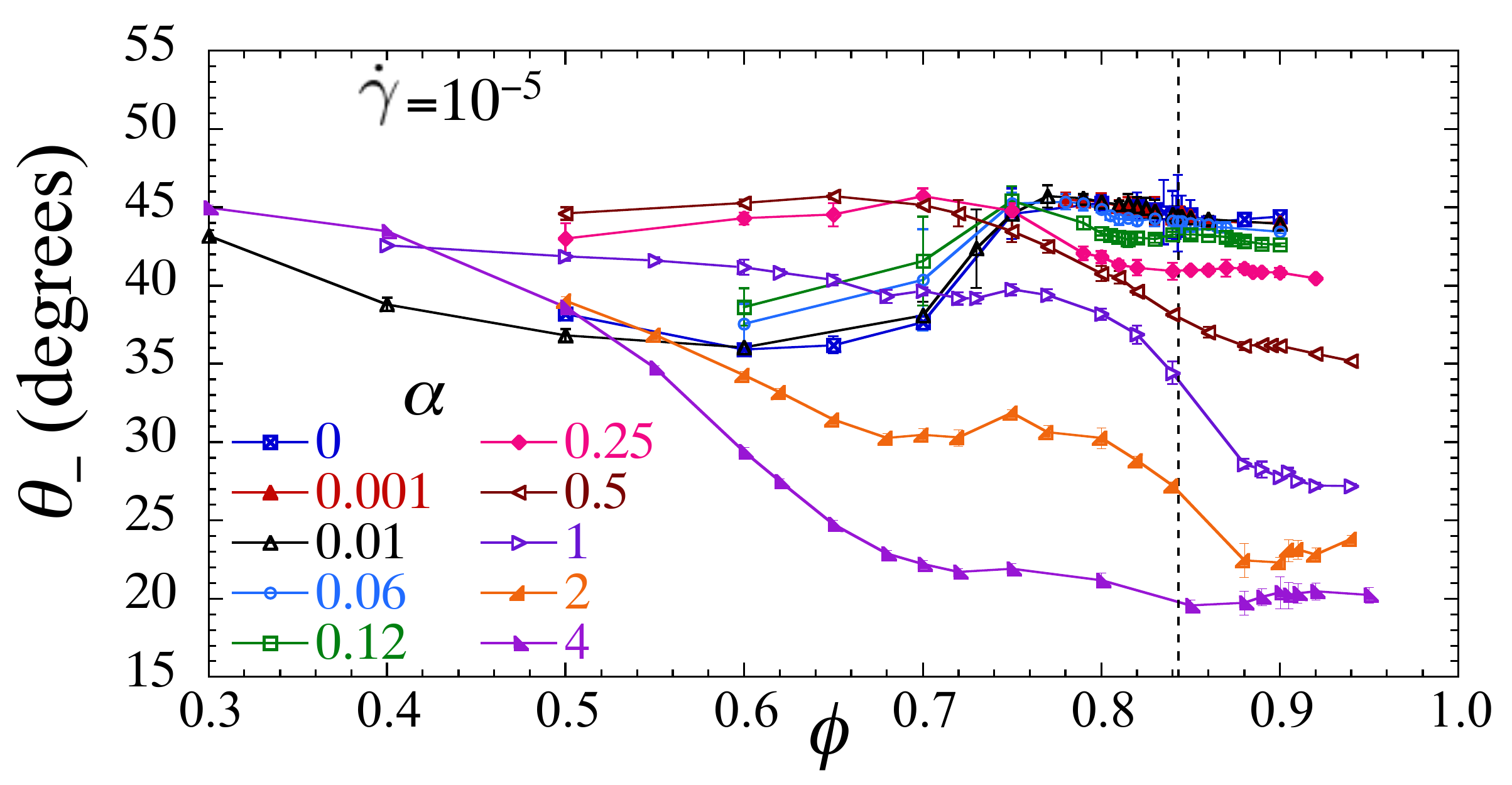}
\caption{Orientation $\theta_-$ of the minimal stress axis of the stress tensor $\mathbf{p}$ vs packing fraction $\phi$ for spherocylinders with various asphericities $\alpha$.  Results are for a strain rate $\dot\gamma=10^{-5}$.  The vertical dashed line locates the jamming transition of circular particles, $\phi_J^{(0)}\approx 0.843$.
}
\label{thetaminus} 
\end{figure}

Finally, in Fig.~\ref{thetaminus} we present results for the direction of the minimal stress axis $\theta_-$ vs $\phi$ for  different  asphericities $\alpha$.  In this plot we use  $\dot\gamma=10^{-5}$, for which we have data at lower values of $\phi$; as noted in connection with Fig.~\ref{thetaminus-v-phi}, the $\dot\gamma$ dependence of $\theta_-$ seems generally quite small.
For small asphericities $\alpha < 0.25$, we see behavior qualitatively similar to that shown previously for $\alpha=0.01$ in Fig.~\ref{thetaminus-v-phi}(a).  There is a non-monotonic variation with $\phi$, with a  peak somewhat below jamming, and with $\theta_-$ becoming close to $45^\circ$ above jamming.  As $\alpha$ increases above 0.25, we see that the height of this peak decreases, the peak becomes less pronounced, and $\theta_-$ at dense $\phi$ steadily decreases as $\alpha$ increases.  We do not have any understanding for this rather complex behavior of $\theta_-$, except to note that as $\alpha$ increases, particles tend to align closer to the flow direction and this presumably influences the eigen-directions of the stress tensor.  This is discussed further in Ref.~\cite{MT2}

\subsection{Contacts}

\subsubsection{Average Contact Number $Z$}

An important concept in jamming is the notion of isostaticity \cite{OHern,LiuNagel,Wyart0}.  For a static, mechanically stable, jammed packing there should be enough particle contacts to constrain the motion of all the $d_f$ degrees of freedom of each of the $\tilde N$ particles that participate in the rigid backbone of the packing.  When the number  of force constraints $N_c$ arising from the particle contacts is exactly equal to the number of particle degrees of freedom, $N_c=\tilde Nd_f$, the system is said to be {\em isostatic}.  For frictionless  particles each contact force is normal to the particle's surface and  gives one force constraint, so the total number of force constraints on the rigid backbone is $N_c = \tilde NZ/2$, where $Z$ is the average number of contacts per backbone particle (the factor of $1/2$ is because each contact is shared by two particles).  For frictionless spheres in $d$ dimensions, rotations of individual particles leave the configuration invariant, and so $d_f=d$, the number of translational degrees of freedom.  Hence the isostatic condition for frictionless spheres is $Z_\mathrm{iso}=2d=4$ in 2D.  For frictionless spherocylinders in 2D one must add one rotational degree of freedom, and so $d_f=3$, giving $Z_\mathrm{iso}=6$.

It has been demonstrated numerically in 2D and 3D  \cite{OHern} that for frictionless spheres, the system is isostatic exactly at the compression-driven jamming transition $\phi_J$, i.e., the average number of contacts at the transition is $Z_J=Z_\mathrm{iso}=2d$.  Numerical studies \cite{Heussinger} of the shear-driven jamming transition of frictionless disks in 2D have also claimed to find $Z_J=Z_\mathrm{iso}$.  However for many non-spherical frictionless particles, and in particular for spherocylinders \cite{VanderWerf,Williams2003PRE,Wouterse.JPCM.2007,MarschallCompress}, packings are  found to be {\em hypostatic} at the compression-driven jamming transition, with $Z_J < Z_\mathrm{iso}$, especially when the particles deviate only slightly from spheres \cite{Donev.Science.2004,Donev.PRE.2007,Zeravcic.EPL.2009,Schreck.PRE.2012,Mailman.PRL.2009,VanderWerf,Wouterse.JPCM.2007,Azema2010,Zhao,MarschallCompress}.  The difference $Z_\mathrm{iso}-Z_J$ has been attributed to eigenmodes of small displacements which are  quartically, rather than quadratically, constrained in an expansion of the elastic energy about the energy minimum of the mechanically stable  configuration at jamming \cite{Donev.PRE.2007,Schreck.PRE.2012,VanderWerf,MarschallCompress}.  Our goal here is to investigate the value of $Z_J$ for the shear-driven jamming of 2D spherocylinders, and compare it to what we have previously found \cite{MarschallCompress} for  compression-driven jamming.

The first step in computing $Z_J$ is to identify the $\tilde N$ particles that participate in the rigid backbone of the packing.  We can write $\tilde N=N-N_r$, where $N_r$ is the number of {\em rattler} particles \cite{OHern}.  A rattler is any particle which is not at a strict local energy minimum, but may move without cost in energy in one or more directions. Such particles may exist in voids created by the rigid backbone.  For circular disks in 2D, an effective algorithm to detect rattlers is to recursively remove any particle with fewer than three contacts.  For 2D spherocylinders, however, the situation is more complicated; because of the flat sides, a particle may have a zero-energy sliding mode in the direction parallel to the spine, while still being important for the stability of the backbone.  We therefore take a particle to be a rattler whenever it has fewer than three contacts, unless there are two contacts that are oriented on opposite flat sides parallel to the spine.
Passing through the configuration to remove such rattlers, we then iterate the process until no further rattlers are found.
We note that for compression-driven jamming, we have found \cite{MarschallCompress} that the fraction of rattlers in the system at jamming decreases significantly as the asphericity $\alpha$ increases, varying from roughly 3.3\% for $\alpha=0.01$ to 0.1\% for $\alpha=4$.

For the current study, with our system sheared at a finite rate $\dot\gamma>0$, there is yet another complication because our flowing configurations are not in mechanically stable states; only in the limit $\dot\gamma\to 0$ do we arrive at mechanically stable states.  The $Z$ that we seek should therefore be taken as the $\dot\gamma\to 0$ limit of the $Z$ computed at finite $\dot\gamma$.
For the purpose of computing $Z$ we count each side-to-side contact (as in Fig.~\ref{rodConfigs}(c)) as two contacts, since  the contact of two flat sides constrains two degrees of freedom: translational motion transverse to the spine, and rotation \cite{MarschallCompress,Azema3}.

\begin{figure}
\centering
\includegraphics[width=3.3in]{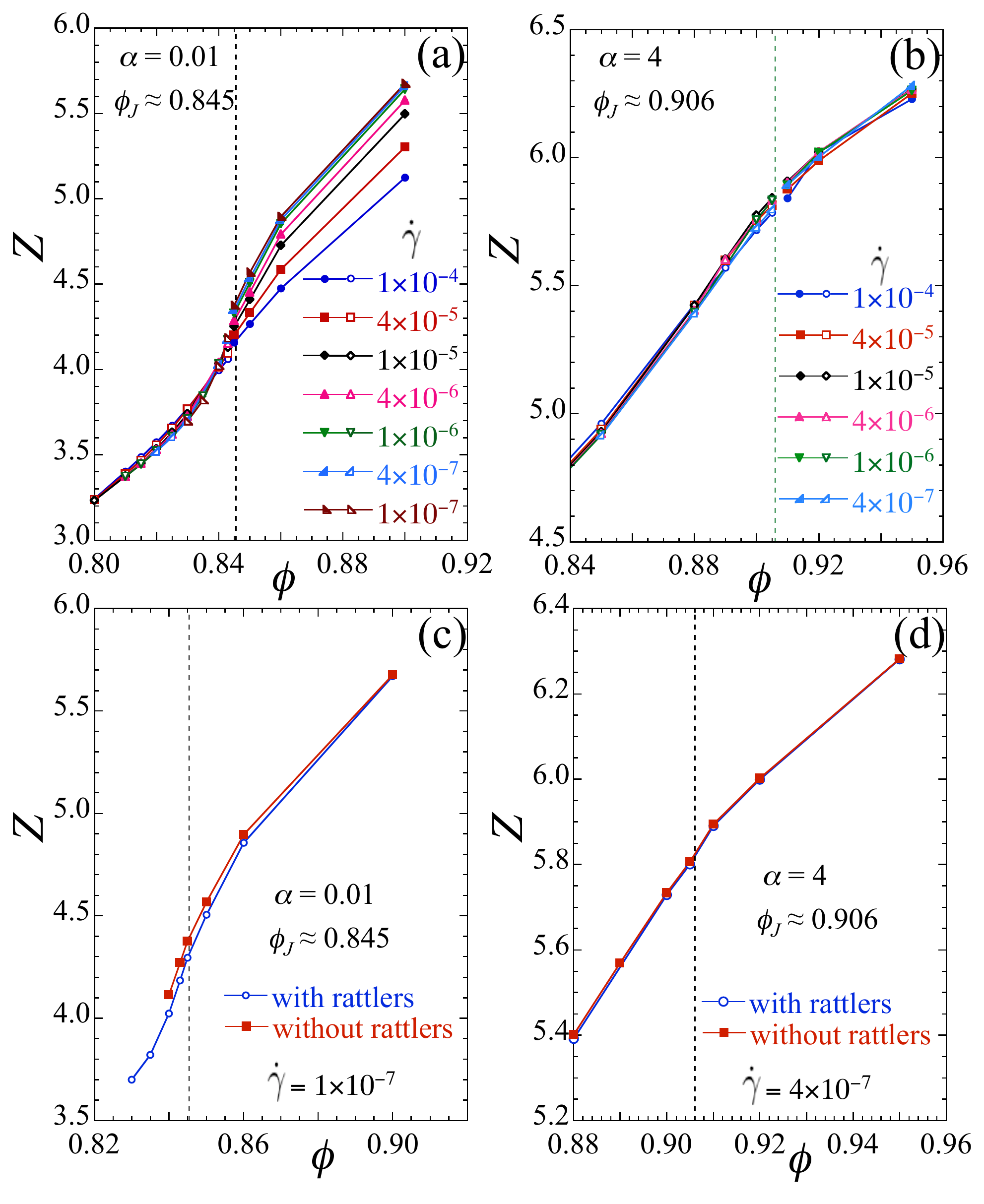}
\caption{Average contact number $Z$ vs $\phi$ at different shear strain rates $\dot\gamma$ for (a) $\alpha=0.01$ and (b) $\alpha=4$; open symbols at $\phi<\phi_J$ include rattler particles in the calculation of $Z$, while solid symbols at $\phi>\phi_J$ exclude rattlers.  Also shown is the average contact number $Z$ vs $\phi$ at the smallest strain rate for (c) $\alpha=0.01$ and (d) $\alpha=4$; open symbols include rattler particles in the calculation of $Z$, while solid symbols exclude rattlers.  Vertical dashed lines denote the jamming density $\phi_J$.
}
\label{Z-vs-phi} 
\end{figure}

When we exclude rattlers from the calculation of $Z$,  we find that most particles become rattlers as $\phi$ decreases below $\phi_J$.  Previous calculations of $Z$ for sheared 2D circular  disks \cite{Heussinger} therefore included rattlers when presenting results for $Z$.  
In Fig.~\ref{Z-vs-phi}(a) and \ref{Z-vs-phi}(b) we plot our results for $Z$ vs $\phi$ at different $\dot\gamma$ for $\alpha=0.01$ and $\alpha=4$, respectively.  
To allow for comparison with previous work \cite{Heussinger}, for $\phi<\phi_J$ we plot the value of $Z$ obtained when including rattlers in the calculation (open symbols);   
for $\phi>\phi_J$, to allow for a more accurate counting of constraints, we exclude ratters when calculating $Z$ (solid symbols).  In both Figs.~\ref{Z-vs-phi}(a) and \ref{Z-vs-phi}(b) the value of $\phi_J$ is indicated by the vertical dashed line.  We see that for small $\alpha=0.01$ the dependence of $Z$ on $\dot\gamma$ is significant as $\phi$ approaches and goes above $\phi_J$; for large $\alpha=4$ the $\dot\gamma$ dependence is significantly reduced.  In both cases we see that $Z$ approaches a limiting curve as $\dot\gamma$ decreases, and the values from our smallest strain rate,  $\dot\gamma=10^{-7}$ for $\alpha=0.01$ and $\dot\gamma=4\times 10^{-7}$ for $\alpha=4$, give a good approximation to this limit.  For $\phi<\phi_J$, $Z$ varies roughly linearly with $\phi$ as found previously for circular disks \cite{Heussinger}. For $\phi>\phi_J$ we see a hint of the $Z-Z_J\propto(\phi-\phi_J)^{1/2}$ dependence found for compression-driven \cite{OHern} and shear-driven \cite{Heussinger} circular disks, however we do not have enough data at $\phi$ close to, but above, $\phi_J$ to check this form in detail.

For completeness, in Fig.~\ref{Z-vs-phi}(c) and \ref{Z-vs-phi}(d) we plot $Z$ vs $\phi$, again for $\alpha=0.01$ and $\alpha=4$ respectively, but this time only for our smallest strain rate.  We show results for $Z$ when both including and excluding rattlers in the calculation at all $\phi$.  For $\alpha=0.01$ we see, as expected, that $Z$ increases slightly near $\phi_J$ when rattlers are excluded, but this difference decreases as $\phi$ increases above $\phi_J$ and the fraction of rattlers decreases.  For $\alpha=4$ the difference is everywhere exceedingly small, reflecting the very small fraction of rattlers even at $\phi_J$.

\begin{figure}
\centering
\includegraphics[width=3.3in]{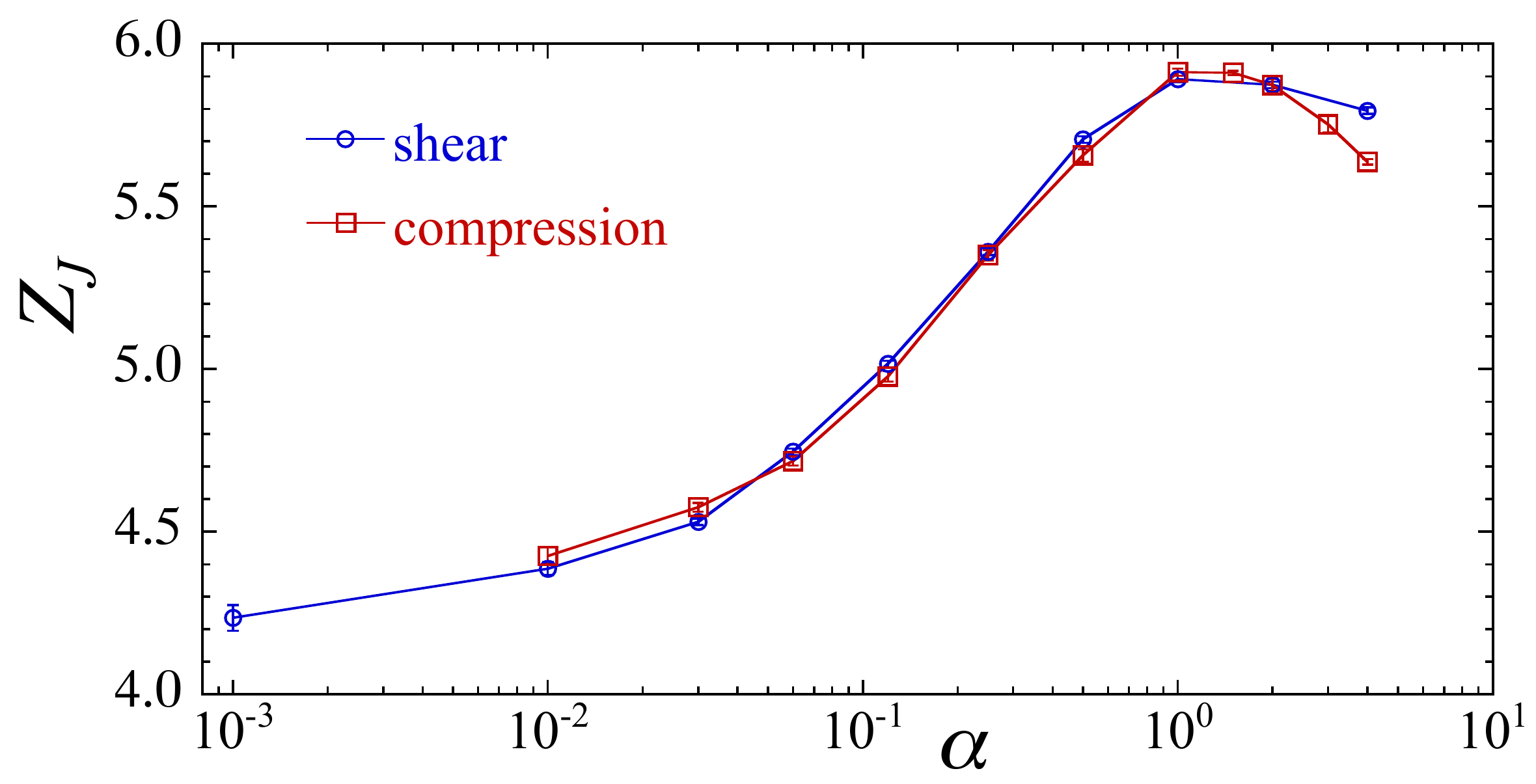}
\caption{Average contact number $Z_J$ at jamming vs particle asphericity $\alpha$.  Here $Z$ is computed excluding rattlers and counting each side-to-side contact twice.  Results for both shear-driven jamming (circles) and compression-driven jamming (squares, from Ref.~\cite{MarschallCompress}) are shown.
}
\label{ZJ-vs-alpha} 
\end{figure}

Using the values of $Z$ at our smallest $\dot\gamma$  as computed excluding rattlers, and the values of $\phi_J$ from Fig.~\ref{phiJ-vs-alpha}, we plot our resulting estimate for $Z_J$ vs $\alpha$ in Fig.~\ref{ZJ-vs-alpha}.  
For comparison, we also show our previous results \cite{MarschallCompress} for $Z_J$ from compression-driven jamming.
We see that $Z_J<Z_\mathrm{iso}=6$ for all $\alpha$, reaching a maximum $Z_J\approx 5.9$ at $\alpha\approx 1$.
Our system, therefore, is hypostatic at the shear-driven jamming transition for all $\alpha>0$.  We see that $Z_J$ from shearing is slightly higher than from compression at large $\alpha>1$, but slightly smaller for small $\alpha<0.03$ (though for our smallest $\alpha$ there may still be finite $\dot\gamma$ effects in our data, which would cause the true $Z_J$ to be slightly higher).
As $\alpha\to 0$ we see that $Z_J$ approaches, but remains slightly larger than, the isostatic value  $Z_\mathrm{iso}=4$ for circular disks.  This is because the fraction of side-to-side contacts remains finite as $\alpha\to 0$ and, as discussed above, we have counted such contacts twice; if we count such contacts only once, then we do find that $Z_J\to 4$ as $\alpha\to 0$.

\begin{figure}
\centering
\includegraphics[width=3.3in]{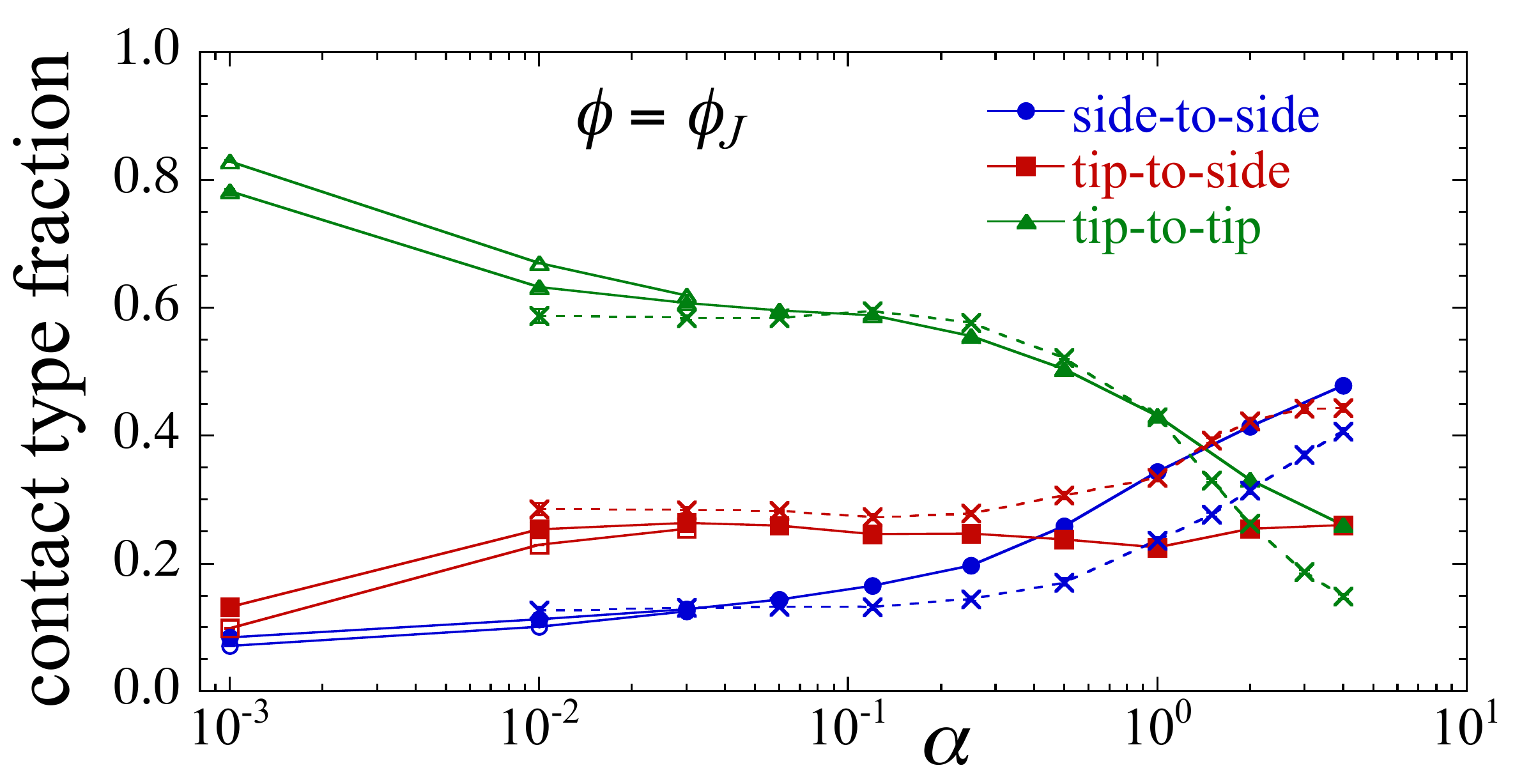}
\caption{Average fraction of side-to-side, tip-to-side, and tip-to-tip contacts at $\phi_J$ vs particle asphericity $\alpha$.  For shear-driven jamming, solid symbols are results from the smallest strain rate $\dot\gamma$ at each $\alpha$, while open symbols are from the next smallest $\dot\gamma$; solid lines connect these data points.  Results for compression-driven jamming are given by crosses connected by dashed lines
}
\label{contacts-vs-alpha} 
\end{figure}

To see this, in Fig.~\ref{contacts-vs-alpha} we plot the fraction of side-to-side, tip-to-side, and tip-to-tip contacts at the jamming $\phi_J$ vs asphericity $\alpha$.  As in our calculation of $Z$,  each side-to-side contact is counted here with a double weight.  We show results from the smallest strain rate $\dot\gamma$ at each $\alpha$ (solid symbols connected by solid lines) and from the next smallest $\dot\gamma$ (open symbols connected by solid lines) to demonstrate that there remains a small but noticeable dependence on $\dot\gamma$ for the smallest $\alpha$.  Our results are qualitatively similar to those of Ref.~\cite{Azema2010}.  For comparison, we show the corresponding results for compression-driven jamming (crosses connected by dashed lines) \cite{MarschallCompress}.  The larger value for the fraction of side-to-side contacts that we see in shear-driven as compared to compression-driven jamming as $\alpha$ increases, is likely due to the increased orientational ordering of particles as $\alpha$ increases \cite{MKOT}.  As $\alpha \to 0$, we see that the fraction of side-to-side contacts appears to be approaching the finite value $\approx 0.1$.
We return to this point further below.

Similar results for  $Z_J$ have been presented for dry granular systems obeying Bagnoldian rheology.  For frictionless 3D spherocylinders, Ref.~\cite{Nagy} finds a $Z_J$ that starts near the 3D spherical isostatic value of $6=2d$ for small $\alpha$, rises to a maximum $\approx 9.5$ that is below the isostatic value of 10 for 3D spherocylinders at $\alpha\approx 0.7$, and then decreases; however these authors do not explain whether they treat side-to-side contacts as a single or double constraint in the counting of $Z_J$.  For frictionless 2D ellipsoids, Ref.~\cite{Trulsson} finds results comparable  to ours, with $Z_J\approx 4$ for nearly circular particles, and then rising monotonically to a value somewhat above 5 as the ellipses become more elongated.  Both these works thus find, in agreement with our results, that shear-driven jamming is hypostatic for all asphericities studied.  When inter-particle friction is added to these models, Ref.~\cite{Trulsson} finds for 2D ellipses that $Z_J$ at small $\alpha$ follows the behavior expected for circular particles,  decreasing from  the frictionless $Z_\mathrm{iso}=2d=4$ to  the infinite frictional $Z_\mathrm{iso}=d+1=3$ as the friction coefficient increases; for large friction, however, it is found that the the variation of $Z_J$ with $\alpha$ is greatly reduced compared to the frictionless case. 

\subsubsection{Contact Location Distribution}

\begin{figure}
\centering
\includegraphics[width=2.in]{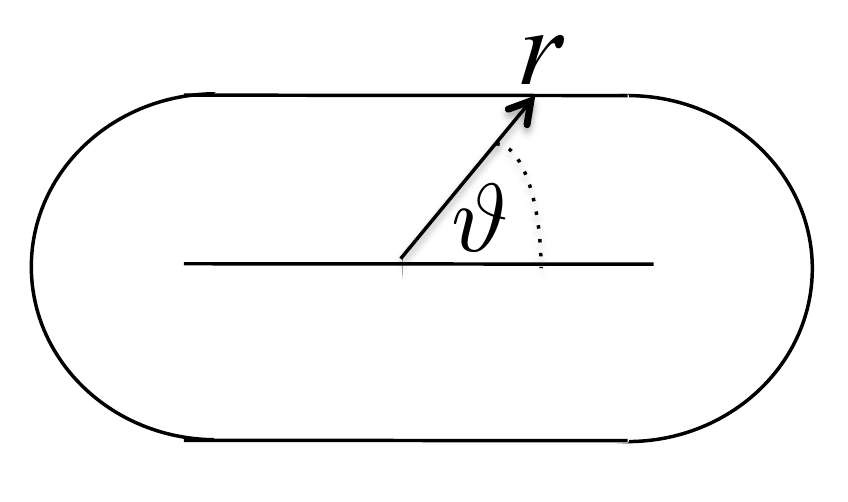}
\caption{Radial distance $r$ and polar angle $\vartheta$ of a point on the surface of a 2D spherocylinder.
}
\label{vartheta}
\end{figure}

Having counted the number of contacts we now turn to ask where these contacts tend to lie on the surface of our particles.
We define $(r,\vartheta)$ as the radial distance and polar angle of a point on the surface with respect to the center of mass of the particle and the direction of the spine, as illustrated in Fig.~\ref{vartheta}.  The probability density {\em per unit length} to find a contact at angle $\vartheta$ is then $\mathcal{P}(\vartheta)$, which is normalized so that,
\begin{equation}
1=\frac{1}{\mathcal{L}}\int_0^{2\pi} \!\!\!d\vartheta\, \sqrt{r^2 + (dr/d\vartheta)^2}\,
\mathcal{P}(\vartheta), 
\label{Pvartheta}
\end{equation}
where $\mathcal{L}=2\pi R_i+4A_i$ is the perimeter length of the spherocylinder, and
$d\ell\equiv d\vartheta\sqrt{r^2+(dr/d\vartheta)^2}$ is the differential surface length subtended by $d\vartheta$ at polar angle $\vartheta$.
A uniform probability per unit surface length is thus characterized by $\mathcal{P}(\vartheta)=1$.  
Note that the angle $\vartheta$ is measured with respect to the  spine of the particle, rather than with respect to the direction of the shear flow.  Since the spine  rotates with the particle, and since our particles have no head nor tail, by symmetry we must have $\mathcal{P}(\vartheta)=\mathcal{P}(\vartheta+\pi)$,
and we therefore restrict our plots below to the range $\vartheta\in [0,\pi]$.
For the purpose of computing $\mathcal{P}(\vartheta)$ we will take a side-to-side contact to have a weight of unity, but  its location distributed uniformly  over the segments of the flat surfaces that are in contact.

\begin{figure}
\centering
\includegraphics[width=3.5in]{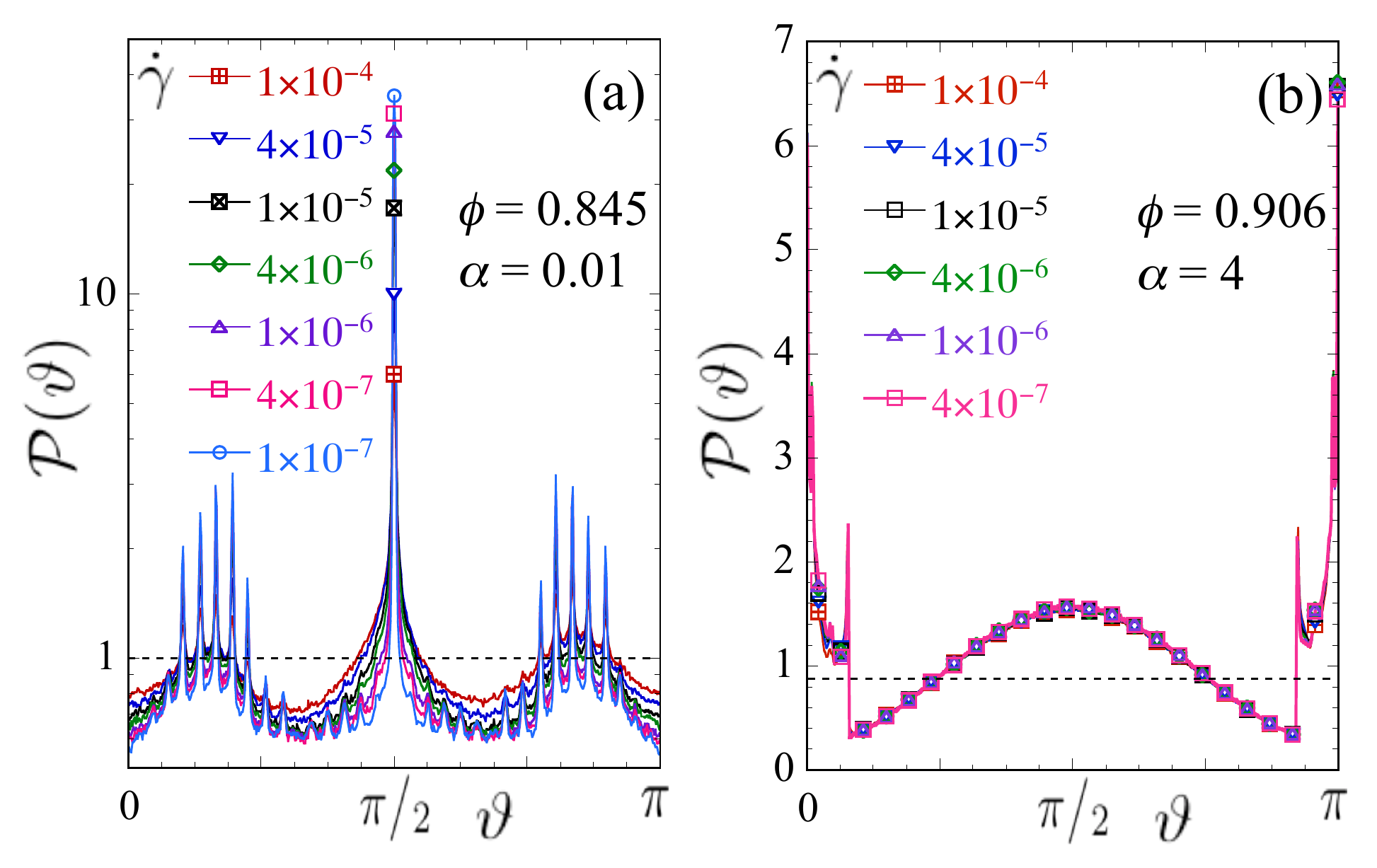}
\caption{Probability per unit length  $\mathcal{P}(\vartheta)$ vs $\vartheta$ for a particle to have a contact at polar angle $\vartheta$ on its surface for different values of the shear strain rate $\dot\gamma$. (a)  Nearly spherical particles with $\alpha=0.01$ at $\phi=0.845$ and (b) elongated particles with $\alpha=4$ at $\phi=0.905$, close to their  jamming fractions $\phi_J=0.8454$ and $0.906$ respectively.  For clarity, in (a) symbols on the different curves are shown only at the central peak at $\vartheta=\pi/2$; in (b) symbols are shown on only every 20th data point.  Note the logarithmic vertical scale in (a).  Dashed horizontal lines represent the value $\mathcal{P}(\vartheta)=1$ that would describe a uniform distribution.  
}
\label{Ptheta-gdot}
\end{figure}

In Figs.~\ref{Ptheta-gdot}(a) and \ref{Ptheta-gdot}(b) we plot $\mathcal{P}(\vartheta)$ vs $\vartheta$ for the cases of a nearly spherical particle with $\alpha=0.01$ and an elongated particle with $\alpha=4$, respectively.   We show results at a fixed packing fraction $\phi$ close to each case's jamming $\phi_J$, for a range of shear strain rates $\dot\gamma$.  For $\alpha=0.01$ in Fig.~\ref{Ptheta-gdot}(a) we see a sharp peak in $\mathcal{P}(\vartheta)$ at $\vartheta=\pi/2$, i.e., the largest probability is along the short flat sides of the spherocylinder, even though the flat sides represent only roughly 0.63\% of the particle perimeter.  This is in stark contrast to the uniform probability distribution expected for a perfectly circular particle.  As $\dot\gamma$ decreases, the heights of the sharp peaks increase, the depths of the minima decrease, and the distribution $\mathcal{P}(\vartheta)$ appears to be approaching a well defined limit as $\dot\gamma\to 0$.  The smaller sharp peaks near $\vartheta\approx \pi/6$ and $5\pi/6$ are shadows of the main peak at $\vartheta=\pi/2$.  In a monodisperse system if a contact exists at $\pi/2$, the next particle contact can be no closer than $\pi/3$ away, i.e. at $\pi/6$ and $5\pi/6$. In a bidisperse system these shadow peaks at $\pi/6$ and $5\pi/6$ get split to allow for contacts between big-big, big-small, and small-small pairs.  
The smaller oscillations in $\mathcal{P}(\vartheta)$ at other angles, which are seen at the smaller values of $\dot\gamma$,  arise from excluded angle effects related to the spacing of additional next-neighbor particle contacts with respect to the reference particle at $\pi/2$.

For elongated particles with $\alpha=4$ Fig.~\ref{Ptheta-gdot}(b) shows  qualitatively different behavior.  While $\vartheta=\pi/2$ remains a local maximum, that maximum is broad, and the largest probability is at the particle tips, $\vartheta=0$ and $\pi$.  Sharp discontinuities are seen at $\vartheta=\pi/2\pm\arctan(\alpha)$, where the flat sides end and the semicircular end-caps begin.  
In contrast to $\alpha=0.01$, we see essentially no dependence of $\mathcal{P}(\vartheta)$ on the strain rate $\dot\gamma$.  

\begin{figure}
\centering
\includegraphics[width=3.5in]{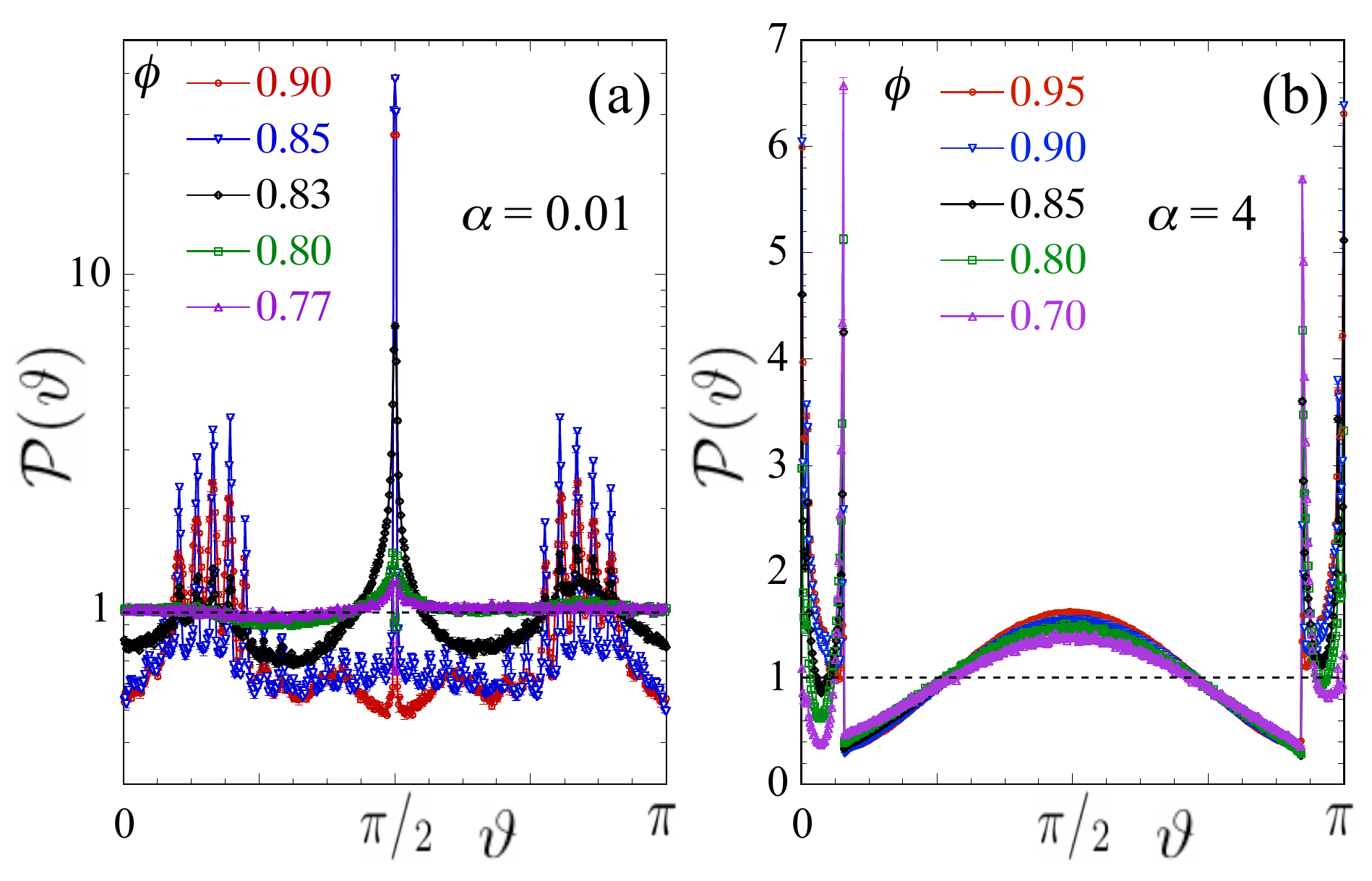}
\caption{Probability per unit length  $\mathcal{P}(\vartheta)$ vs $\vartheta$ for a particle to have a contact at polar angle $\vartheta$ on its surface for different values of the packing fraction $\phi$ at our lowest strain rate $\dot\gamma$. (a)  Nearly spherical particles with $\alpha=0.01$ at $\dot\gamma=10^{-7}$ and (b) elongated particles with $\alpha=4$ at $\dot\gamma=4\times 10^{-7}$.  Note the logarithmic vertical scale in (a).  Dashed horizontal lines represent the value $\mathcal{P}(\vartheta)=1$ that would describe a uniform distribution.  
}
\label{Ptheta-phi}
\end{figure}

In Fig.~\ref{Ptheta-phi} we plot $\mathcal{P}(\vartheta)$ vs $\vartheta$ for different values of the packing fraction $\phi$, at our smallest value of $\dot\gamma$; in Fig.~\ref{Ptheta-phi}(a) we show results for $\alpha=0.01$ at $\dot\gamma=10^{-7}$, while in Fig.~\ref{Ptheta-phi}(b) we show results for $\alpha=4$ at $\dot\gamma=4\times 10^{-7}$. 
For $\alpha=0.01$ above $\phi_J\simeq 0.845$ we see that the peak at $\vartheta=\pi/2$ is exceedingly sharp and there are sharp shadow peaks at $\vartheta\approx \pi/6$ and $5\pi/6$.  In the region near $\vartheta=\pi/2$, but to the sides of the peak, the distribuition decreases as $\phi$ increases.  Slightly below $\phi_J$ the peak broadens and the distribution starts to flatten.  Further below $\phi_J$ ($\phi=0.80$ and 0.77 in Fig.~\ref{Ptheta-phi}(a)) the distribution gets rather flatter, but at $\vartheta=\pi/2$ there now develops a sharp minimum with nearby peaks on either side (one must enlarge the figure in order to see this); the distributions also become slightly asymmetrical about $\vartheta=\pi/2$.
For $\alpha=4$ the main variation as $\phi$ decreases is a slight decrease in the local maximum at $\vartheta=\pi/2$,  a sharpening of the discontinuity at the end of the flat sides $\vartheta=\pi/2\pm\arctan(\alpha)$, and a decrease of the peaks at the particle tips $\vartheta=0$ and $\pi$.

\begin{figure}
\centering
\includegraphics[width=3.5in]{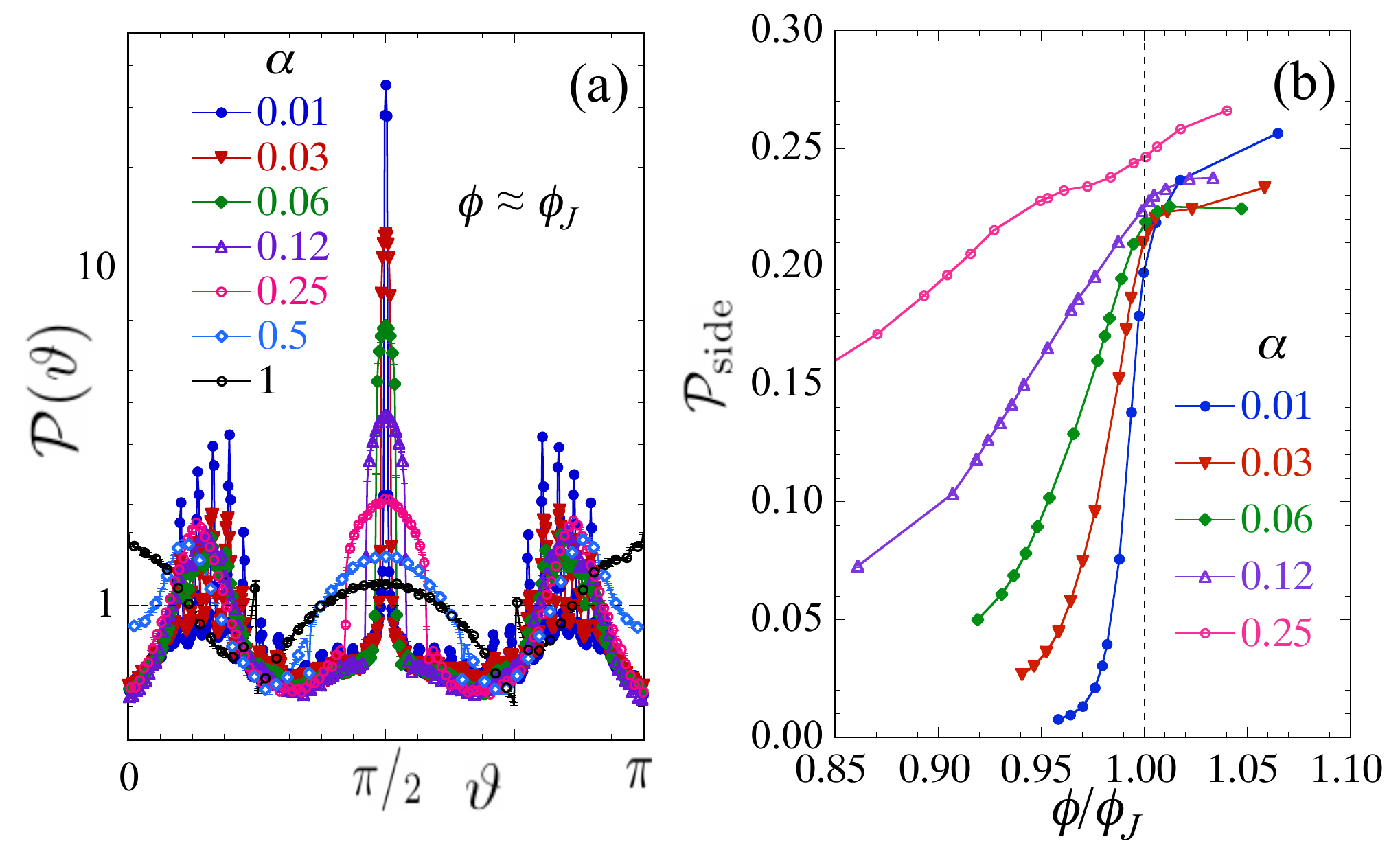}
\caption{(a) Probability per unit length  $\mathcal{P}(\vartheta)$ vs $\vartheta$ for a particle to have a contact at polar angle $\vartheta$ on its surface for different values of particle asphericity $\alpha$, at $\phi\approx\phi_J(\alpha)$ and our lowest strain rate $\dot\gamma$ for each $\alpha$.  Note the logarithmic vertical scale.  The dashed horizontal line represents the value $\mathcal{P}(\vartheta)=1$ that would describe a uniform distribution.  
(b) Total probability $\mathcal{P}_\mathrm{side}$ for a particle to have a contact anywhere on its flat sides vs relative packing fraction $\phi/\phi_J$ for several small $\alpha$.
}
\label{Pside-phi}
\end{figure}

Next, to compare different $\alpha$, in Fig.~\ref{Pside-phi}(a) we plot $\mathcal{P}(\vartheta)$ vs $\vartheta$  for different $\alpha$, at fixed $\phi$  close to the $\alpha$-specific jamming fraction $\phi\approx\phi_J(\alpha)$, for the lowest strain rate $\dot\gamma$ that we have simulated at that $\alpha$; for each $\alpha$ this $\dot\gamma$ is small enough that $\mathcal{P}(\vartheta)$ is close to its $\dot\gamma\to 0$ limiting form.  We see (as reported earlier by us for sheared 2D spherocylinders and 3D ellipsoids \cite{MKOT}) that as $\alpha$ decreases, the peak on the flat side at $\vartheta=\pi/2$ increases in magnitude, while the width of this peak $\Delta\vartheta=2\arctan(\alpha)$ decreases.  
Similar results have been previously reported for static jammed configurations of 2D spherocylinders and ellipses obtained by isotropic compression \cite{VanderWerf,MarschallCompress}.  
Evidence suggesting such an effect has also been reported for both frictionless and frictional  2D ellipses with a Bagnoldian rheology \cite{Trulsson}, though the effect seems to be reduced as the friction coefficient increases; similar conclusions were found for Bagnoldian 3D spherecylinders \cite{Nath}.

To measure the likelihood of a contact on a flat side, we can compute the total probability $\mathcal{P}_\mathrm{side}$ to have a contact anywhere on one of the flat sides of the particle,
\begin{equation}
\mathcal{P}_\mathrm{side}=\frac{2}{\mathcal{L}}\int_{\vartheta_1}^{\vartheta_2}\!\!\! d\vartheta \sqrt{r^2+(dr/d\vartheta)^2}\,\mathcal{P}(\vartheta),
\end{equation}
with $\vartheta_{2,1}=\pi/2\pm\arctan(\alpha)$.  
In Fig.~\ref{Pside-phi}(b) we plot $\mathcal{P}_\mathrm{side}$ vs the relative packing fraction $\phi/\phi_J$ for several smaller values of $\alpha$.  We see that as $\alpha$ decreases, $\mathcal{P}_\mathrm{side}$ stays roughly constant at $\phi_J$.  Thus, as $\alpha\to 0$ and particles approach a circular shape, the flat sides of the spherocylinders become a negligible fraction of the total perimeter but the probability for a contact to lie on a flat side remains constant.

We note that as $\alpha$ gets smaller, we must go to smaller  strain rates $\dot\gamma$ for $\mathcal{P}(\vartheta)$ to approach its $\dot\gamma\to 0$ limit.  If $\alpha$ is decreased keeping $\dot\gamma$ fixed to a constant, and one measured the peak height $\mathcal{P}(\pi/2)$ at  $\phi=\phi_J(\alpha)$, one would see  $\mathcal{P}(\pi/2)$ first increase, then reach a maximum, and then decrease.  We believe this may explain the results of Fig.~5 in Ref.~\cite{Nagy}, which studies sheared frictionless 3D spherocylinders with a Bagnoldian rheology.  For a fixed $\dot\gamma$ at $\alpha=0.05$ they find that collisions strongly peak along the narrow  cylindrical surface of their particles, but for $\alpha=0.01$ this effect is greatly reduced.  We believe this is because they keep $\dot\gamma$ fixed.  In order to see $\mathcal{P}(\pi/2)$ continually grow as $\alpha$ decreases, as in our Fig.~\ref{Pside-phi}(a), it is necessary to similarly decrease $\dot\gamma$ so that $\mathcal{P}(\vartheta)$ is close to its $\dot\gamma\to 0$ limit.  This is why we do not include results for $\alpha=0.001$ in Fig.~\ref{Pside-phi}; for this case our smallest $\dot\gamma=10^{-7}$ is not sufficiently small for $\mathcal{P}(\vartheta)$ to be close to its $\dot\gamma\to 0$ limit.

It is interesting to note that recent experiments have observed a possible confirmation of the behavior of $\mathcal{P}(\vartheta)$ shown in Fig.~\ref{Pside-phi}(a).  Carrying out  experiments in a split-bottom shear cell, Harrington et al. \cite{Losert} have studied the shear behavior of a 3D system of gravity compacted, hole-drilled, spheres.   Their spherical particles have a small hole drilled through them, passing through the center and exiting on opposite sides of the surface, so as to enable imaging of the orientation of the particles.  However this hole also introduces a small but finite asphericity to the particles, due to the absence of the curved surface where the hole exits the sphere.  One can parametrize this asphericity in terms of the cross-sectional area of the hole vs the surface area of the sphere, $2A_\mathrm{hole}/A_\mathrm{sphere}=0.031$, or in terms of the anisotropy of the eigenvalues of the moment of inertial tensor, $I_{2,3}/I_1=0.929$.  
To make a comparison with our work, we note that for spherocylinders the fraction of flat sides to the perimeter is $1/(1+\pi/2\alpha)$, so if we set this fraction to the value 0.031 of the hole-drilled spheres, we get an equivalent of $\alpha\approx 0.05$.  Using the moment of inertial anisotropy would give $(I_1-I_2)/(I_1+I_2)\approx 0.03$, and the results in our Appendix would give an equivalent value $\alpha\approx 0.04$.  Thus by both measures the asphericity is  small, but the results of our Fig.~\ref{Pside-phi}(a) suggest that one would see a $\mathcal{P}(\vartheta)$ that is strongly peaked at the point where the asphericity lies, i.e., at the location of the hole in the experimental hole-drilled spheres, and the height of that peak is about one order of magnitude larger than the value expected if $\mathcal{P}(\vartheta)$ was uniform, as for a perfect sphere.  Figure~6(c) of Ref.~\cite{Losert} shows exactly that behavior.

Finally, another way to characterize the contact distribution $\mathcal{P}(\vartheta)$ is in terms of the orientational ordering of a director-like quantity.  We define $\mathbf{\hat c}$ as the unit vector pointing from the center of the spherocylinder to the  point of contact on the surface at angle $\vartheta$.  Noting the symmetry $\mathcal{P}(\vartheta)=\mathcal{P}(\vartheta+\pi)$, so that $\mathbf{\hat c}$ and $-\mathbf{\hat c}$ are equally likely, the order parameter measuring the orientation of $\mathbf{\hat c}$ should be regarded as a director-like quantity (i.e. a head-less, tail-less, vector) similar to the order parameter of a nematic liquid crystal (note that the orientation of $\mathbf{\hat c}$ we are considering here is defined with respect to axes fixed on the spherocylinder, and so gives no information about the orientation of the spherocylinder itself).  We can then compute an order parameter measuring the $m$-fold orientational order of $\mathbf{\hat c}$, which for a 2D system \cite{Torquato} has magnitude $C_m$ given by,
\begin{equation}
C_m = \sqrt{\langle \cos m\vartheta\rangle^2 +\langle\sin m\vartheta\rangle^2},
\end{equation}
and  is oriented at angle $\vartheta_m$ given by,
\begin{equation}
\tan m\vartheta_m = \dfrac{\langle\sin m\vartheta\rangle}{\langle\cos m\vartheta\rangle},
\end{equation}
where $\langle q(\vartheta)\rangle\equiv (1/\mathcal{L})\int_0^{2\pi} d\vartheta\, \sqrt{r^2+(dr/d\vartheta)^2}\,\mathcal{P}(\vartheta)q(\vartheta)$.  
The magnitudes $C_m$ measure the degree of anisotropy in the contact locations ($C_m=0$ for isotropic and  $C_m=1$ for perfect alignment in a particular direction), while $\vartheta_m$ give the directions in which the density of contacts have maxima.  Note that the angles $\vartheta_m$ are meaningful only modulo $2\pi/m$.

The quantities $C_m$ and $\vartheta_m$  also give the $m$-th Fourier coefficient in a Fourier series expansion of $\mathcal{P}(\vartheta)$,
\begin{equation}
\frac{1}{\mathcal{L}}\sqrt{r^2+\left(\frac{dr}{d\vartheta}\right)^2}\mathcal{P}(\vartheta)=\frac{1}{2\pi}+\frac{1}{\pi}\sum_{m\text{ even}} C_m\cos m(\vartheta-\vartheta_m).
\end{equation}
Since $\mathcal{P}(\vartheta)$ has period $\pi$, only terms with even integer $m$  appear in the sum.

\begin{figure}
\centering
\includegraphics[width=3.5in]{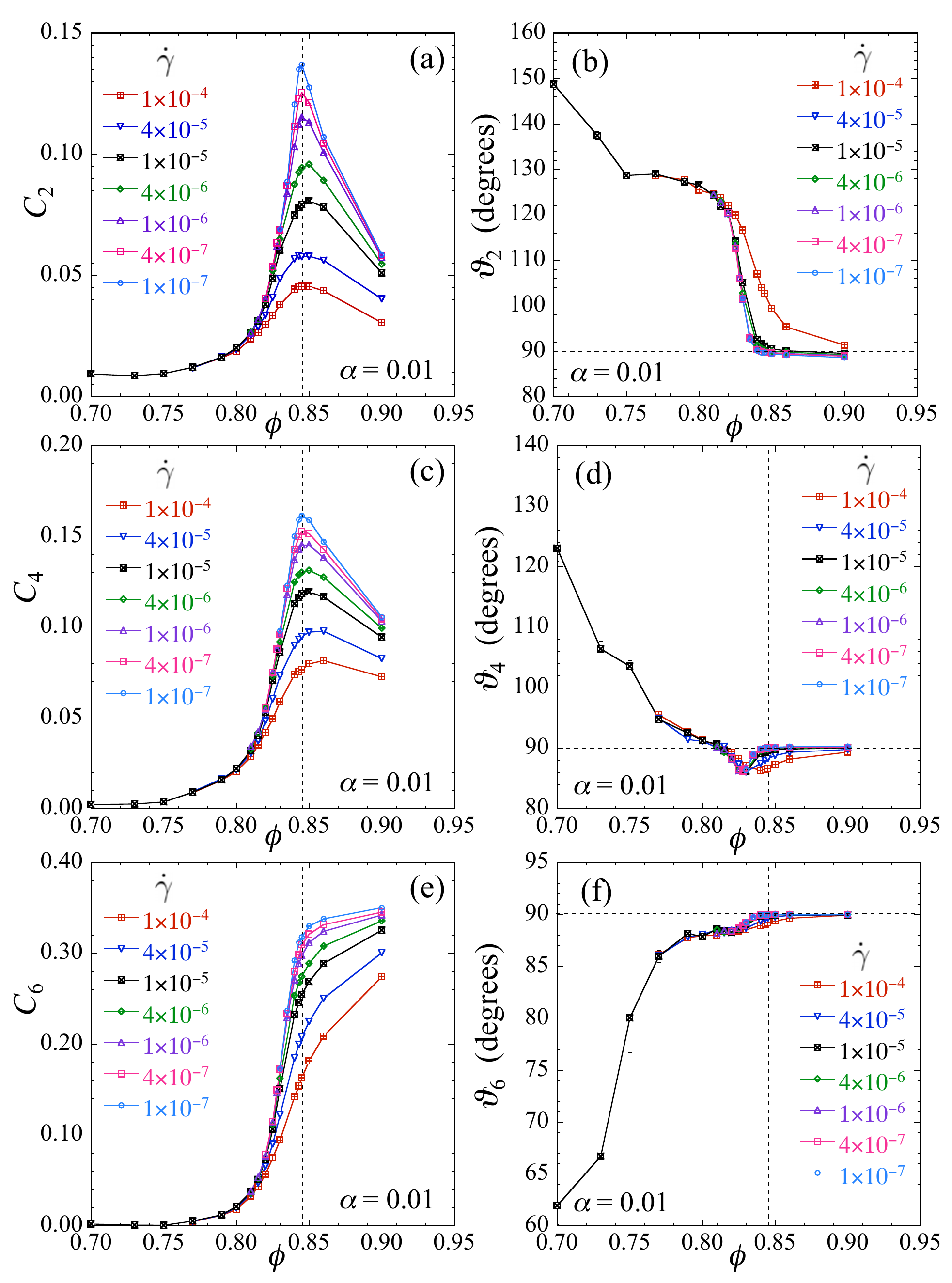}
\caption{Contact orientational order parameter magnitude $C_m$ and director angle $\vartheta_m$ vs $\phi$ for (a) and (b) $m=2$, (c) and (d)  $m=4$, and (e) and (f) $m=6$.  Results are for particles with asphericity $\alpha=0.01$ at different shear strain rates $\dot\gamma$.  The  vertical dashed lines locate the jamming transition $\phi_J\approx 0.845$.  The horizontal dashed lines at $\vartheta_m=90^\circ$ denote a director oriented towards the flat sides of the particle.  Angles $\vartheta_m$ are meaningful only modulo $360^\circ/m$.
}
\label{C-vs-phi-a01}
\end{figure}

In Fig.~\ref{C-vs-phi-a01} we consider nearly circular particles with $\alpha=0.01$ and plot $C_m$ and $\theta_m$ vs $\phi$ for $m=2$, 4 and 6 at different $\dot\gamma$.  We see that $C_2$, $C_4$, and $C_6$ all increase rapidly as one approaches jamming, indicating an increase in the anisotropy of contact locations.  $C_2$ and $C_4$ show a peak at $\phi_J$ that sharpens as $\dot\gamma$ decreases, while $C_6$ levels off but continues to slowly grow as $\phi$ increases above $\phi_J$.  The larger value of $C_6$ compared to $C_2$ and $C_4$, as well as its different dependence on $\phi$ above $\phi_J$, is a consequence of the increasing weight of the distribution $\mathcal{P}(\vartheta)$ in the shadow peaks at $\vartheta\approx\pi/6=30^\circ$ and $5\pi/6=150^\circ$ as $\phi$ increases, see Fig.~\ref{Ptheta-phi}(a).  
Considering the orientation angles, we see that all the $\vartheta_m$ lock onto the value $\pi/2=90^\circ$ once $\phi>\phi_J$.  
Thus, once the system jams, particles show a marked preference to have contacts on their flat sides, consistent with the results shown in Fig.~\ref{Pside-phi}(b), even though these sides form a small fraction of the particle perimeter.  

We note  that, as we presented in an earlier work \cite{MKOT} and  report on in more detail elsewhere  \cite{MT2}, the nematic order parameter $S_2$ that describes the orientational orienting of the spherocylinder spines with respect to the flow direction $\mathbf{\hat x}$ shows a similar qualitative behavior as $C_2$ in Fig.~\ref{C-vs-phi-a01}(a), rising rapidly a $\phi_J$ is approached from below, and then decreasing as $\phi$ increases above jamming.  We believe that the behavior of the $\vartheta_m$ of Fig.~\ref{C-vs-phi-a01} is strongly correlated with the orientational ordering of the nematic order parameter $S_2$.  As found in \cite{MT2}, for $\alpha=0.01$ at lower densities $\phi\lesssim 0.80$, although $S_2$ is small, the particles on average align with their spines parallel to the direction of the  shear flow $\mathbf{\hat x}$.  In this case particle contacts tend to occur along the direction of maximum  stress $\theta_+\approx 135^\circ$, which similarly corresponds to $\vartheta\approx 135^\circ$ as measured with respect to the spine direction.  However as the particles jam, $S_2$ is found to align parallel to the direction of minimum stress $\theta_-\approx 45^\circ$; the direction of maximum stress, measured relative to the direction of the spine, is then $\vartheta\approx 135^\circ - 45^\circ=90^\circ$, corresponding to the location of the flat sides.  However, we believe that it is more a matter of increasing density and the energetics of minimizing particle overlaps, rather than a global alignment of particles, that causes contacts to proliferate on the small flat sides as $\phi$ increases above $\phi_J$.  Recall that results similar to those in Fig.~\ref{Pside-phi}(a) have also been reported for compression-driven jamming \cite{VanderWerf,MarschallCompress}, even though there is no nematic ordering of the particle spine directions, and so $S_2=0$, in that case.

\begin{figure}
\centering
\includegraphics[width=3.5in]{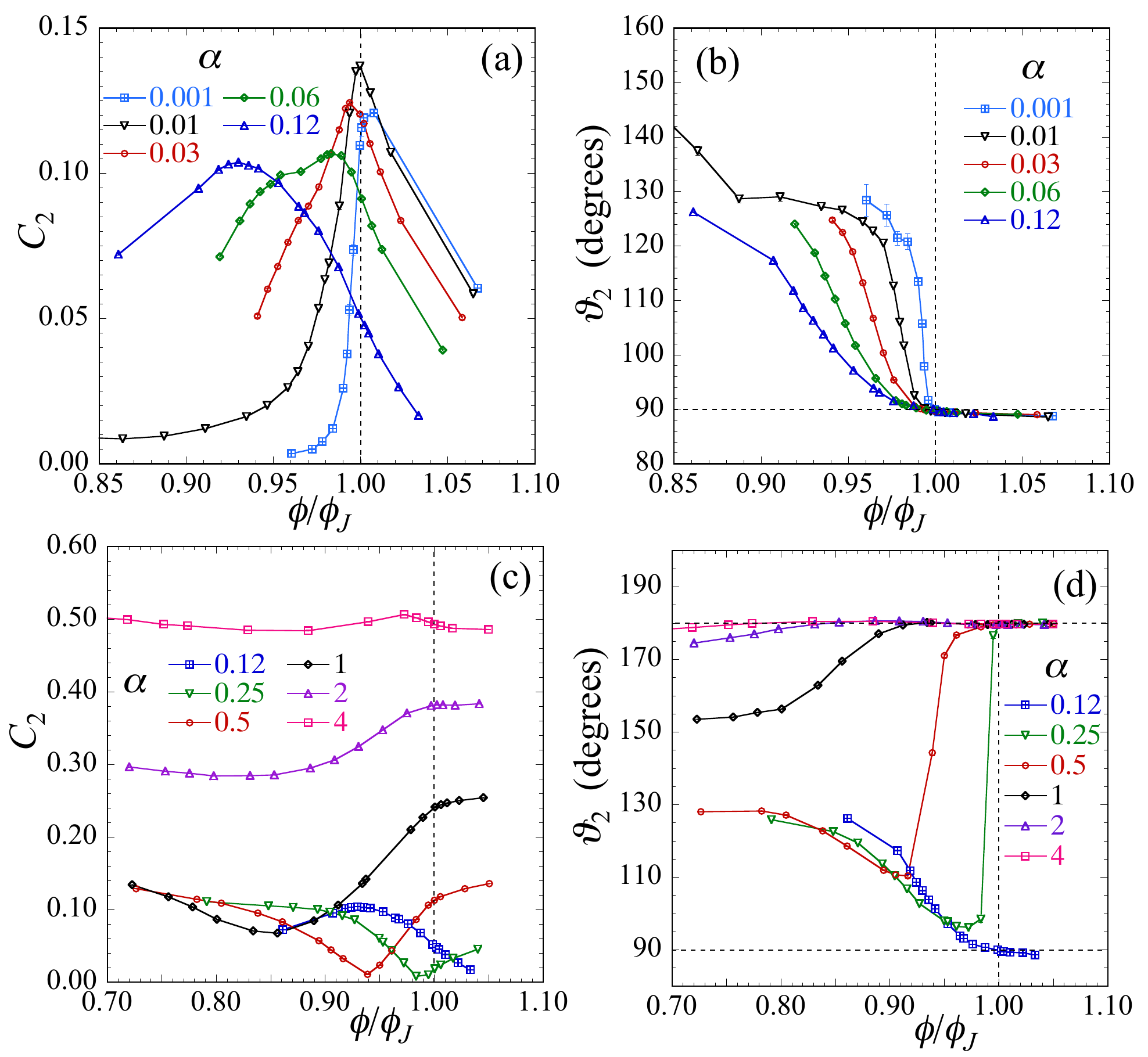}
\caption{Contact orientational order parameter magnitude $C_2$ and director angle $\vartheta_2$ vs $\phi/\phi_J$ for (a) and (b) particles with small asphericity $\alpha\le 0.12$ and (c) and (d)  particles with larger asphericity $\alpha \ge 0.12$.  For each $\alpha$ the results are for the lowest strain rate $\dot\gamma$ that we have simulated.  The  vertical dashed lines locate the jamming transition $\phi/\phi_J=1$.  The horizontal dashed lines at $\vartheta_2=90^\circ$ denote a director oriented towards the flat sides of the particle.  In (d) the horizontal dashed line at $\vartheta_2=180^\circ$ denotes a director oriented towards the tips of the particles.  Angles $\vartheta_2$ are meaningful only modulo $180^\circ$.
}
\label{C2-vs-phi-all}
\end{figure}

In Fig.~\ref{C2-vs-phi-all} we consider the behavior at other values of $\alpha$, plotting $C_2$ and $\vartheta_2$ vs the normalized packing fraction $\phi/\phi_J$.  We show results only from our lowest value of the strain rate $\dot\gamma$ at each $\alpha$.  For nearly spherical particles with $\alpha\le 0.06$, Figs.~\ref{C2-vs-phi-all}(a) and \ref{C2-vs-phi-all}(b) show that results are qualitatively similar to what was shown for $\alpha=0.01$ in Figs.~\ref{C-vs-phi-a01}(a) and \ref{C-vs-phi-a01}(b); $C_2$ peaks near, or just a bit below, $\phi_J$ and $\vartheta_2$ locks onto the value $90^\circ$ above $\phi_J$; the width over which $C_2$ rises to its peak value decreases as $\alpha$ decreases.  Our results for larger $\alpha\ge 0.12$ are shown in Figs.~\ref{C2-vs-phi-all}(c) and \ref{C2-vs-phi-all}(d).  For $\alpha=0.12$ the behavior is similar to the smaller $\alpha=0.06$ in that $S_2$ peaks somewhat below $\phi_J$ and $\vartheta_2=90^\circ$ above jamming.  However for larger $\alpha$ we see a qualitative change in behavior.  For $\alpha=0.25$ and 0.5 as $\phi$ increases, $\vartheta_2$ follows the same behavior as that of $\alpha=0.12$, but upon approaching $\phi_J$, $\vartheta_2$ shows an abrupt increase to $\vartheta=180^\circ$ and stays locked into that value as $\phi$ increases above jamming; as $\alpha$ increases, the location of this abrupt change decreases to lower $\phi$.  
Corresponding to this abrupt change in $\vartheta_2$, the magnitude $C_2$ takes a dip almost to zero.
The value $\vartheta_2=180^\circ$ indicates that the contact distribution $\mathcal{P}(\vartheta)$ is now peaking at the particle tips rather than the sides, as is evident in Fig.~\ref{Ptheta-phi}(b) for the larger value $\alpha=4$.   
For larger values of $\alpha=1$, 2 and 4, we see a similar behavior but the variations in $\vartheta_2$ and $C_2$ are more gradual.
The relatively large values of $C_2$ that we find at low $\phi$ for these larger values of $\alpha$ is a result of the sizable nematic ordering of the particle spine orientations with respect to the shear flow direction (with large $S_2$) that we find for such elongated rods even at low $\phi$ \cite{MKOT}.


\subsubsection{Contract Force Distribution}

Having found the distribution of the location of the contacts $\mathcal{P}(\vartheta)$, we now wish to investigate the relative magnitude of these contact forces as $\vartheta$ varies.  We  define the average magnitude of the force  per unit length on the particle surface at polar angle $\vartheta$ to be $\mathcal{F}(\vartheta)$.
The force per unit length is normalized so that,
\begin{equation}
\int_0^{2\pi}\! d\vartheta \sqrt{r^2+(dr/d\vartheta)^2}\,\mathcal{F}(\vartheta)=F^\mathrm{total},
\end{equation}
where $F^\mathrm{total}$ is just the average pressure on a particle's surface multiplied by the surface perimeter $\mathcal{L}$,
\begin{equation}
F^\mathrm{total}=\frac{1}{N}\sum_{i=1}^N\sideset{}{^\prime} \sum_{j}  \left|\mathbf{F}_{ij}^\mathrm{el}\right| = Z\langle |\mathbf{F}_{ij}^\mathrm{el}|\rangle.
\end{equation}
Here the second sum is over all particles $j$ in contact with a given particle $i$, and we average over all particles $i$.

\begin{figure}
\centering
\includegraphics[width=3.5in]{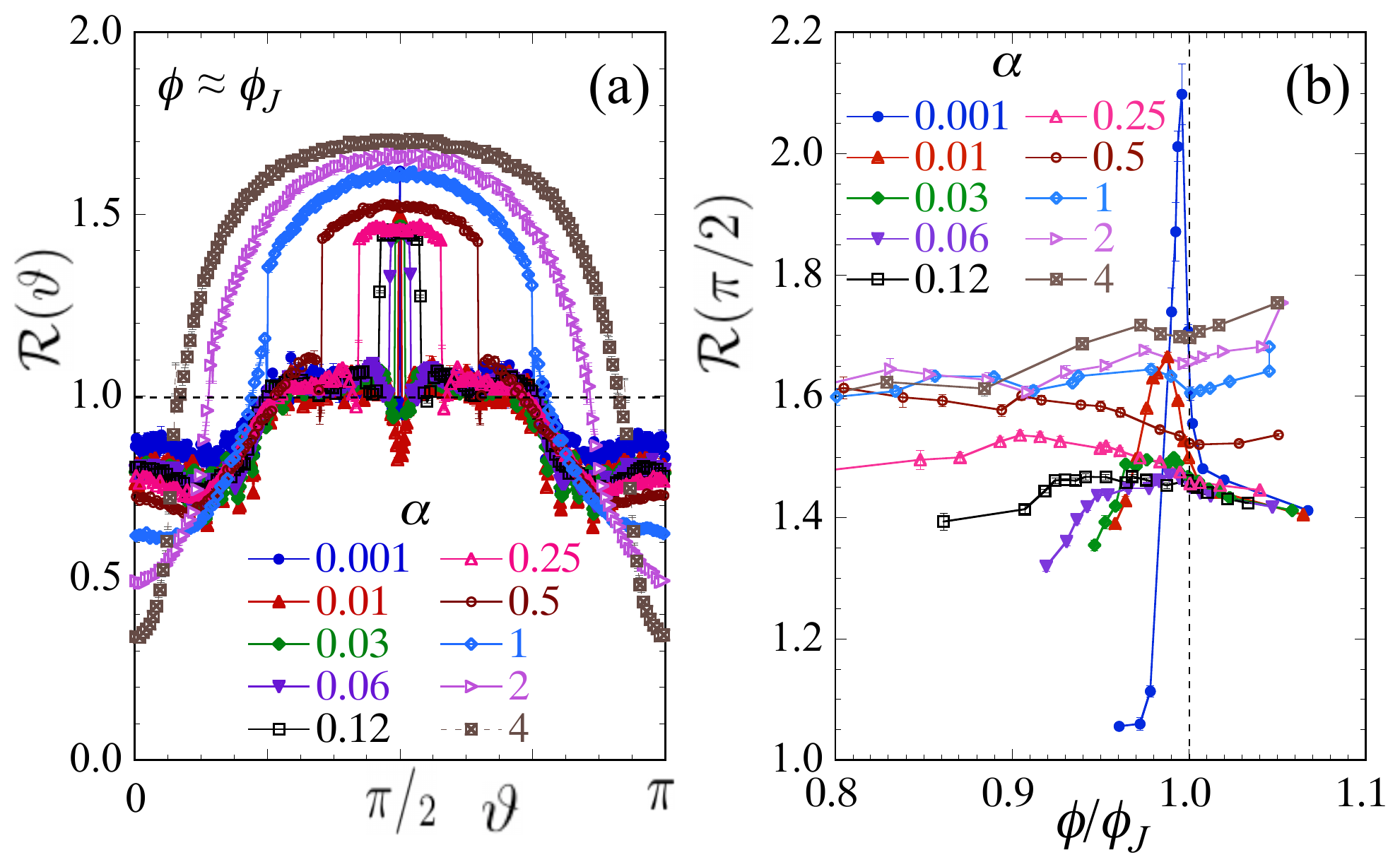}
\caption{(a) Force ratio  $\mathcal{R}(\vartheta)$ of Eq.~(\ref{Ratio}) vs $\vartheta$  for different values of particle asphericity $\alpha$, at $\phi\approx\phi_J(\alpha)$.   The dashed horizontal line represents the value $\mathcal{R}(\vartheta)=1$ that would describe a uniform force distribution.  
(b) Peak value $\mathcal{R}(\pi/2)$  vs relative packing fraction $\phi/\phi_J$ for different $\alpha$.  Results in both (a) and (b) are from our lowest strain rate $\dot\gamma$ for each $\alpha$.
}
\label{Rtheta}
\end{figure}

If the average magnitude of the contact force  $|\mathbf{F}_{ij}^\mathrm{el}|$ was independent of where on the surface of the particle the contact lies, we would expect to have, $\mathcal{F}(\vartheta)=\mathcal{P}(\vartheta)F^\mathrm{total}/\mathcal{L}$, so that the force on the surface at $\vartheta$ would simply be  determined by the probability to have a contact at $\vartheta$.  To look for deviations from this we therefore plot in Fig.~\ref{Rtheta}(a) the ratio,
\begin{equation}
\mathcal{R}(\vartheta)=\dfrac{\mathcal{F}(\vartheta)\mathcal{L}}{\mathcal{P}(\vartheta)F^\mathrm{total}},
\label{Ratio}
\end{equation}
vs $\vartheta$ for different values of $\alpha$.  For each $\alpha$ we show results close the the $\alpha$-specific jamming packing $\phi\approx \phi_J(\alpha)$, at the smallest strain rate $\dot\gamma$ that we have for that $\alpha$; we include results for $\alpha=0.001$ even though our smallest strain rate for that case, $\dot\gamma=10^{-7}$, is still not close to the $\dot\gamma\to 0$ limit.  When $\mathcal{R}(\vartheta)>1$ then the average contact force at that $\vartheta$ is larger than  the average contact force.  We see  from Fig.~\ref{Rtheta}(a) that forces located on the flat sides of the particles tend to be larger than the average, while forces on the semi-circular end caps are generally smaller than the average.  

In Fig.~\ref{Rtheta}(b) we plot the peak value $\mathcal{R}(\pi/2)$ vs the relative packing $\phi/\phi_J$ for different $\alpha$, at the smallest strain rate $\dot\gamma$ that we have for that $\alpha$.  For the larger $\alpha$ we see that $\mathcal{R}(\pi/2)$ varies little as $\phi$ passes through the jamming $\phi_J$.  However for small $\alpha$ there is a clear peak somewhat below $\phi_J$, that moves closer to $\phi_J$ as $\alpha$ decreases.  Reference \cite{MT2} shows that the behavior of $\mathcal{R}(\pi/2)$ behaves qualitatively similarly to the behavior of the nematic order parameter $S_2$; when particles are more aligned, the average force on the flat sides increases.

\section{Conclusions}

We have studied the behavior of an athermal, bidisperse, distribution of frictionless soft-core spherocylinders in two dimensions, driven by a uniform steady-state shear strain applied at a fixed rate.  Energy dissipation in our model is via a viscous drag with respect to a uniformly sheared host medium, thus modeling flow in a non-Brownian suspension and resulting in Newtonian rheology.  We have studied behavior as a function of particle packing fraction $\phi$, shear strain rate $\dot\gamma$, and particle asphericity $\alpha$, focusing on behavior near the jamming transition $\phi_J$.  Unlike  compression-driven jamming, where  $\phi_J(\alpha)$ is a non-monotonic function of $\alpha$ peaking near $\alpha=1$, we find for shear-driven jamming that $\phi_J(\alpha)$ is monotonically increasing in $\alpha$, at least to the largest $\alpha=4$ that we have studied.  We believe this difference  is due to the nematic orientational ordering of particles that takes place in shear-driven flow, allowing particles to pack more densely; no such ordering was observed in isotropically compressed states.  However, as  was found for compression-driven jamming, we found for shear-driven jamming that the average number of contacts at jamming, $Z_J$, is always hypostatic, varying non-monotonically in $\alpha$ with a peak near $\alpha=1$.

Concerning the stress in the system, we found that the stress tensor is in general not co-aligned with the strain tensor, except for the case of nearly circular particles with small $\alpha$.  Considering the viscosity transport coefficients $\eta_p=p/\dot\gamma$ and $\eta=\sigma/p$ for different $\alpha$, we found that these behave qualitatively the same as a function of the packing, provided one plots as a function of a normalized packing $\phi/\phi_J(\alpha)$.   
However, a scaling analysis of pressure for our most elongated particles with $\alpha=4$  suggests that the universality class of the jamming transition for spherocylinders may be different from  that of circular disks  ($\alpha=0$), with the exponent $\beta$ that describes the divergence of $\eta_p$ being larger for spherocylinders than for circular disks.

We have also examined the Herschel-Bulkley rheology, $p=p_0+c\dot\gamma^n$ (and similarly $\sigma=\sigma_0+c^\prime\dot\gamma^{n^\prime}$) above jamming,  fitting to this form for the entire range of $\dot\gamma\le 10^{-4}$ that we have simulated.   We found that the empirically determined exponents $n$ and $n^\prime$ take a range of values $0.2\lesssim n\lesssim 0.5$ as $\phi$ and $\alpha$ vary, and that $n$ obtained from pressure $p$ is generally not the same as $n^\prime$ obtained from shear stress $\sigma$.  Thus we believe that the values of $n$ and $n^\prime$ which we found here are just effective values obtained for our particular range of $\dot\gamma$ and do not necessarily reflect the true asymptotic values that would  describe behavior in the limit  $\dot\gamma\to 0$.

Finally we have considered the probability per unit surface length $\mathcal{P}(\vartheta)$ for a particle to have a contact on its surface at angle $\vartheta$, as measured with respect to the particle's spine (see Fig.~\ref{vartheta}).  We found that $\mathcal{P}(\vartheta)$ approaches a limiting form as the strain rate $\dot\gamma\to 0$.  As $\alpha\to 0$, and particles approach circular, this limiting form develops a sharp peak at $\vartheta=\pi/2$ (i.e., along the flat sides of the spherocylinders) as $\phi$ approaches and goes above the jamming $\phi_J$.  This is in contrast to the uniform distribution that would be expected for a perfectly circular particle.  Moreover, in this small $\alpha$ limit, we found that the total probability $\mathcal{P}_\mathrm{side}$ for a particle to have a contact anywhere along its flat sides appears to approach an $\alpha$-independent constant  at jamming.  Thus, as $\alpha\to 0$ and the length of the flat sides is shrinking to a negligible fraction of the particle perimeter, we found that the probability for a contact to be on the flat sides is nevertheless staying constant.  This signals that the $\alpha\to 0$ limit is in some sense singular.  We have found similar results for ellipsoids in three dimensions \cite{MKOT}, suggesting that this conclusion may hold for more generally aspherical particles rather than being specific to spherocylinders.
We further found that, for all $\alpha$, the  magnitude of the elastic force for contacts located on the flat sides is larger than the average; for forces at the particle tips it is smaller than the average.

In Ref.~\cite{MT2} we present our results for the orientational and translational order in the system.  That analysis provides other indicators that the $\alpha\to 0$ limit is singular.

\section*{Acknowledgements}

This work was supported in part by National Science Foundation Grants  CBET-1435861 and DMR-1809318. Computations were carried out at the Center for Integrated Research Computing at the University of Rochester. 


\setcounter{figure}{0}
\setcounter{equation}{0}
\renewcommand{\theequation}{A\arabic{equation}}
\renewcommand{\thefigure}{A\arabic{figure}}

\section*{Appendix}

In this Appendix we derive the force-moment tensor $\boldsymbol{\Sigma}_i^\mathrm{dis}$ for particle $i$, arising from the dissipative drag force in a uniform shear flow.  We treat a generally shaped particle.  Having found $\boldsymbol{\Sigma}_i^\mathrm{dis}$, we will then use it to compute the dissipative torque on the particle, $\tau_i^\mathrm{dis}$.  

If $\mathbf{r}_i$ is the center of mass of particle $i$, then we can write for a general position $\mathbf{r}$ on the particle,
\begin{equation}
\mathbf{r}=\mathbf{r}_i+\mathbf{\delta r}.
\end{equation}
We then have
\begin{equation}
\boldsymbol{\Sigma}_i^\mathrm{dis}=\int_i d^2 \delta r\, \mathbf{\delta r}\otimes \mathbf{f}_i^\mathrm{dis}(\mathbf{r}),
\end{equation}
where the integral is over the area of particle $i$ and from Eq.~(\ref{efrdis})
\begin{equation}
\mathbf{f}_i^\mathrm{dis}(\mathbf{r})=-k_d\left[{\mathbf{v}}_i+\dot\theta_i\mathbf{\hat z}\times\mathbf{\delta r}-\mathbf{v}_\mathrm{host}(\mathbf{r})\right],
\end{equation}
with $\mathbf{v}_i$  the center of mass velocity and $\dot\theta_i$ the angular velocity about the center of mass.  We are interested  in linear deformations of the host medium  for which
\begin{equation}
\mathbf{v}_\mathrm{host}=\boldsymbol{\dot\gamma}\cdot\mathbf{r},
\end{equation}
where $\boldsymbol{\dot\gamma}$ is the strain rate tensor.

Taking a uniform  mass density for the particle,  the definition of the center of mass gives,
\begin{equation}
\int_i d^2\delta r\, \mathbf{\delta r}=0,
\end{equation}
and the force-moment tensor reduces to,
\begin{equation}
\boldsymbol{\Sigma}_i^\mathrm{dis}=-k_d\int_i d^2 \delta r\, \mathbf{\delta r}\otimes\left[  \dot\theta_i\mathbf{\hat z}\times \mathbf{\delta r}-\boldsymbol{\dot\gamma}\cdot\mathbf{\delta r}
\right].
\end{equation}

In this work we are interested in simple shear with flow in the $\mathbf{\hat x}$ direction, $\mathbf{v}_\mathrm{host}=\dot\gamma y \mathbf{\hat x}$, for  which
\begin{equation}
\boldsymbol{\dot\gamma}_\mathrm{ss}=\left[
\begin{array}{cc}
0 &\dot\gamma\\[10pt]
0 & 0
\end{array}
\right],
\end{equation}
and so we get

\begin{equation}
\boldsymbol{\Sigma}_i^\mathrm{dis}=k_d\int_i d^2\delta r\left[
\begin{array}{cc}
 (\dot\theta_i+\dot\gamma)\delta x\delta y &  -\dot\theta_i \delta x^2    \\[10pt]
 (\dot\theta_i+\dot\gamma)\delta y^2 & -\dot\theta_i \delta x\delta y     
\end{array}
\right].
\end{equation}
Assuming a uniform unit mass density for all particles, the moment of inertia tensor for particle $i$ is,
\begin{equation}
\mathbf{I}_i
=\frac{1}{\mathcal{A}_i}\int_i d^2\delta r\,\left[
\begin{array}{cc}
\delta y^2 & -\delta x\delta y    \\[10pt]
 -\delta x\delta y & \delta x^2     
\end{array}
\right],
\end{equation}
where $\mathcal{A}_i$ is the area of the particle,
\begin{equation}
\mathcal{A}_i=\int_i d^2\delta r.
\end{equation}
Hence
\begin{equation}
\boldsymbol{\Sigma}_i^\mathrm{dis}=k_d\mathcal{A}_i\left[
\begin{array}{cc}
 -(\dot\theta_i+\dot\gamma)I_{ixy} & -\dot\theta_i I_{iyy}   \\[10pt]
(\dot\theta_i+\dot\gamma)I_{ixx} & \dot\theta_i I_{ixy}     
\end{array}
\right].
\label{eSigdisAppend}
\end{equation}
Since $\mathbf{I}_i$ is a symmetric tensor, it may be diagonalized.  Labeling its two eigenvalues as $I_{i1}$ and $I_{i2}$, with $I_{i2}\ge I_{i1}>0$, and the corresponding orthonormal eigenvector directions as $\mathbf{\hat e}_{i1}$ and $\mathbf{\hat e}_{i2}$, we can denote the orientation of the  axis $\mathbf{\hat e}_{i1}$ with respect to the flow direction $\mathbf{\hat x}$ by the angle $\theta_i$.  For a spherocylinder, $\mathbf{\hat e}_{i1}$ is just the direction along the spine.  Using a rotation of coordinates transformation, one can then express $I_{ixx}$, $I_{iyy}$, and $I_{ixy}$ in terms of $I_{i1}$, $I_{i2}$, and $\theta_i$.  Defining 
\begin{equation}
I_i=I_{i1}+I_{i2}\quad \mathrm{and}\quad \Delta I_i=I_{i2}-I_{i1},
\end{equation} 
we have,
\begin{align}
I_{ixx}&=\frac{1}{2}\left(I_i-\Delta I_i\cos 2\theta_i\right)\label{eIxx}\\[5pt]
I_{iyy}&=\frac{1}{2}\left(I_i+\Delta I_i\cos 2\theta_i\right)\label{eIyy}\\[5pt]
I_{ixy}&=-\frac{1}{2}\Delta I_i\sin 2\theta_i.\label{eIxy}
\end{align}
Inserting Eqs.~(\ref{eIxx}-\ref{eIxy}) into Eq.~(\ref{eSigdisAppend}), and using $\kappa=k_d\mathcal{A}_i I_i/2$, we obtain the result for $\boldsymbol{\Sigma}_i^\mathrm{dis}$ stated earlier as Eq.~(\ref{eSigdisi}).

Using Eq.~(\ref{eSigdisi}) we then get  the net dissipative torque on particle $i$,
\begin{align}
\tau_i^\mathrm{dis}&=\int_i d^2r\,[xf_{iy}^\mathrm{dis}-yf_{ix}^\mathrm{dis}]
=\boldsymbol{\Sigma}_{ixy}^\mathrm{dis}-\boldsymbol{\Sigma}_{iyx}^\mathrm{dis}\\[5pt]
&=-k_d\mathcal{A}_iI_i\left[\dot\theta +\frac{\dot\gamma}{2}-\frac{\dot\gamma}{2}\frac{\Delta I_i}{I_i}\cos 2\theta_i \right],\label{e1234}
\end{align}
which is the same result stated earlier as Eq.~(\ref{etaudisShear}).

It is interesting to note that one can decompose a simple shear transformation into a pure shear plus a rotation,
$\boldsymbol{\dot\gamma}_\mathrm{ss}=\boldsymbol{\dot\gamma}_\mathrm{ps}+\boldsymbol{\dot\gamma}_\mathrm{rot}$,
\begin{equation}
\boldsymbol{\dot\gamma}_\mathrm{ss}=\left[
\begin{array}{cc}
0 & \dot\gamma    \\[10pt]
0 &  0   
\end{array}
\right]=\left[
\begin{array}{cc}
0 & \dot\gamma/2    \\[10pt]
\dot\gamma/2 &  0   
\end{array}
\right]+\left[\
\begin{array}{cc}
0 & \dot\gamma/2    \\[10pt]
-\dot\gamma/2 &  0   
\end{array}
\right].
\end{equation}
Here the first term $\boldsymbol{\dot\gamma}_\mathrm{ps}$ on the right corresponds to a pure shear with compression along the $(1,-1)$ diagonal and expansion along the $(1,1)$ diagonal, both at rate $\dot\gamma/2$ so as to keep the system area fixed; the second term $\boldsymbol{\dot\gamma}_\mathrm{rot}$ on the right  corresponds to a rotation with angular velocity $-(\dot\gamma/2)\mathbf{\hat z}$.
It is straightforward to show that it is the pure shear contribution $\boldsymbol{\dot\gamma}_\mathrm{ps}$ that gives the orientation dependent $\sim\cos 2\theta_i$ term in  Eq.~(\ref{e1234}), while it is the rotational contribution $\boldsymbol{\dot\gamma}_\mathrm{rot}$ that gives the constant driving term $\dot\gamma/2$. It is this term that results in a steady-state rotation of particles under simple shear, while there is no such steady-state rotation for a pure shear deformation.

For the spherocylinders of the present work, it is easiest to compute the moment of inertial tensor in a coordinate frame aligned with the spherocylinder spine and with origin at the center of mass.  In this frame $\mathbf{I}$ is diagonal, and so readily gives the eigenvalues $I_1$ and $I_2$.  Taking the spine as the direction of the $x$-axis,
\begin{equation}
I_1=\frac{1}{\mathcal{A}}\int d^2r\, y^2,\qquad I_2=\frac{1}{\mathcal{A}}\int d^2r\,x^2,
\end{equation}
where the integrals are over the area of the spherocylinder.  To do these integrals it is convenient to integrate  over the rectangular body and the semicircular endcaps separately.  For the rectangular part we have,
\begin{align}
\int\displaylimits_\mathrm{rectangle}\!\! \!\!\!d^2r\,y^2 &= \int_{-R}^{R}dy\int_{-A}^{A}dx\, y^2 =\frac{4R^3A}{3}\\
\int\displaylimits_\mathrm{rectangle}\!\! \!\!\!d^2r\,x^2 &= \int_{-R}^{R}dy\int_{-A}^{A}dx\, x^2 =  \frac{4RA^3}{3}.
\end{align}
To integrate over the end-caps we parametrize the coordinates $x$ and $y$ in terms of polar coordinates $s$ and $\varphi$ centered about the spine tip:  $x=A+s\cos\varphi$ and $y=s\sin\varphi$.  For one endcap we then have,
\begin{align}
\int\displaylimits_\text{end-cap}\!\!\!\!\!d^2r\,y^2 &= \int_{-\pi/2}^{\pi/2}\!\!\!d\varphi\int_0^R\!\!\!ds\, s (s\sin\varphi)^2
=\frac{\pi R^4}{8}\\[10pt]\nonumber
\int\displaylimits_\text{end-cap}\!\!\!\!\!d^2r\,x^2 &= \int_{-\pi/2}^{\pi/2}\!\!\!d\varphi\int_0^R\!\!\!ds\, s (A+s\cos\varphi)^2\\
&=\frac{\pi R^4}{8}+\frac{\pi R^2 A^2}{2} +\frac{4R^3 A}{3}.
\end{align}
Collecting terms, and noting that there are two end-caps, we then get,
\begin{align}
I_1&=\frac{4R^3A}{3\mathcal{A}}+\frac{\pi R^4}{4\mathcal{A}}\\[10pt]
I_2&=\frac{4RA^3}{3\mathcal{A}}+\frac{\pi R^4}{4\mathcal{A}}+\frac{\pi R^2 A^2}{\mathcal{A}} +\frac{8R^3 A}{3\mathcal{A}}.
\end{align}
Finally, using $\alpha=A/R$ and the spherocylinder area $\mathcal{A}=\pi R^2+4RA=(\pi +4\alpha)R^2$, we get,
\begin{align}
I&=I_1+I_2=\left[\frac{3\pi+24\alpha+6\pi\alpha^2+8\alpha^3}{6(\pi+4\alpha)}\right]R^2\\[10pt]
\Delta I&=I_2-I_1=\left[\frac{4+3\pi\alpha+4\alpha^2}{3(\pi+4\alpha)}\right]\alpha R^2.
\end{align}

\bibliographystyle{apsrev4-1}

\end{document}